\documentclass[12pt]{report}
\newcommand{\linha}{\enlargethispage{1\baselineskip}} 
\usepackage{epsfig}
\usepackage{amsfonts}
\usepackage{amssymb}
\usepackage{amsmath}
\usepackage{newcent}
\newcommand{\slb}[1]{\textbf{\textsl{#1}}}
 
\newcommand{\nin}{\noindent}
\def\text#1{{\scriptstyle\mathrm{#1}}} 
\renewcommand{\lesssim}{\stackrel{<}{\sim}}
\renewcommand{\gtrsim}{\stackrel{>}{\sim}}                              
\newcommand{\gc}{\gamma_5}
\newcommand{\g}{\gamma}
\newcommand{\sla}[1]{\not \!\! #1}
\newcommand{\ld}{\mbox{\sf L}}
\newcommand{\rs}{\mbox{\sf R}}
\newcommand{\lag}{{\cal L}}

\newcommand{\del}{\partial}
\newcommand{\dels}{\not \! \partial}
\newcommand{\et}{{\it et al.\/}}

\newcommand{\beqar}{\begin{eqnarray*}}
\newcommand{\eeqar}[1]{\label{#1} \end{eqnarray*}}            
\newcommand{\ba}{\begin{array}}
\newcommand{\ea}{\end{array}}
\newcommand{\rar}{$\longrightarrow$}
\newcommand{\lar}{$\longleftarrow$}
\newcommand{\uar}{$\uparrow$}
\newcommand{\dar}{$\downarrow$}
\newcommand{\ner}{$\nearrow$}
\newcommand{\nor}{$\nwarrow$}
\newcommand{\ser}{$\searrow$}
\newcommand{\sor}{$\swarrow$}
\newcommand{\btab}{\begin{tabular}{ll}}
\newcommand{\etab}{\end{tabular}}

\begin{document}
\thispagestyle{empty}

\hfill IFT--P/010/2000

\vskip 1.5cm    
\begin{center}
{\LARGE \slb{Standard Model: An Introduction}}
\footnote{To be published in {\it Particle and Fields}, Proceedings of the X
~J.\ A.\ Swieca Summer School (World Scientific, Singapore, 2000)}

\vskip 1.5cm
{\large\slb{S.\ F.\ Novaes}}

\vskip 0.8cm
\textsl{Instituto de F\'{\i}sica Te\'orica}

\textsl{Universidade  Estadual Paulista}

\textsl{Rua Pamplona 145, 01405--900, S\~ao Paulo}

\textsl{Brazil}
\end{center}
\vspace{2cm}

\begin{minipage}[h]{12cm}
\centerline{\large\bf Abstract}

\vspace{1.2cm}

\hskip 1cm 
We present a primer on the Standard Model of the electroweak
interaction. Emphasis is given to the historical aspects of the
theory's formulation.  The radiative corrections to the Standard
Model are presented and its predictions for the electroweak
parameters are compared with the precise experimental data obtained
at the $Z$ pole. Finally, we make some remarks on the perspectives
for the discovery of the Higgs boson, the most important challenge of
the Standard Model. 
\end{minipage}

\tableofcontents

\newpage
\chapter{Introduction} \indent

The joint description of the electromagnetic and the weak
interaction  by a single theory certainly is one of major
achievements of the physical science in this century. The model
proposed by Glashow, Salam and Weinberg in the middle sixties, has
been extensively tested during the last 30 years. The discovery of
neutral weak interactions and the production of intermediate vector
bosons ($W^\pm$ and $Z^0$) with the expected properties increased
our confidence in the model. Even after the recent precise
measurements of the electroweak parameters in electron--positron
collisions at the $Z^0$ pole, there is no experimental result that
contradicts the Standard Model predictions. 

The description of the electroweak interaction is implemented by a
gauge theory based on the $SU(2)_L \otimes U(1)_Y$ group, which is
spontaneously broken via the Higgs mechanism. The matter fields ---
leptons and quarks --- are organized in families, with the
left--handed fermions belonging to weak isodoublets while the
right--handed components transform as weak isosinglets. The vector
bosons, $W^\pm$, $Z^0$ and $\gamma$, that mediate the interactions
are introduced via minimal coupling to the matter fields. An
essential ingredient of the model is the scalar potential that is
added to the Lagrangian to generate the vector--boson (and fermion)
masses in a gauge invariant way, via the Higgs mechanism. A remnant
scalar field, the Higgs boson, is part of the physical spectrum. This
is the only missing piece of the Standard Model that still awaits
experimental confirmation.  

In this course, we intend to give a quite pedestrian introduction to
the main concepts involved in the construction of the Standard Model
of electroweak interactions. We should not touch any subject ``beyond
the Standard Model''. This primer should provide the necessary
background for the lectures on more advanced topics that were 
covered in this school, such as $W$ physics and extensions of the
Standard Model. A special emphasis will be given to the historical
aspects of the formulation of the theory. The interplay of new ideas 
and experimental results make the history of weak interactions a very
fruitful laboratory for understanding how the development of a
scientific theory works in practice. More formal aspects and details
of the model can be found in the vast literature on this subject,
from textbooks
\cite{Quigg:83,Cheng:84,Halzen:84,Aitchison:90,Renton:90,Donoghue:92,Leader:96}
to reviews \cite{Abers:73,Kim:81,Beg:82,Chanowitz:88}.  

We start these lectures  with a chronological account of the ideas
related to the development of electromagnetic and weak theories
(Section \ref{chrono}). The gauge principle (Sec.\ \ref{gau:pri})
and the concepts of spontaneous symmetry breaking (Sec.\ \ref{ssb})
and the Higgs mechanism (Section \ref{higgs}) are presented. In the
Chapter \ref{sm}, we introduce the Standard Model, following the
general principles that should guide the construction of a gauge
theory. We discuss topics like the mass matrix of the neutral bosons,
the measurement of the Weinberg angle, the lepton mass, anomaly
cancelation, and the introduction of quarks in the model. We finalize
this chapter giving an overview on the Standard Model Lagrangian in
Sec.\ \ref{sm:lag}. In Chapter \ref{loops}, we give an introduction
to the radiative corrections to the Standard Model. Loop
calculations are important to compare the predictions of the
Standard Model with the precise experimental results of $Z$ physics
that are presented in Sec.\ \ref{z:phy}. We finish our lectures with
an account on the most important challenge to the Standard Model: the
discovery of the Higgs boson. In Chapter \ref{higgs:bos}, we discuss
the main properties of the Higgs, like mass, couplings and decay
modes and discuss the phenomenological prospects for the search of the
Higgs in different colliders.

Most of the material covered in these lectures can be found in a
series of very good textbook on the subject.  Among them we can point
out the books from Quigg \cite{Quigg:83}, Aitchison and Hey
\cite{Aitchison:90}, and Leader and Predazzi \cite{Leader:96}.

\section{A Chronology of the Weak Interactions} \label{chrono} \indent

We will present in this section the main steps given towards a
unified description of the electromagnetic and weak interactions. In
order to give a historical flavor to the presentation, we will
mention some parallel achievements in Particle Physics in this
century, from theoretical developments and predictions to
experimental confirmation and surprises. The topics closely related
to the evolution and construction of the model will be worked with
more details.

\linha

The chronology of the developments and discoveries in Particle
Phy\-sics can be found in the books of Cahn and Goldhaber
\cite{Cahn:89} and the annotated bibliography from COMPAS and
Particle Data Groups \cite{AIP:96}. An extensive selection of
original papers on Quantum Electrodynamics can be found in the book
edited by Schwinger \cite{Schwinger:58}. Original papers on gauge
theory of weak and electromagnetic interactions appear in Ref.\
\cite{Lai:81}.

\vskip 1cm

\nin
\fbox{\slb{1896}} $^\star$\footnotetext{${}^{\star}$The ``star''
($\star$) means that the author(s) have received the Nobel Prize in
Physics for this particular work.}  Becquerel \cite{Becquerel:96}:
evidence for spontaneous radioactivity effect in uranium decay, using
photographic film.

\nin
\fbox{\slb{1897}} $^\star$ Thomson: discovery of the electron in cathode
rays.

\nin
\fbox{\slb{1900}} $^\star$ Planck: start of the quantum era.

\nin
\fbox{\slb{1905}} Einstein: start of the relativistic era. 

\nin
\fbox{\slb{1911}} $^\star$ Millikan: measurement of the electron charge.

\nin
\fbox{\slb{1911}} Rutherford: evidence for the atomic nucleus.

\nin 
\fbox{\slb{1913}} $^\star$ Bohr: invention of the quantum theory of atomic
spectra.

\nin
\fbox{\slb{1914}} Chadwick \cite{Chadwick:14}: first observation
that the $\beta$ spectrum is continuous. Indirect evidence on the
existence of neutral penetrating particles.

\nin
\fbox{\slb{1919}} Rutherford: discovery of the proton, constituent of the
nucleus.

\nin 
\fbox{\slb{1923}} $^\star$ Compton: experimental confirmation
that the photon is an elementary particle in $\gamma + C \to \gamma +
C$.

\linha

\nin
\fbox{\slb{1923}} $^\star$ de Broglie: corpuscular--wave dualism for electrons.

\nin
\fbox{\slb{1925}} $^\star$ Pauli: discovery of the exclusion principle.

\nin
\fbox{\slb{1925}} $^\star$ Heisenberg: foundation of quantum mechanics.

\nin
\fbox{\slb{1926}} $^\star$ Schr\"odinger: creation of wave quantum mechanics.

\nin
\fbox{\slb{1927}} Ellis and Wooster \cite{Ellis:27}: confirmation
that the $\beta$ spectrum is continuous.


\nin
\fbox{\slb{1927}} Dirac \cite{Dirac:27}: foundations of Quantum 
Electrodynamics (QED).

\nin
\fbox{\slb{1928}} $^\star$ Dirac: discovery of the relativistic wave
equation for electrons; prediction of the magnetic moment of the
electron.

\nin
\fbox{\slb{1929}} Skobelzyn: observation of cosmic ray showers
produced by energetic electrons in a cloud chamber.

\nin
\fbox{\slb{1930}} Pauli \cite{Pauli:30}: first proposal, in an open
letter, of the existence of a light, neutral and feebly interacting 
particle emitted in $\beta$ decay.

\nin
\fbox{\slb{1930}} Oppenheimer \cite{Oppenheimer:30}: self--energy of
the electron: the first ultraviolet divergence in QED.

\nin
\fbox{\slb{1931}} Dirac: prediction of the positron and anti--proton.

\nin
\fbox{\slb{1932}} $^\star$ Anderson: first evidence for the positron.

\nin
\fbox{\slb{1932}} $^\star$ Chadwick: first evidence for the neutron
 in $\alpha + Be \to C + n$.

\nin
\fbox{\slb{1932}} Heisenberg: suggestion that nuclei are composed of
protons and neutrons. 

\nin
\fbox{\slb{1934}} Pauli \cite{Pauli:34}: explanation of continuous electron
spectrum of $\beta$ decay --- proposal for the neutrino.
\[
n \to p + e^- + {\bar{\nu}_e} \; .
\]

\nin
\fbox{\slb{1934}} Fermi \cite{Fermi:34}: field theory for $\beta$
decay, assuming the existence of the neutrino. In analogy to ``the
theory of radiation that describes the emission of a quantum of light
from an excited atom'', $e J_\mu A^\mu$, Fermi proposed a
current--current Lagrangian to describe the $\beta$ decay:
\[ 
\lag_{\text{weak}} = \frac{G_F}{\sqrt{2}}
\left( \bar{\psi}_p \, \g_\mu \, \psi_n \right)
\left( \bar{\psi}_e \, \g^\mu \, \psi_\nu \right) \; .
\]


\nin
\fbox{\slb{1936}} Gamow and Teller \cite{Gamow:36}: proposed an
extension of the Fermi theory to describe also transitions with
$\Delta J^{\text{nuc}} \neq 0$. The vector currents proposed by Fermi
are generalized to:
\[
\lag_{\text{weak}} = \frac{G_F}{\sqrt{2}}
\sum_i C_i \left( \bar{\psi}_p \, \Gamma^i \, \psi_n \right)
\left( \bar{\psi}_e \, \Gamma^i \, \psi_\nu \right) \; ,
\]
with the scalar, pseudo--scalar, vector, axial and tensor structures:
\[
\Gamma^S = 1  \, , \;\;\;   
\Gamma^P = \gc \, , \;\;\;   
\Gamma^V_\mu = \g_\mu \, , \;\;\;   
\Gamma^A_\mu = \g_\mu \gc \, , \;\;\;   
\Gamma^T_{\mu\nu} = \sigma_{\mu\nu} \; .
\]
Nuclear transitions with $\Delta J = 0$ are described by the
interactions $S.S \;\; \mbox{and/or} \;\; V.V$, while $\Delta J = 0,
\pm 1 \; (0 \not \! \to 0)$ transitions can be taken into account by
$A.A \;\; \mbox{and/or} \;\; T.T$ interactions ($\Gamma^P \to 0$ in
the non--relativistic limit).  However, interference between them are
proportional to $m_e/E_e$ and should increase the emission of low energy
electrons. Since this behavior was not observed, the weak Lagrangian
should contain, 
\[
S.S \;\; \mbox{\underline{or}} \;\;\; V.V  \;\;\;
\mbox{\underline{and}} \;\;\; 
A.A \;\; \mbox{\underline{or}} \;\;\; T.T \; .
\]

\nin
\fbox{\slb{1937}} Neddermeyer and Anderson: first evidence for the muon.

\nin
\fbox{\slb{1937}} Majorana: Majorana neutrino theory.

\nin
\fbox{\slb{1937}} Bloch and Nordsieck \cite{Block:37}: treatment of
infrared divergences.

\nin
\fbox{\slb{1940}} Williams and Roberts \cite{Williams:40}: first
observation of muon decay
\[
\mu^- \to e^- + (\bar{\nu}_e + \nu_\mu) \; .
\]


\nin
\fbox{\slb{1943}} Heisenberg: invention of the S--matrix formalism.

\nin
\fbox{\slb{1943}} $^\star$ Tomonaga \cite{Tomonaga:43}: creation of
the covariant quantum electrodynamic theory.

\nin
\fbox{\slb{1947}} Pontecorvo \cite{Pontecorvo:47}: first idea about
the universality of the Fermi weak interactions {\it i.e.} decay and 
capture processes have the same origin.

\nin
\fbox{\slb{1947}} Bethe \cite{Bethe:47}: first theoretical
calculation of the Lamb shift in non--relativistic QED.

\nin
\fbox{\slb{1947}} $^\star$  Kusch and Foley \cite{Kusch:47}: first
measurement of $g_e-2$ for the electron using the Zeeman effect: $g_e =
2 (1 + 1.19 \times 10^{-3})$.

\nin
\fbox{\slb{1947}} $^\star$ Lattes, Occhialini and Powell:
confirmation of the $\pi^-$ and first evidence for pion decay
$\pi^\pm \to \mu^\pm + (\nu_\mu)$.

\nin
\fbox{\slb{1947}} Rochester and Butler: first evidence for $V$
events (strange particles).

\nin
\fbox{\slb{1948}} Schwinger \cite{Schwinger:48a}: first theoretical
calculation of $g_e-2$ for the electron: $g_e = 2 (1 + \alpha/2\pi) = 
2 (1 + 1.16 \times 10^{-3})$. The high--precision measurement of the
anomalous magnetic moment of the electron is the most stringent QED test.
The present theoretical and experimental value of $a_e = (g_e - 2)/2$,
are \cite{pdg:98},
\begin{eqnarray*}
a_e^{\text{thr}}&=& (115 \; 965 \; 215.4 \pm 2.4) \times 10^{-11}\; ,
\\ 
a_e^{\text{exp}}&=& (115 \; 965 \; 219.3 \pm 1.0) \times 10^{-11}\; ,
\end{eqnarray*}
where we notice the impressive agreement at the 9 digit level!

\nin
\fbox{\slb{1948}} $^\star$ Feynman \cite{Feynman:48}; Schwinger
\cite{Schwinger:48b}; Tati and Tomonaga \cite{Tati:48}:
creation of the covariant theory of QED. 

\nin
\fbox{\slb{1949}} Dyson \cite{Dyson:49}: covariant QED and
equivalence of Tomonaga, Sch\-winger and Feynman methods. 


\nin
\fbox{\slb{1949}} Wheeler and Tiomno \cite{Wheeler:49};  Lee,
Rosenbluth and Yang \cite{Lee:49}: proposal of the universality of
the Fermi weak interactions. Different processes like,
\begin{eqnarray*}
\beta-\mbox{decay} \; &:& \; n \to p + e^- + \bar{\nu}_e  \; ,\\
\mu-\mbox{decay}   \; &:& \; \mu^- \to e^- + \bar{\nu}_e + \nu_\mu \; , \\ 
\mu-\mbox{capture} \; &:& \; \mu^- + p \to \nu_\mu + n \; ,
\end{eqnarray*}
must have the same nature and should share the same coupling
constant, 
\[
G_F = \frac{1.03 \times 10^{-5}}{M_p^2} \; ,
\]
the so--called Fermi constant.

\nin
\fbox{\slb{50's}} A large number of new particles where discovered in
the 50's: $\pi^0$,  $K^\pm$, $\Lambda$, $K^0$, $\Delta^{++}$,
$\Xi^-$, $\Sigma^\pm$,  $\bar{\nu}_e$, $\bar{p}$, $K_{L,S}$,
$\bar{n}$, $\Sigma^0$, $\bar{\Lambda}$, $\Xi^0$,  $\cdots$

\nin
\fbox{\slb{1950}} Ward \cite{Ward:50}: Ward identity in QED.

\nin
\fbox{\slb{1953}} St\"uckelberg; Gell--Mann: invention and
exploration of renormalization group.

\nin
\fbox{\slb{1954}} Yang and Mills \cite{Yang:54}: introduction of
local gauge isotopic invariance in quantum field theory. This was one
of the key theoretical developments that lead to the invention of
non--abelian gauge theories.

\nin
\fbox{\slb{1955}} Alvarez and Goldhaber \cite{Alvarez:55}; Birge \et\
\cite{Birge:55}:
$\theta-\tau$ puzzle: The ``two'' particles seem to be a single state
since they have the  same width ($\Gamma_\theta = \Gamma_\tau$), and
the same mass ($M_\theta = M_\tau$). However the observation of
different decay modes, into states with opposite parity: 
\begin{eqnarray*}
\theta^+ & \to & \pi^+ + \pi^0 \; , \;\;\;\;\;\;\;\;\;\;\;  J^P = 0^+
\; , 
\\
\tau^+   & \to & \pi^+ + \pi^+ + \pi^-  \; , \;\;   J^P = 0^- \; , 
\end{eqnarray*}
suggested that parity could be violated in weak transitions.

\nin
\fbox{\slb{1955}} Lehmann, Symanzik and Zimmermann: beginnings
of the axiomatic field theory of the S--matrix.

\nin
\fbox{\slb{1955}} Nishijima: classification of strange
particles and prediction of $\Sigma^0$ and $\Xi^0$.


\nin
\fbox{\slb{1956}} $^\star$ Lee and Yang \cite{Lee:56}: proposals to
test spatial parity conservation in weak interactions.

\nin
\fbox{\slb{1957}} Wu \et\ \cite{Wu:57}: obtained the first evidence
for parity nonconservation in weak decays. They measured the angular
distribution of the electrons in $\beta$ decay,
\[
^{60}\mbox{Co} \; (\mbox{polarized}) \to \; 
^{60}\mbox{Ni} + e^- + \bar{\nu}_e \; ,
\]
and observed that the decay rate depend on the pseudo--scalar
quantity: $<\vec{J}_{\text{nuc}}> . \; \vec{p}_e$.

\nin
\fbox{\slb{1957}} Garwin, Lederman and Weinrich \cite{Garwin:57};
Friedman and Telegdi \cite{Friedman:57}: confirmation of parity
violation in weak decays. They make the measurement of the electron
asymmetry (muon polarization) in the decay chain, 
\begin{eqnarray*}
\pi^+  & \to & \mu^+  + \nu_\mu \\
       && \;\; \hookrightarrow e^+ + \nu_e + \bar{\nu}_\mu \; .
\end{eqnarray*}

\nin
\fbox{\slb{1957}} Frauenfelder \et\ \cite{Frauenfelder:57}: further
confirmation of parity nonconservation in weak decays. The
measurement of the longitudinal polarization of the electron
($\vec{\sigma}_e . \vec{p}_e$) emitted in $\beta$ decay,
\[
^{60}\mbox{Co} \to \; e^- \; (\mbox{long.\ polar.}) + \bar{\nu}_e + X
\; , 
\]
showed that the electrons emitted in weak transitions are mostly
left--handed.

The confirmation of the parity violation by the weak interaction
showed that it is necessary to have a term containing a $\gc$ in the
weak current: 
\[
\lag_{\text{weak}} \; \to \; \frac{G_F}{\sqrt{2}}
\sum_i C_i \left( \bar{\psi}_p \, \Gamma^i \, \psi_n \right)
\left[ \bar{\psi}_e \, \Gamma^i \, (1 \pm \gc) \, \psi_\nu \right] 
\; .
\]

Note that $CP$ remains conserved since $C$ is also violated.


\nin
\fbox{\slb{1957}} Salam \cite{Salam:57} ; Lee and Yang
\cite{Lee:57a}; Landau \cite{Landau:57}: two--component theory of
neutrino. This requires that the neutrino is either right
\underline{or} left--handed. 

Since it was known that electrons (positrons) involved in weak
decays are left (right) handed, the leptonic current should be written
as:
\[
J^i_{\text{lept}} \equiv 
\left[ \bar{\psi}_e \, \Gamma^i \, (1 \pm \gc) \, \psi_\nu \right] \to 
\left[ \bar{\psi}_e \, \frac{(1 + \gc)}{2} \, \Gamma^i \, (1 \pm \gc) \, 
\psi_\nu \right] \; .
\]

Therefore the measurement of the neutrino helicity is crucial to
determine the structure of the weak current. If $\Gamma^i$ = $V$  or
$A$ then $\{\gc, \Gamma^i\} =0$ and the neutrino should be
left--handed, otherwise the current is zero.  On the other hand, if
$\Gamma^i$ = $S$ or $T$, then $[\gc, \Gamma^i] = 0$, and the neutrino
should be right--handed. 

\nin
\fbox{\slb{1957}} Schwinger \cite{Schwinger:57}; Lee and Yang
\cite{Lee:57b}: development of the idea of the intermediate vector
boson in weak interaction. The four--fermion Fermi interaction is 
``point--like'' {\it i.e.} a $s$--wave interaction.  Partial wave
unitarity requires that such interaction must give rise to a cross
section that is bound by $\sigma < 4\pi / p^2_{\text{cm}}$. However,
since $G_F$ has dimension of $M^{-2}$, the cross section for the
Fermi weak interaction should go like $\sigma \sim G_F^2
p^2_{\text{cm}}$. Therefore the Fermi theory violates unitarity for
$p_{\text{cm}} \simeq 300$ GeV. 

This violation can be delayed by imposing that the interaction is
transmitted by a intermediate vector boson (IVB) in analogy, once
again, with the quantum electrodynamics. Here, the IVB should have
quite different characteristics, due to the properties of the weak
interaction. The IVB should be charged since the $\beta$ decay
requires charge--changing currents. They should also be very massive
to account for short range of the weak interaction and they should not
have a definite parity to allow, for instance, a $V-A$ structure for
the weak current.

With the introduction of the IVB, the Fermi Lagrangian for leptons,
\[
\lag_{\text{weak}} = \frac{G_F}{\sqrt{2}} 
\left[ J^\alpha (\ell) J_\alpha^\dagger (\ell^\prime) + \mbox{h.c.} \right]
\; , 
\]
where $J^\alpha (\ell) = \bar{\psi}_{{\nu}_\ell} \Gamma^\alpha \psi_\ell$, 
becomes:
\begin{equation}
\lag_{\text{weak}}^W = G_W 
\left( J^\alpha W_\alpha^+ +  J^{\dagger\,\alpha} W_\alpha^- \right)
\; ,
\label{l:w}
\end{equation}
with a new coupling constant $G_W$.

Let us compare the invariant amplitude for $\mu$--decay, in the
low--energy limit in both cases. For the Fermi Lagrangian, we have,
\begin{equation}
{\cal M}_{\text{weak}} = i
{ \frac{G_F}{\sqrt{2}} } J^{\alpha}{(\mu)} J_{\alpha} (e) \; .
\label{amp:fer}
\end{equation}

On the other hand, when we take into account the exchange of the IVB,
the invariant amplitude should include the vector boson propagator,
\[
{\cal M}_{\text{weak}}^W = \left[ i \, G_W J^{\alpha}{(\mu)}
\right]
\left[\frac{-i}{k^2 - M_W^2}\left( g_{\alpha\beta} -
\frac{k_\alpha k_\beta}{M_W^2}\right)
\right] \left[ i \, G_W J^{\beta}{(e)} \right] \; .
\]
At low energies, {\it i.e.} for $k^2 \ll M_W^2$,
\begin{equation}
{\cal M}_{\text{weak}}^W \; \longrightarrow \; 
i \frac{G_W^2}{M_W^2} J^{\alpha}{(\mu)} J_{\alpha} (e) \; , 
\label{amp:ivb}
\end{equation}
and, comparing (\ref{amp:ivb}) with (\ref{amp:fer}) we obtain the relation
\begin{equation}
\fbox{$ \; \displaystyle{G_W^2 = \frac{M_W^2 G_F}{\sqrt{2}}} \; $} \; ,
\label{gw}
\end{equation}
which shows that $G_W$ is dimensionless.

However, at high energies, the theory of IVB still violates unitarity,
for instance, in the cross section for  $\nu \bar{\nu} \to W^+ W^-$
(see Fig.\ 1).

Let us consider the $W^\pm$ polarization states. At the $W^\pm$ rest
frame, we can define the transversal and longitudinal polarizations as
\begin{eqnarray*} 
\epsilon^\mu_{T_1} (0) = (0,1,0,0) \; , \\
\epsilon^\mu_{T_2} (0) = (0,0,1,0) \; , \\
\epsilon^\mu_{L} (0)   = (0,0,0,1) \; . 
\end{eqnarray*}

\begin{figure}[ht]
\protect
\epsfxsize=10cm
\begin{center}
\centerline{\epsfig{figure=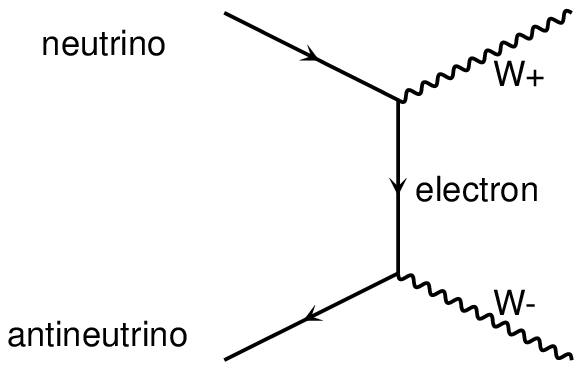,width=1.2\textwidth}}
\end{center}
\end{figure}
\nobreak 
\begin{minipage}[h]{12cm}
\begin{center}
\centerline{\it Fig.\ 1: Feynman diagram for the process $\nu +
\bar{\nu} \to W^+ + W^-$.}
\end{center}
\end{minipage}

After a boost along the $z$ direction, {\it i.e.} for $p^\mu = (E,
0,0, p)$, the transversal states remain unchanged while the
longitudinal state becomes,
\[
\epsilon^\mu_{L} (p) = \left(\frac{|\vec{p}|}{M_W},
\frac{E}{M_W}\hat{p} \right) 
\simeq \frac{p^\mu}{M_W} \; .
\]
Since the longitudinal polarization is proportional to the vector
boson momentum, at high energies the longitudinal amplitudes should
give rise to the worst behavior. 

In fact, in high energy limit, the polarized cross section for $\nu
\bar{\nu} \to W^+ W^-$ behaves like,
\begin{eqnarray*}
\sigma (\nu\bar{\nu} \to W^+_T W^-_T) & \longrightarrow & 
\mbox{constant}
\\
\sigma (\nu\bar{\nu} \to W^+_L W^-_L) & \longrightarrow & 
\frac{G_F^2 s}{3 \pi} \; ,
\end{eqnarray*}
which still violates unitarity for large values of $s$.

\nin
\fbox{\slb{1958}} Feynman and Gell--Mann \cite{Feynman:58};  Marshak
and Sudarshan \cite{Marshak:58}; Sakurai \cite{Sakurai:58}: universal
$V-A$ weak interactions.
\begin{equation}
J^{+ \; \mu}_{\text{lept}} = 
\left[ \bar{\psi}_e \, \g^\mu (1 - \gc) \,  \psi_\nu \right] \; .
\label{wea:cur}
\end{equation}

\nin
\fbox{\slb{1958}} Leite Lopes \cite{Lopes:58}: hypothesis of neutral
vector mesons exchanged in weak interaction. Prediction of its mass of
$\sim 60 \; m_{\text{proton}}$.

\nin
\fbox{\slb{1958}} Goldhaber, Grodzins and Sunyar \cite{Goldhaber:58}:
first evidence for the negative $\nu_e$ helicity. As mentioned
before, this result requires that the structure of the weak
interaction is $V-A$.

\nin
\fbox{\slb{1959}} $^\star$ Reines and Cowan: confirmation of the
detection of the $\bar{\nu}_e$ in $\bar{\nu}_e + p \to e^+ + n$.

\nin
\fbox{\slb{1961}} Goldstone \cite{Goldstone:61}: prediction of
unavoidable massless bosons if global symmetry of the Lagrangian is
spontaneously broken.

\nin
\fbox{\slb{1961}} Salam and Ward \cite{Salam:61}: invention of the
gauge principle as basis to construct quantum field theories of
interacting fundamental fields.

\nin
\fbox{\slb{1961}} $^\star$ Glashow \cite{Glashow:61}: first
introduction of the neutral intermediate weak boson ($Z^0$).

\nin
\fbox{\slb{1962}} $^\star$ Danby \et: first evidence of $\nu_\mu$
from $\pi^\pm \to \mu^\pm + (\nu/\bar{\nu})$.

\nin
\fbox{\slb{1963}} Cabibbo \cite{Cabibbo:63}: introduction of the
Cabibbo angle and hadronic weak currents.

It was observed experimentally that weak decays with change of
strangeness ($\Delta s = 1$) are strongly suppressed in nature.
For instance, the width of the neutron is much larger than the
$\Lambda$'s,
\[
\Gamma_{\Delta s = 0}  \left( n_{udd} \to p_{uud} \; e  \bar{\nu}
\right) \gg
\Gamma_{\Delta s = 1}  \left( \Lambda _{uds} \to p_{uud}\; e 
\bar{\nu} \right) \; ,
\]
which yield a branching ratio of 100\% in the case of neutron and
just $\sim 8 \times 10^{-4}$ for the $\Lambda$.  

The hadronic current, in analogy with leptonic current
(\ref{wea:cur}), can be written in terms of the $u$, $d$, and $s$ quarks,
\begin{equation}
J_\mu^H =  \bar{d}  \g_\mu (1 - \gc) u  +  \bar{s} \g_\mu (1 - \gc) u
\; ,
\label{had:cur:0}
\end{equation}
where the first term is responsible for the $\Delta s =0$ transitions
while the latter one gives rise to the $\Delta s =1$ processes. In order
to make the hadronic current also universal, with a common coupling
constant $G_F$, Cabibbo introduced a mixing angle to give the right
weight to the $\Delta s =0$ and $\Delta s =1$ parts of the hadronic current, 
\begin{eqnarray}
\left( \ba{c}
{ d^\prime}\\
{ s^\prime}
\ea \right) & = &
\left( \ba{cc}
\cos\theta_C & \sin\theta_C \\
-\sin\theta_C & \cos\theta_C
\ea \right) 
\left( \ba{c}
{ d} \\
{ s}
\ea \right)  \; ,
\label{cab:ang}
\end{eqnarray}
where $d^\prime$, $s^\prime$ ($d$, $s$) are interaction (mass)
eigenstates. Now the transition $\bar{d}  \leftrightarrow u$ is
proportional to $G_F \cos\theta_C \simeq 0.97 \; G_F$ and  the 
$\bar{s}  \leftrightarrow u$ goes like $G_F \sin\theta_C \simeq 0.24
\; G_F$.

The hadronic current should now be given in terms of the new
interaction eigenstates,
\begin{eqnarray}
J_\mu^H &=& \bar{d} {}^\prime  \g_\mu (1 - \gc) u 
\nonumber \\
&=& \cos\theta_C \; \bar{d} \g_\mu (1 - \gc) u  + 
    \sin\theta_C \; \bar{s} \g_\mu (1 - \gc) u \; .
\label{had:cur}
\end{eqnarray}

\nin
\fbox{\slb{1964}} Bjorken and Glashow \cite{Bjorken:64}: proposal for
the existence of a charm\-ed fundamental fermion ($c$). 

\nin
\fbox{\slb{1964}} Higgs \cite{Higgs:64}; Englert and Brout
\cite{Englert:64}; Guralnik, Hagen and Kibble \cite{Guralnik:64}:
example of a field theory with spontaneous symmetry breakdown, no
massless Goldstone boson, and massive vector boson.

\nin
\fbox{\slb{1964}} $^\star$ Christenson, Cronin, Fitch and Turlay
\cite{Christenson:64}: first evidence of CP violation in the decay of
$K^0$ mesons.

\nin
\fbox{\slb{1964}} $^\star$ Salam and Ward \cite{Salam:64}: Lagrangian
for the electroweak synthesis, estimation of the $W$ mass.

\nin
\fbox{\slb{1964}} $^\star$ Gell--Mann; Zweig: introduction of quarks
as fundamental building blocks for hadrons.

\nin
\fbox{\slb{1964}} Greenberg; Han and Nambu:  introduction of color quantum
number and colored quarks and gluons.


\nin
\fbox{\slb{1967}} Kibble \cite{Kibble:67}: extension of the Higgs
mechanism of mass generation for non--abelian gauge field theories.

\nin
\fbox{\slb{1967}} $^\star$ Weinberg \cite{Weinberg:67}: Lagrangian for the
electroweak synthesis and estimation of $W$ and $Z$ masses.

\nin
\fbox{\slb{1967}} Faddeev and Popov \cite{Faddeev:67}: method for
construction of Feynman rules for Yang--Mills gauge theories.

\nin
\fbox{\slb{1968}} $^\star$ Salam \cite{Salam:68}: Lagrangian for the
electroweak synthesis.

\nin
\fbox{\slb{1969}} Bjorken: invention of the Bjorken scaling behavior.

\nin
\fbox{\slb{1969}} Feynman: birth of the partonic picture of hadron
collisions.

\nin
\fbox{\slb{1970}} Glashow, Iliopoulos and Maiani \cite{Glashow:70}:
introduction of lepton-- quark symmetry and the proposal of
charmed quark (GIM mechanism).

\nin
\fbox{\slb{1971}} $^\star$ 't Hooft \cite{Hooft:71}: rigorous proof
of renormalizability of the massless and massive Yang-- Mills quantum
field theory with spontaneously broken gauge invariance.

\nin
\fbox{\slb{1973}} Kobayashi and Maskawa \cite{Kobayashi:73}: CP
violation is  accommodated in the Standard Model with six favours.

\nin
\fbox{\slb{1973}} Hasert \et\ (CERN) \cite{Hasert:73a}: first experimental
indication of the existence of weak neutral currents.
\[
\bar{\nu}_\mu + e^-   \to   \bar{\nu}_\mu + e^-
\;\; , \;\;\;\; 
\nu_\mu + N   \to   \nu_\mu + X \; .
\]
This was a dramatic prediction of the Standard Model and its discovery
was a major success for the model. They also measured the ratio of
neutral--current to charged--current events giving a estimate for the
Weinberg angle $\sin^2\theta_W$ in the range 0.3 to 0.4.

\nin
\fbox{\slb{1973}} Gross and Wilczek; Politzer: discovery of
asymptotic freedom property of interacting  Yang--Mills field
theories.

\nin
\fbox{\slb{1973}} Fritzsch, Gell--Mann and Leutwyler: invention of the QCD
Lagrangian.


\nin
\fbox{\slb{1974}} Benvenuti \et\ (Fermilab) \cite{Benvenuti:74}:
confirmation of the existence of weak neutral currents in the reaction
\[
\nu_\mu + N \to \nu_\mu + X \; .
\]

\nin
\fbox{\slb{1974}} $^\star$ Aubert \et\ (Brookhaven); Augustin \et\
(SLAC): evidence for the $J/\psi$ ($c\bar{c}$).

\nin
\fbox{\slb{1975}} $^\star$ Perl \et\ (SLAC) \cite{Perl:75}: first
indication of the $\tau$ lepton.

\nin
\fbox{\slb{1977}} Herb \et\ (Fermilab) \cite{Herb:77}: first evidence
of $\Upsilon$ ($b\bar{b}$).

\newpage
\nin
\fbox{\slb{1979}} Barber \et\ (MARK J Collab.); Brandelik \et\ (TASSO
Collab.); Berger \et\ (PLUTO Collab.); W. Bartel (JADE Collab.):
evidence for the gluon jet in $e^+e^- \to 3$ jet.

\nin
\fbox{\slb{1983}} $^\star$ Arnison \et\ (UA1 Collab.) \cite{Arnison:83a};
Banner \et\ (UA2 Collab.) \cite{Banner:83}:  evidence for the charged
intermediate bosons $W^\pm$ in the reactions 
\[
p + \bar{p} \to W  (\to \ell + \nu) + X \; .
\]
They were able to estimate the $W$ boson mass ($M_W = 81 \pm 5$ GeV)
in good agreement with the predictions of the Standard Model.

\nin \fbox{\slb{1983}} $^\star$ Arnison \et\  (UA1 Collab.)
\cite{Arnison:83b};  Bagnaia \et\ (UA2 Collab.) \cite{Bagnaia:83}:
evidence for the neutral intermediate boson $Z^0$ in the reaction
\[
p + \bar{p} \to Z  (\to \ell^+ + \ell^-) + X \; .
\]
This was another important confirmation of the electroweak theory.

\nin
\fbox{\slb{1986}} $^\star$ Van Dyck, Schwinberg and Dehmelt
\cite{Dyck:86}: high precision measurement of the electron $g_e-2$
factor.

\nin
\fbox{\slb{1987}} Albrecht \et\ (ARGUS Collab.) \cite{Albrecht:87}:
first evidence of $B^0-\bar{B}^0$ mixing.

\nin
\fbox{\slb{1989}} Abrams \et\ (MARK-II Collab.) \cite{Abrams:89}:
first evidence that the number of light neutrinos is 3.

\nin  
\fbox{\slb{1992}} LEP Collaborations (ALEPH, DELPHI, L3 and OPAL)
\cite{Lep:92}: precise determination of the $Z^0$ parameters.

\nin
\fbox{\slb{1995}} Abe \et\ (CDF Collab.) \cite{Abe:95}; Abachi \et\
(D\O\ Collab.) \cite{Abachi:95}: observation of the top quark
production.


\newpage
\section{The Gauge Principle} 
\label{gau:pri}
\indent

As it is well known, symmetry has always played a very important
r\^ole in the development of physics. From the spacetime symmetry of
special relativity, up to the internal and gauge invariances, the
symmetries have mapped out the route to most of the physical theories
in this last century.  

An important result for field theory and particle physics is provided
by the Noether's theorem. If an action is invariant under some group
of transformations (symmetry), then there exist one or more conserved
quantities (constants of motion) which are associated to these
transformations. In this sense, Noether's theorem establishes that
symmetries imply conservation laws. 

A natural question to ask would be: upon imposing to a given
Lagrangian the invariance under a certain symmetry, would it be
possible to  determine the form of the interaction among the
particles? In other words, could symmetry also imply dynamics?

In fact, this happens in Quantum Electrodynamics (QED), the best
theory ever built to describe Nature, which had become a prototype
of a successful quantum field theory. In QED the existence and some of
the properties of the gauge field --- the photon ---  follow from a
principle of invariance under {\it local gauge transformations} of
the $U(1)$ group.

Could this principle be generalized to other interactions? For Salam
and Ward \cite{Salam:61}, who invented the gauge principle as the
basis to construct the quantum field theory of interacting fields,
this was a possible dream:

\begin{center}
\begin{minipage}[h]{12cm}
{\it ``Our basic postulate is that it should be possible to generate
strong, weak and electromagnetic interaction terms (with all their
correct symmetry properties and also with clues regarding their
relative strengths) by making local gauge transformations on the
kinetic--energy terms in the free Lagrangian for all particles.''} 
\end{minipage}
\end{center}

In fact, those ideas could be accomplished just after some new and
important ingredients were introduced to describe short distance
(weak) and strong interactions. In the case of weak interactions the
presence of very heavy weak gauge bosons require the new concept of
spontaneous breakdown of the gauge symmetry and the Higgs mechanism
\cite{Higgs:64,Englert:64,Guralnik:64}. On the other hand, the
concept of asymptotic freedom \cite{Gross:73,Politzer:73} played a
crucial r\^ole to describe perturbatively the strong interaction at
short distances, making the strong gauge bosons trapped. The Quantum
Chromodynamics (QCD), the gauge theory for strong interactions, is
the subject of Mangano's lecture at this school.


\subsection{Gauge Invariance in Quantum Mechanics} \indent

The gauge principle and the concept of gauge invariance are already
present in Quantum Mechanics of a particle in the presence of an
electromagnetic field \cite{Aitchison:90}. Let us start from the
classical Hamiltonian that gives rise to the Lorentz force ($\vec{F}
= q \vec{E} + q \vec{v}\times\vec{B}$),
\begin{equation}
{\cal H} =  \frac{1}{2m} \left(\vec{p} - q \vec{A}\right)^2 + q \phi
\; ,
\label{h:em}
\end{equation}
where the electric and magnetic fields can be described in terms of 
the potentials  $A^\mu = (\phi, \vec{A})$, 
\[
\vec{E} = - \vec{\nabla} \phi - \frac{\del \vec{A}}{\del t} ;\;,
\;\;\;\;
\vec{B} = \vec{\nabla} \times \vec{A} \; .
\]
These fields remain exactly the same when we make the {\it gauge
transformation} ($G$) in the potentials:
\begin{equation}
{ \phi \to \phi^\prime = \phi - \frac{\del \chi}{\del t}
\;\;, \;\;\;\;
\vec{A} \to \vec{A}^\prime = \vec{A} + \vec{\nabla} \chi } \; .
\label{gauge:em}
\end{equation}

When we quantize the Hamiltonian (\ref{h:em}) by applying the
usual prescription $\vec{p} \to -i \vec{\nabla}$, we get the
Schr\"odinger equation for a particle in an electromagnetic field,
\[
\left[ \frac{1}{2m} \left(-i \vec{\nabla} - q \vec{A}\right)^2 
+ q \phi \right] \psi(x,t) = i \frac{\del \psi(x,t)}{\del t} \; ,
\]
which can be written in a compact form as
\begin{equation}
\frac{1}{2m} (-i \vec{D})^2  \psi = i D_0 \psi \; ,
\label{schr}
\end{equation}
The equation (\ref{schr}) is equivalent to make the substitution
\[
\vec{\nabla} \to \vec{D} = \vec{\nabla} - i q \vec{A}
\;\;, \;\;\;\;
\frac{\del}{\del t} \to D_0 = \frac{\del}{\del t} + i q
\phi \; .
\]
in the free Schr\"odinger equation.

If we make the gauge transformation, $(\phi, \vec{A})
\stackrel{G}{\longrightarrow} (\phi^\prime, \vec{A}^\prime)$, given
by (\ref{gauge:em}), does the new field $\psi^\prime$ which is
solution of
\[
\frac{1}{2m} (-i \vec{D}^\prime)^2 \; \psi^\prime = i D_0^\prime \;
\psi^\prime  \;\; ,
\]
describe the same physics?

The answer to this question is {\it no}. However, we can recover the
invariance of our theory by making, at the same time, the phase 
transformation in the matter field
\begin{equation}
 \psi^\prime  = \exp\left(i q \chi \right) \psi 
\label{gauge:psi}
\end {equation}
with the same function $\chi = \chi (x,t)$ used in the transformation
of electromagnetic fields (\ref{gauge:em}). The derivative of
$\psi^\prime$ transforms as,
\begin{eqnarray}
 \vec{D}^\prime \psi^\prime &=& 
\left[\vec{\nabla} - i q ( \vec{A}  + \vec{\nabla}
\chi ) \right] \exp\left(i q \chi\right)  \psi 
\nonumber \\
&=& \exp\left(i q \chi\right) ( \vec{\nabla} \psi )
{ + i q (\vec{\nabla} \chi ) \exp\left(i q \chi\right)\psi }
\nonumber\\
&&- i q  \vec{A} \exp\left(i q \chi\right) \psi  
{ - i q (\vec{\nabla} \chi ) \exp\left(i q \chi\right) \psi }
\nonumber \\
&=&   \exp\left(i q \chi\right) \vec{D} \psi \; ,
\label{gauge:d}
\end{eqnarray}
and in the same way, we have for $D_0$,
\begin{equation}
 D_0^\prime \psi^\prime = 
\exp\left(i q \chi\right) D_0 \psi  \; .
\label{gauge:d0}
\end{equation}

We should mention that now the field $\psi$ (\ref{gauge:psi}) and its
derivatives $\vec{D}\psi$ (\ref{gauge:d}), and $D_0 \psi$
(\ref{gauge:d0}), all transform exactly in the same way: they
are all multiplied by the same phase factor.

Therefore, the Schr\"odinger equation (\ref{schr}) for $\psi^\prime$
becomes 
\begin{eqnarray*}
{ \frac{1}{2m} (-i\vec{D}^\prime)^2  \psi^\prime } &=& 
\frac{1}{2m} (-i \vec{D}^\prime) ( -i \vec{D}^\prime \psi^\prime) 
\\
&=&\frac{1}{2m} (-i \vec{D}^\prime) 
\left[ -i \exp\left(i q \chi\right) \vec{D} \psi \right] 
\\
&=& \exp\left(i q \chi\right) { \frac{1}{2m} (-i \vec{D})^2 \psi }
\\
&=& 
\exp\left(i q \chi\right) { ( i D_0 ) \psi} = 
{ i D_0^\prime \psi^\prime} \; .
\end{eqnarray*}
and now both $\psi$ and $\psi^\prime$ describe the same physics,
since $|\psi|^2 = |\psi'|^2$. In order to get the invariance for all
observables, we should assure that the following substitution is made:
\[
\vec{\nabla} \to \vec{D}  \;\;, \;\;\;\; \frac{\del}{\del t} \to D_0
\; ,
\] 
For instance, the current
\[
\vec{J} \propto \psi^* (\vec{\nabla} \psi) - (\vec{\nabla} \psi)^* \psi
\; , 
\]
becomes also gauge invariant with this substitution since
\[
\psi^{* \, \prime} (\vec{D}^\prime \psi^\prime) = 
\psi^* \exp\left(-i q \chi\right) 
\exp\left(i q \chi\right) (\vec{D} \psi) = \psi^* (\vec{D} \psi)
\; .
\]

After we have shown how to obtain a gauge invariant quantum
description of a particle in an electromagnetic field, could we
reverse the argument? That is: when we demand that a theory is
invariant under a spacetime dependent phase transformation, can
this procedure impose the specific form of the interaction with the
gauge field? In other words, can the {\it symmetry imply dynamics}?

Let us examine what happens when we start from the Dirac free Lagrangian
\[
\lag_\psi = \bar{\psi} (i \dels - m) \psi \; , 
\]
that is not invariant under the {\it local gauge transformation}, 
\[
\psi \to \psi^\prime  = \exp\left[-i \alpha(x) \right] \psi \; ,
\]
since
\[
\lag_\psi \to \lag_\psi^\prime = \lag_\psi 
{ \; + \; \bar{\psi} \g_\mu \psi (\del^\mu \alpha)} \; ,
\]

However, if we introduce the {\it gauge field} $A_\mu$ through
the {\it minimal coupling}
\[
D_\mu \equiv \del_\mu + i e A_\mu \; ,
\]
and, at the same time, require that $A_\mu$ transforms like
\begin{equation}
A_\mu \to A_\mu^\prime =  A_\mu + \frac{1}{e} \del_\mu \alpha \; .
\label{gauge:a}
\end{equation}
we have
\begin{eqnarray}
\lag_\psi \to \lag_\psi^\prime &=& 
\bar{\psi}^\prime  \left[(i \dels  - e \sla{A}^\prime) - m \right] \psi^\prime 
\nonumber \\
&=& \bar{\psi} \exp(+i \alpha )  
\left[i \dels  - e \left(\sla{A} + \frac{1}{e}\dels\alpha\right) - m\right]
\exp(-i \alpha) \psi 
\nonumber \\
&=& \lag_\psi \; {  - \; e \bar{\psi} \g_\mu \psi A^\mu} \; .
\label{l:psi}
\end{eqnarray}
The coupling between $\psi$ ({\it e.g.} electrons) and the gauge
field $A_\mu$ (photon) arises naturally when we require the
invariance under local gauge transformations of the kinetic--energy
terms in the free fermion Lagrangian. 

Since, the electromagnetic strength tensor 
\begin{equation}
F_{\mu\nu} \equiv \del_\mu A_\nu - \del_\nu A_\mu \; ,
\label{abe:f}
\end{equation}
is invariant under the gauge transformation (\ref{gauge:a}), so is
the Lagrangian for free gauge field,
\begin{equation}
\lag_A = - \frac{1}{4} F_{\mu\nu}  F^{\mu\nu} \; ,
\label{abe:l}
\end{equation}
This Lagrangian together with (\ref{l:psi}) describes the Quantum
Electrodynamics. 

We should point out that a hypothetical mass term for the gauge field, 
\[
\lag_A^m  =  - \frac{1}{2} A_\mu  A^{\mu} \; ,
\]
is not invariant under the transformation (\ref{gauge:a}). 
Therefore, something else should be necessary to describe massive
vector bosons in a gauge invariant way, preserving the
renormalizability of the theory.


\subsection{Gauge Invariance for Non--Abelian Groups} \indent

As suggested by Heisenberg \cite{Heisenberg:32} in 1932, under
nuclear interactions, protons and neutron can be regarded as
degenerated since their mass are quite similar and electromagnetic
interaction is negligible. 

Therefore any arbitrary combination of their wave function would be
equivalent,
\[
\psi \equiv      \left( \ba{c}
                        \psi_p\\
                        \psi_n
              \ea \right) \to \psi^\prime = U \psi \; ,
\]
where $U$ is unitary transformation ($U^\dagger U = U U^\dagger =1$)
to preserve normalization (probability). Moreover, if det$|U|$ = 1,
$U$ represents the Lie group $SU(2)$:
\[
U = \exp\left(-i \frac{\tau^a}{2} \alpha^a \right) 
\simeq 1 - i  \frac{\tau^a}{2} \alpha^a \; ,
\]
where $\tau^a$, $a = 1,2,3$ are the Pauli matrices.

In 1954, Yang and Mills \cite{Yang:54} introduced the idea of local
gauge isotopic invariance in quantum field theory. 
\begin{center}
\begin{minipage}[h]{12cm}
{\it  ``The differentiation between a neutron and a proton
is then a purely arbitrary process. As usually conceived, however,
this arbitrariness is subject to the following limitation: once
one chooses what to call a proton, what a neutron, at one
spacetime point, one is then not free to make any choices at
other spacetime points. It seems that this is not consistent
with the localized field concept that underlies the usual
physical theories.''}
\end{minipage}
\end{center}

Following their argument, we should preserve our freedom to choose
what to call a proton or a neutron {\it no matter when or where we
are}. This can be implemented by requiring that the gauge parameters 
depend on the spacetime points, {\it i.e.} $\alpha^a \to \alpha^a
(x)$.

This idea was generalized by Utiyama \cite{Utiyama:56} in 1956 for
any non--Abelian group $G$ with generators $t_a$ satisfying the Lie
algebra \cite{Abers:73},
\[
\left[ t_a, t_b \right] = i \; C_{abc} \; t_c \; ,
\]
with $C_{abc}$ being the structure constant of the group.

The Lagrangian $\lag_\psi$ should be invariant under the {\it matter
field transformation}
\[
\psi \to \psi^\prime  = \Omega \; \psi \; ,
\]
with 
\[ 
\Omega \equiv \exp\left[-i \, T^a \alpha^a(x)\right] \; ,
\]
where $T^a$ is a convenient representation ({\it i.e.} according to 
the fields $\psi$) of the generators $t^a$.

Introducing one gauge field for each generator, and defining the
{\it  covariant derivative} by 
\[
D_\mu \equiv \del_\mu - i g T^a A_\mu^a \; ,
\]

Since the covariant derivative transforms just like the matter field,
{\it i.e.}  $D_\mu \psi \to \Omega \; (D_\mu \psi)$, this will ensure
the invariance under the local non--Abelian gauge transformation for
the terms containing the fields and its gradients as long as the
{\it gauge field transformation} is
\[
T^a A^a_\mu \to   \Omega \left(T^a A^a_\mu + 
\frac{i}{g} \del_\mu \right) \Omega^{-1} \; , 
\]
or, in infinitesimal form, {\it i.e.} for 
$\Omega \simeq 1 - i \, T^a \alpha^a(x)$, 
\[
A^{a \, \prime}_\mu = A^a_\mu - \frac{1}{g} \del_\mu \alpha^a +
C_{abc} \, \alpha^b A^c_\mu \; .
\]

Finally, we should generalize the {\it  strength tensor} (\ref{abe:f})
for a non--abelian Lie group,
\begin{equation}
F_{\mu\nu}^a \equiv \del_\mu A_\nu^a - \del_\nu A_\mu^a + g \, 
C_{abc} \, A_\mu^b A_\nu^c \; ,
\label{nonab:f}
\end{equation}
which transforms like $F_{\mu\nu}^{a \; \prime} \to F_{\mu\nu}^a +
C_{abc} \alpha^b F_{\mu\nu}^c$. Therefore, the invariant kinetic term
for the gauge bosons, can be written as 
\begin{equation}
\lag_A = - \frac{1}{4} F_{\mu\nu}^a F^{a \; \mu\nu} \; ,
\label{nonab:l}
\end{equation}
and is invariant under the local gauge transformation. However, a
{\it mass term} for the gauge bosons like 
\[
A_\mu^a A^{a \; \mu} \to 
\left( A^a_\mu - \frac{1}{g} \del_\mu \alpha^a + C_{abc} \alpha^b
A^c_\mu \right) 
\left( A^{a \; \mu} - \frac{1}{g} \del_\mu \alpha^a + C_{ade}
\alpha^d A^{e \; \mu} \right) \; ,
\] 
is still {\it not} gauge invariant.

Note that since
\[
F \propto (\del A - \del A) + g A A \; , 
\]
unlike the Abelian case, there is a new feature: the gauge fields
have triple and quartic {\it self--couplings}, 
\[
\ba{ccccc}
\lag_A & \propto & (\del A - \del A)^2 \; + & g (\del A - \del A) A A
\; + & g^2 A A A A \\
&&\mbox{propagator} & \mbox{triple} & \mbox{quartic}\\
\ea \; .
\]

\section{Spontaneous Symmetry Breaking} \label{ssb} \indent

Exact symmetries give rise, in general, to exact conservation laws.
In this case both the Lagrangian and the vacuum (the ground state of
the theory) are invariant. However, there are some conservation laws
which are not exact, {\it e.g.} isospin, strangeness, etc.  These
situations can be  described by adding to the invariant Lagrangian
($\lag_{\text{sym}}$) a small term that breaks this symmetry
($\lag_{\text{sb}}$),
\[
\lag = \lag_{\text{sym}} + \varepsilon \; \lag_{\text{sb}} \; .
\]

Another situation occurs when the system has a Lagrangian that is
invariant and a non--invariant vacuum. A classic example of
the situation is provided by a ferromagnet where the Lagrangian
describing the spin--spin interaction is invariant under
tridimensional rotations. For temperatures above the ferromagnetic 
transition temperature ($T_C$) the spin system is completely
disordered (paramagnetic phase), and therefore the vacuum is also
$SO(3)$ invariant [see Fig.\ 2(a)]. 

However, for temperatures below $T_C$ (ferromagnetic phase)  a 
spontaneous magnetisation of the system occurs, aligning the spins in
some specific direction [see Fig.\ 2(b)]. In this case, the vacuum is
not invariant under the $SO(3)$ group. This symmetry is broken to
$SO(2)$, representing the rotation of the whole system around the
spin directions.

\vskip 0.3cm
\begin{center}
\begin{tabular}{cc}
{ \rar ~ \nor ~ \ser ~ \uar ~ \rar ~ \sor } ~~~~~~~~ & ~~~~~~~~ 
{ \ner ~ \ner ~ \ner ~ \ner ~ \ner ~ \ner }\\[0.2cm]
{ \nor ~ \dar ~ \ner ~ \nor ~ \sor ~ \sor }~~~~~~~~ & ~~~~~~~~ 
{ \ner ~ \ner ~ \ner ~ \ner ~ \ner ~ \ner }\\[0.2cm]
{ \uar ~ \rar ~ \sor ~ \rar ~ \nor ~ \ser }~~~~~~~~ & ~~~~~~~~  
{ \ner ~ \ner ~ \ner ~ \ner ~ \ner ~ \ner }\\[0.2cm]
{ \lar ~ \ner ~ \nor ~ \nor ~ \sor ~ \sor }~~~~~~~~ & ~~~~~~~~  
{ \ner ~ \ner ~ \ner ~ \ner ~ \ner ~ \ner }\\[0.2cm]
{ \rar ~ \rar ~ \sor ~ \nor ~ \rar ~ \ner }~~~~~~~~ & ~~~~~~~~ 
{ \ner ~ \ner ~ \ner ~ \ner ~ \ner ~ \ner }\\[0.2cm]
& \\
{\it (a)} ~~~~~~~~&~~~~~~~~ {\it (b)} \\
\end{tabular}
\end{center}

\begin{center}
\begin{minipage}[h]{12cm}
\begin{center}
{\it Fig.\ 2: Representation of the spin orientation in the paramagnetic
(a) and ferromagnetic (b) phases.}
\end{center}
\end{minipage}
\end{center}


Let us analyze the simple example of a scalar self--interacting real
field with Lagrangian,
\begin{equation}
\lag = \frac{1}{2} \del_\mu \phi \; \del^\mu \phi -  V(\phi) \; ,
\label{lag}
\end{equation}
with
\begin{equation}
V(\phi) = \frac{1}{2} \mu^2 \phi^2 + \frac{1}{4} \lambda \phi^4 \; .
\label{sca:pot}
\end{equation}
In the theory of the phase transition of a ferromagnet, the
Gibbs free energy density is analogous to $V(\phi)$ with  $\phi$
playing the r\^ole of the average spontaneous magnetisation $M$.

The whole Lagrangian (\ref{lag}) is invariant under the discrete 
transformation
\begin{equation}
\phi \to - \phi \; .
\label{sym:phi}
\end{equation}
Is the vacuum also invariant under this transformation? The
vacuum ($\phi_0$) can be obtained from the Hamiltonian
\[
{\cal H} = \frac{1}{2} \left[(\del_0 \phi)^2 + (\nabla \phi)^2 \right]
+ V(\phi) \; .
\]

We notice that $\phi_0 = constant$ corresponds to the minimum of $V(\phi)$
and consequently of the energy: 
\[
\phi_0 (\mu^2 + \lambda \phi_0^2) = 0 \; .
\]

Since $\lambda$ should be positive to guarantee that ${\cal H}$ is
bounded, the minimum depends on the {\it sign of $\mu$}. For $\mu^2
>0$, we have just one vacuum at $\phi_0 = 0$ and it is also invariant
under (\ref{sym:phi}) [see Fig.\ 3 (a)]. However, for $\mu^2 < 0$, we
have two vacua states corresponding to $\phi_0^\pm = \pm
\sqrt{-\mu^2/\lambda}$ [see Fig.\ 3 (b)]. This case corresponds to a
wrong sign for the $\phi$ mass term. 

\begin{figure}[ht]
\protect
\epsfxsize=12cm
\begin{center}
\mbox{\psfig{file=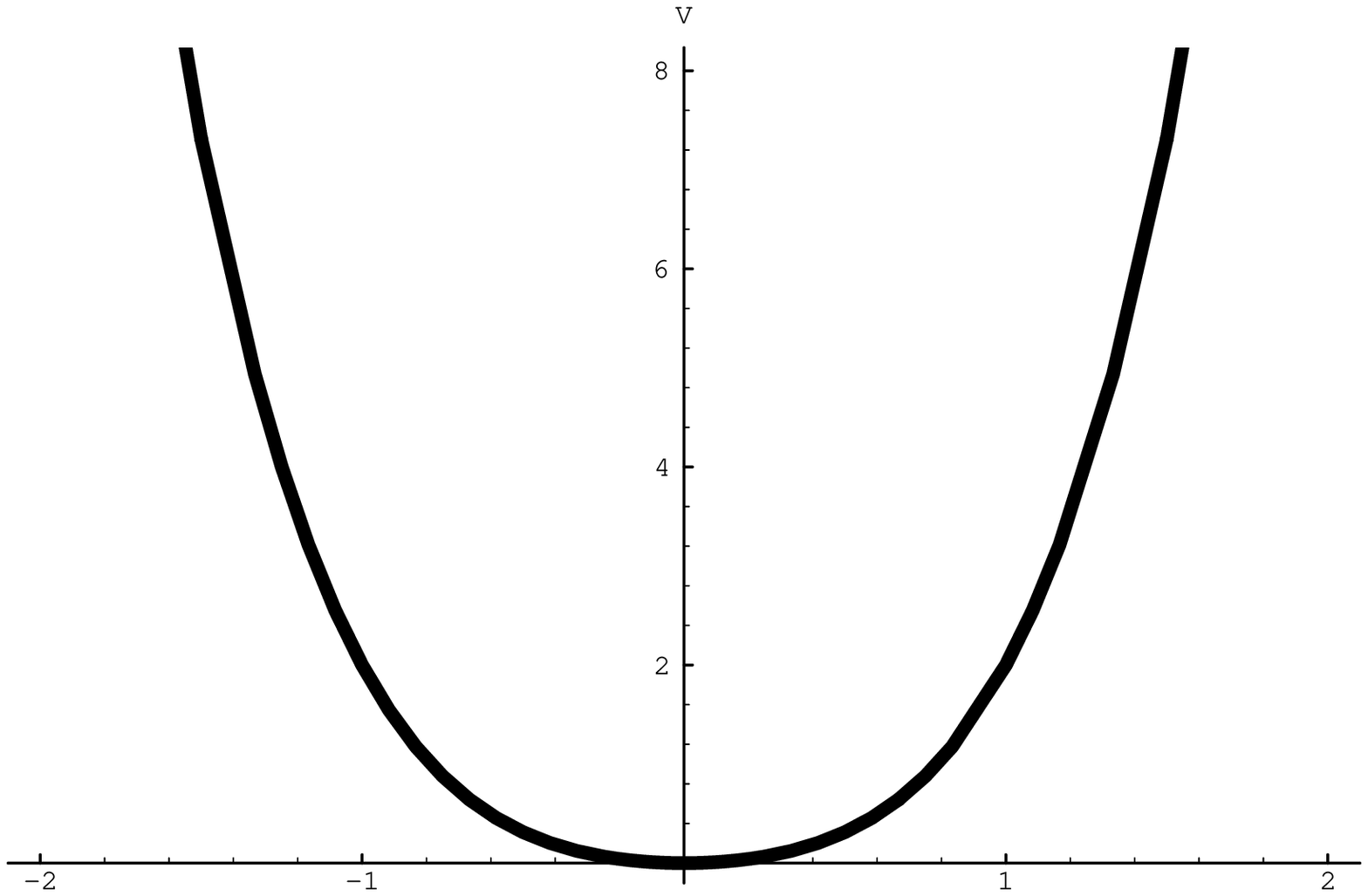,width=0.35\textwidth}} ~~~~~~~~~~~~~
\mbox{\psfig{file=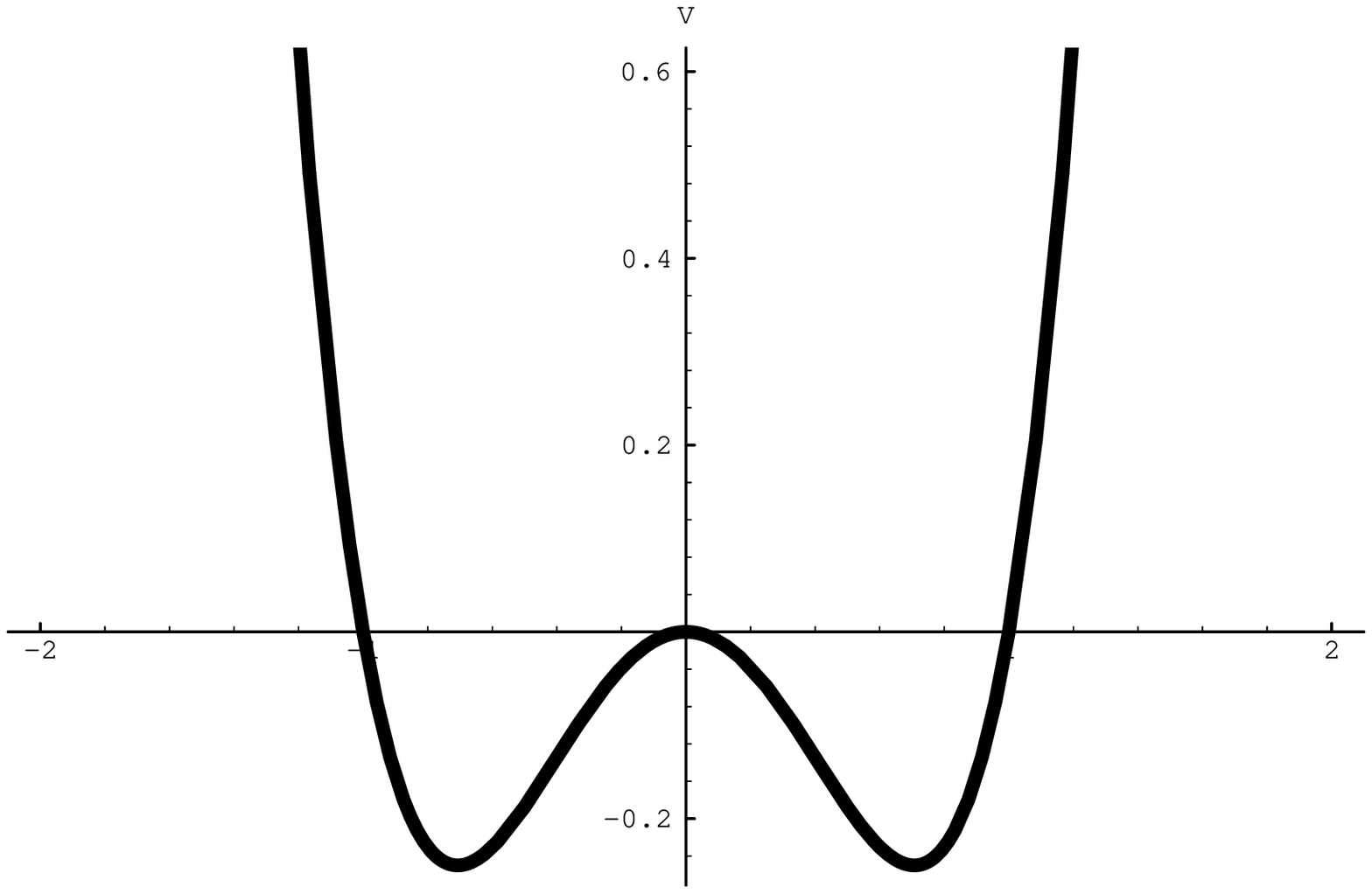,width=0.4\textwidth}}
\end{center}
\end{figure}
\begin{center}
\vspace{-1cm}
\begin{tabular}{cc}
{\it (a)} ~~~~~~~~~~~~~~~~~~~~~~~~~~ & 
~~~~~~~~~~~~~~~~~~~~~~~ {\it (b)} \\
\end{tabular}
\end{center}
\begin{center}
\begin{minipage}[h]{12cm}
\begin{center}
{\it Fig.\ 3: Scalar potential (\ref{sca:pot}) for $\mu^2 > 0$ (a) and
for  $\mu^2 < 0$ (b).}
\end{center}
\end{minipage}
\end{center}

Since the Lagrangian is invariant under (\ref{sym:phi}) the choice
between $\phi_0^+$ or $\phi_0^-$ is irrelevant \footnote{For an 
interesting discussion discarding the invariant state $(\phi_0^+ \pm
\phi_0^-)$ as the true vacuum see Ref.\ \cite{Weinberg:96}}.
Nevertheless, once one choice is made ({\it e.g.} $v = \phi_0^+$) the
symmetry is {\it spontaneously broken} since $\lag$ is invariant but
the vacuum is {\it not}.

Defining a new field $\phi^\prime$ by shifting the old field by $v =
\sqrt{-\mu^2/\lambda}$, 
\[
\phi^\prime \equiv \phi - v \; ,
\]
we verify that the vacuum of the new field is $\phi^\prime_0 = 0$,
making the theory suitable for small oscillations around the vacuum
state. The Lagrangian becomes:
\[
\lag = \frac{1}{2} \del_\mu \phi^\prime \del^\mu \phi^\prime -
\frac{1}{2} \left( { \sqrt{-2 \mu^2}} \right)^2 \phi^{\prime \; 2} -
\lambda \, v \, \phi^{\prime \; 3} - \frac{1}{4} \lambda
\phi^{\prime \; 4} \; .
\]
This Lagrangian describes a scalar field $\phi^\prime$ with real and 
positive mass, $M_{\phi^\prime}=\sqrt{-2 \mu^2}$, but it lost the
original symmetry due to the $\phi^{\prime \; 3}$ term. 

A new interesting phenomenon happens when a {\it continuous}
symmetry is spontaneously broken. Let us analyze the case of  a
charged self--interacting scalar field,
\begin{equation}
\lag = \del_\mu \phi^\ast \del^\mu \phi -  V(\phi^\ast\phi) \; ,
\label{lag:com}
\end{equation}
with a similar potential,
\begin{equation}
V(\phi^\ast\phi) =  \mu^2 (\phi^\ast\phi) + \lambda (\phi^\ast\phi)^2
\; .
\label{pot:com}
\end{equation}

Notice that the Lagrangian (\ref{lag:com}) is invariant under the {\it global
phase} transformation
\[
\phi \to \exp(-i\theta) \phi \; .
\]

When we redefine the complex field in terms of two real fields by
\[
\phi = \frac{(\phi_1 + i \phi_2)}{\sqrt{2}} \; ,
\]
the Lagrangian (\ref{lag:com}) becomes
\begin{equation}
\lag = \frac{1}{2} \left( \del_\mu \phi_1 \del^\mu \phi_1 
+ \del_\mu \phi_2 \del^\mu \phi_2 \right) -  
V(\phi_1, \phi_2 ) \; ,
\label{lag2:com}
\end{equation}
which is invariant under $SO(2)$ rotations,
\[
\left( \ba{c}
\phi_1 \\ 
\phi_2 
\ea \right) \longrightarrow 
\left( \ba{cc}
\cos\theta & -\sin\theta\\ 
\sin\theta & \cos\theta  
          \ea \right) \; 
\left( \ba{c}
\phi_1\\ 
\phi_2  
\ea \right) \; .
\]

For $\mu^2 > 0$ the vacuum is at $\phi_1 = \phi_2 = 0$, and
for small oscillations,
\[
\lag = \sum_{i = 1}^2 \frac{1}{2} 
\left(\del_\mu \phi_i \del^\mu \phi_i - \mu^2 \phi_i^2 \right) \; ,
\]
which means that we have two scalar fields $\phi_1$ and $\phi_2$ with 
mass $m^2 = \mu^2 > 0$. 

In the case of $\mu^2 < 0$ we have a continuum of distinct vacua [see
Fig.\ 4 (a)] located at
\begin{equation}
<|\phi|^2>= \frac{(<\phi_1>^2 + <\phi_2>^2)}{2} 
=\frac{ - \mu^2}{2 \lambda} \equiv \frac{v^2}{2}  \; . 
\label{vac:com}
\end{equation}

\begin{figure}[ht]
\protect
\epsfxsize=12cm
\begin{center}
\begin{tabular}{cc}
\mbox{\psfig{file=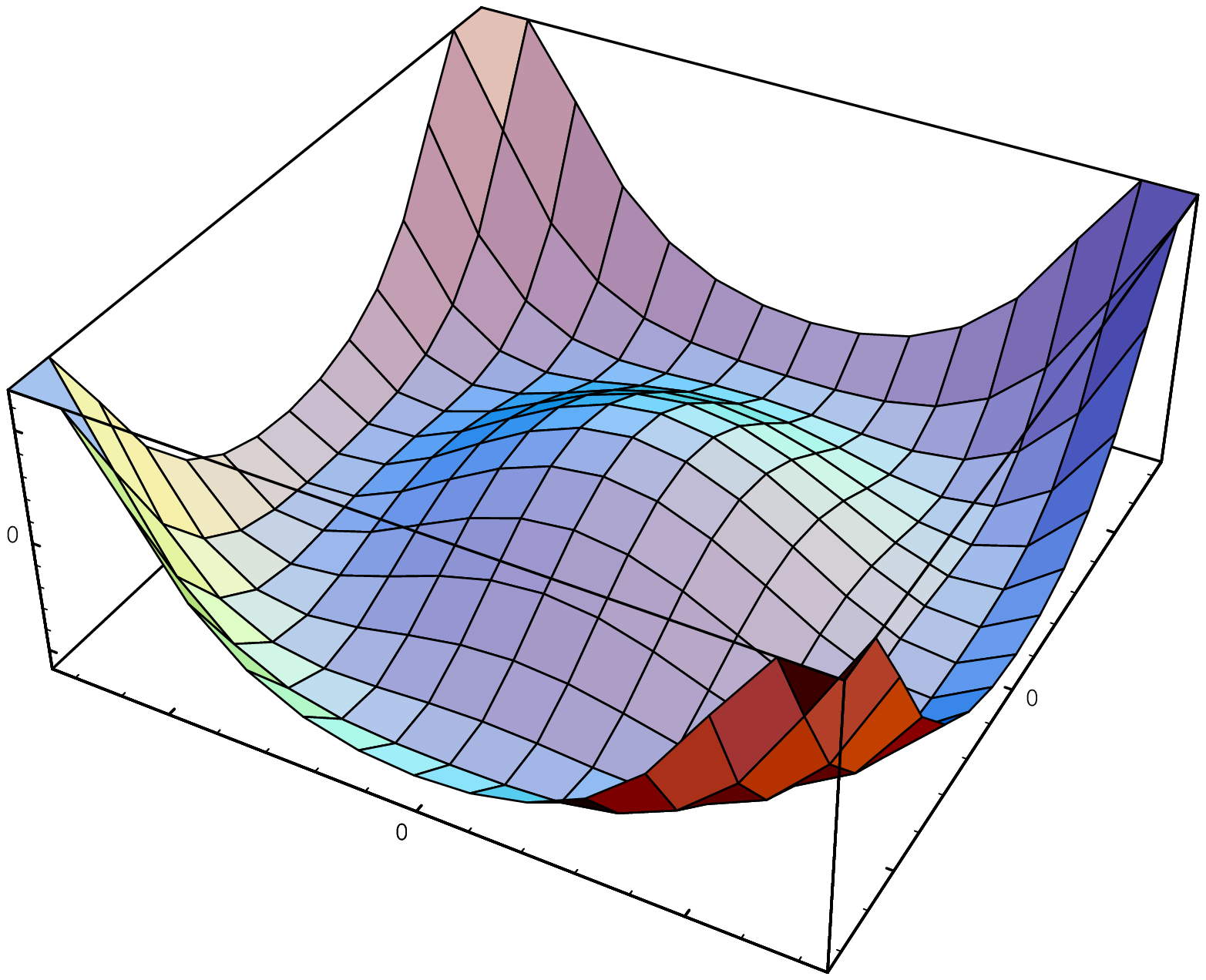,width=0.5\textwidth}}  ~~~ & ~~~ 
\mbox{\psfig{file=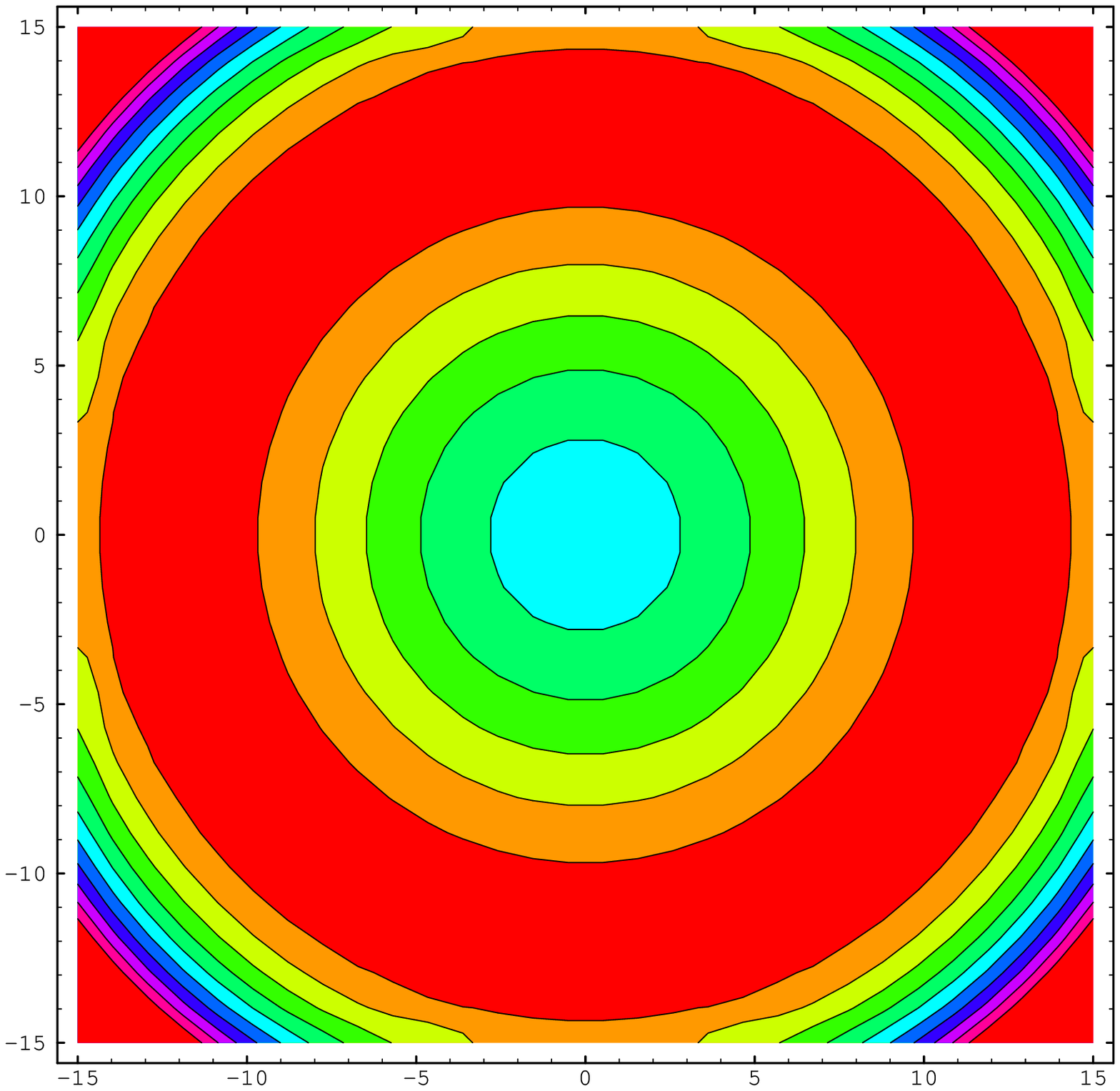,width=0.35\textwidth}} \\
{\it (a)} ~~~~~~~ & 
~~~~~~~ {\it (b)} \\
\end{tabular}
\end{center}
\begin{center}
\begin{minipage}[h]{12cm}
\begin{center}
{\it Fig.\ 4: The potential $V(\phi_1, \phi_2 )$ (a) and its contour
plot (b)}
\end{center}
\end{minipage}
\end{center}
\end{figure}

We can see from the contour plot [Fig.\ 4 (b)] that the vacua are
also invariant under $SO(2)$. However, this symmetry is spontaneously
broken when we choose a particular vacuum. Let us choose, for
instance, the configuration,
\begin{eqnarray*}
\phi_1 & = & v \; ,
\\
\phi_2 & = & 0 \; .
\end{eqnarray*}

The new fields, suitable for small perturbations, can be defined as,
\begin{eqnarray*}
\phi_1^\prime & = &  \phi_1 - v \; ,\\
\phi_2^\prime & = &  \phi_2 \; .
\end{eqnarray*}

In terms of these new fields the Lagrangian (\ref{lag2:com}) becomes,
\[ 
\lag = \frac{1}{2} \del_\mu \phi_1^\prime \del^\mu \phi_1^\prime
-  \frac{1}{2} ({ -2 \mu^2}) \phi_1^{\prime \; 2} +  \frac{1}{2}
\del_\mu \phi_2^\prime \del^\mu \phi_2^\prime  + \; \mbox{interaction
terms} \; .
\] 

Now we identify in the particle spectrum a scalar field
$\phi_1^\prime$ with real and positive mass and a massless scalar
boson ($\phi_2^\prime$). This could be seen from Fig.\ 4 (b), when we
consider the mass matrix in tree approximation,
\[
M_{ij}^2 = \left . \frac{\partial^2 V(\phi_1^\prime, \phi_2^\prime)}
              {\partial \phi_i^\prime\partial\phi_j^\prime} \right
              |_{\phi^\prime = \phi_0^\prime} \; .
\]
The second derivative of $V(\phi_1^\prime, \phi_2^\prime)$ in the
$\phi_2^\prime$ direction corresponds to the zero eigenvalue of the
mass matrix, while for $\phi_1^\prime$ it is positive.

This is an example of the prediction of the so called Goldstone
theorem \cite{Goldstone:61}  which states that when an exact
continuous global symmetry is spontaneously broken, {\it i.e.} it is
not a symmetry of the physical vacuum, the theory contains one
massless scalar particle for each broken generator of the original
symmetry group. 

The Goldstone theorem can be proven as follows. Let us consider a
Lagrangian of $N_G$ real scalar fields $\phi_i$, belonging to a
$N_G$--dimensional vector $\Phi$,
\[
\lag = \frac{1}{2} (\partial_\mu \Phi)(\partial^\mu \Phi) - V(\Phi) 
\; .
\]
Suppose that $G$ is a continuous group that let the Lagrangian
invariant and that $\Phi$ transforms like
\[
\delta \Phi = - i \; \alpha^a \; T^a \; \Phi \; .
\]
Since the potential is invariant under $G$, we have
\[
\delta V(\Phi) = \frac{\partial V(\Phi)}{\partial \phi_i} \; \delta
\phi_i = -i \; \frac{\partial V(\Phi)}{\partial \phi_i} \; \alpha^a
\; (T^a)_{ij} \; \phi_j = 0 \; .
\]
The gauge parameters $\alpha^a$ are arbitrary, and we have $N_G$
equations
\[
\frac{\partial V(\Phi)}{\partial \phi_i} \;  (T^a)_{ij} \; \phi_j = 0
\; ,
\]
for $a = 1, \cdots, N_G$. Taking another derivative of this equation, we
obtain
\[
\frac{\partial^2 V(\Phi)}{\partial \phi_k \partial\phi_i} \;
(T^a)_{ij} \; \phi_j + 
\frac{\partial V(\Phi)}{\partial \phi_i} \; (T^a)_{ik}  = 0 \; .
\]

If we evaluate this result at the vacuum state, $\Phi = \Phi_0$, which
minimizes the potential, we get
\[
\left . \frac{\partial^2 V(\Phi)}{\partial \phi_k \partial\phi_i} 
\right |_{\Phi = \Phi_0} (T^a)_{ij} \; \phi_{j}^0 =  0 \; ,
\]
or, in terms of the mass matrix,
\begin{equation}
M^2_{ki} \; (T^a)_{ij} \; \phi_{j}^0 = 0 \; .
\label{mm}
\end{equation}

If, after we choose a ground state, a sub-group $g$ of $G$, with
dimension $n_g$, remains a symmetry of the vacuum, then for each
generator of $g$, 
\[
(T^a)_{ij} \; \phi_{j}^0 = 0 \;\; \mbox{for} \;\; a = 1, \cdots , n_g
\leq N_G \; ,
\]
while for the $(N_G - n_g)$ generators that break the symmetry,
\[
(T^a)_{ij} \; \phi_{j}^0 \neq 0 \;\; \mbox{for} \;\; a = n_g + 1, \cdots
, N_G \; .
\]

Therefore,  the relation (\ref{mm}) shows that there are $(N_G - n_g)$ zero 
eigenvalues  of the mass matrix: the massless Goldstone bosons.


\section{The Higgs Mechanism} \label{higgs} \indent

\subsection{The Abelian Higgs Mechanism}

The Goldstone theorem implies the existence of massless scalar
particle(s). However, we do not have any experimental evidence in
nature of these particles. In 1964 several authors independently 
\cite{Higgs:64,Englert:64,Guralnik:64} were able to provide a way out
to the Goldstone theorem, that is, a field theory with spontaneous
symmetry breakdown, but with no massless Goldstone boson(s). The so
called Higgs mechanism has an extra bonus: the gauge boson(s) becomes
massive. This is accomplished by requiring that the Lagrangian that
exhibits the spontaneous symmetry breakdown is also invariant under 
{\it local}, rather than global, gauge transformations.  This feature
fits very well in the requirements for a gauge theory of electroweak
interactions where the short range character of this interaction
requires a very massive intermediate particle. 

In order to see how this works let us consider again the charged
self--inter\-act\-ing scalar Lagrangian (\ref{lag:com}) with the
potential (\ref{pot:com}), and let us require a invariance under
the {\it local} phase transformation,
\begin{equation}
\phi \to \exp\left[i \, q \, \alpha (x)\right] \phi \; .
\label{gau:par}
\end{equation}

In order to make the Lagrangian invariant, we introduce a {\it  gauge
boson} ($A_\mu$) and the {\it covariant derivative} ($D_\mu$),
following the same principles of Section \ref{gau:pri}

We introduce a {\it  gauge boson} ($A_\mu$) and the {\it covariant
derivative} ($D_\mu$), so that the Lagrangian becomes invariant,
following the same principles of Section \ref{gau:pri}
\[
\del_\mu \longrightarrow D_\mu = \del_\mu + i q A_\mu 
\;\; , \;\;\; \mbox{with} \;\;\;
A_\mu \longrightarrow A_\mu^\prime = A_\mu - \del_\mu \alpha(x) \; .
\]

The spontaneous symmetry breaking occurs for $\mu^2 < 0$, with the
vacuum  $<|\phi|^2>$ given by (\ref{vac:com}). There is a very
convenient way of parametrizing the new fields, $\phi^\prime$, that
are suitable  for small perturbations, {\it i.e.},
\begin{equation}
\phi = \exp\left(i \frac{\phi_2^\prime}{v} \right) 
        \frac{(\phi_1^\prime + v)}{\sqrt{2}} 
 \simeq  \frac{1}{\sqrt{2}} 
\left( \phi_1^\prime + v + i \phi_2^\prime \right) 
= \phi^\prime + \frac{v}{\sqrt{2}} \; .
\label{par:phi}
\end{equation}

Therefore the Lagrangian (\ref{lag:com}) becomes,
\begin{eqnarray}
\lag & = & { \frac{1}{2} \del_\mu \phi_1^\prime \del^\mu \phi_1^\prime - 
\frac{1}{2} (-2 \mu^2) \phi_1^{\prime \; 2} } + 
\frac{1}{2} \del_\mu \phi_2^\prime \del^\mu \phi_2^\prime + \;
{\mbox{interact.}}
\nonumber \\
&& { - \frac{1}{4} F_{\mu\nu} F^{\mu\nu} + 
\frac{q^2 v^2}{2} A_\mu A^\mu } + { q v A_\mu \del^\mu \phi_2^\prime }
\; .
\label{lag:com:2}
\end{eqnarray}

This Lagrangian presents a scalar field $\phi_1^\prime$ with mass
$M_{\phi_1^\prime} = \sqrt{-2 \mu^2}$, a  massless scalar boson
$\phi_2^\prime$ (the Goldstone boson) and a massive vector boson
$A_\mu$, with mass $M_{A} = q v$. 

However the presence of the last term in (\ref{lag:com:2}), which is
proportional to $A_\mu \del^\mu \phi_2^\prime$ is quite inconvenient
since it mixes the propagators of $A_\mu$ and $\phi_2^\prime$
particles. In order to eliminate this term,  we can choose the gauge
parameter in (\ref{gau:par}) to be proportional to $\phi_2^\prime$
as
\[
\alpha (x) = - \frac{1}{qv} \phi_2^\prime (x) \; .
\]
In this way, the field $\phi$ (\ref{par:phi}) becomes,
\[
\phi = { \exp\left[i q \left( - \frac{\phi_2^\prime }{qv} \right)\right]} 
               \exp\left(i \frac{\phi_2^\prime}{v} \right) 
        \frac{(\phi_1^\prime + v)}{\sqrt{2}}  
= \frac{1}{\sqrt{2}}  \left( \phi_1^\prime + v  \right) \; .
\]

With this choice of gauge (called unitary gauge) the Goldstone boson
disappears, and we get the Lagrangian
\begin{eqnarray}
\lag & = & { \frac{1}{2} \del_\mu \phi_1^\prime \del^\mu \phi_1^\prime - 
\frac{1}{2} (-2 \mu^2) \phi_1^{\prime \; 2}}
{  - \frac{1}{4} F_{\mu\nu} F^{\mu\nu} + 
\frac{q^2 v^2}{2} A_\mu^\prime A^{\mu \; \prime} } 
\nonumber \\
&&  
+ \frac{1}{2} q^2 \left( \phi_1^\prime + 2 v \right) \phi_1^\prime
A_\mu^\prime A^{\mu \; \prime} 
- \frac{\lambda}{4} \phi_1^{\prime \; 3} \left( \phi_1^\prime + 4 v
\right) \; .
\label{lag:com:3}
\end{eqnarray}

Where is $\phi_2^\prime$, the Goldstone boson? To answer this 
question, it is convenient to count the total number of degrees of
freedom from the initial (\ref{lag:com}) and final (\ref{lag:com:3})
Lagrangians:

\vskip 1.2cm
\begin{center}
\begin{tabular}{cc}
{ Initial $\lag$ ~ (\ref{lag:com})}  ~~~~~~ & ~~~~~~ { Final $\lag$ ~
(\ref{lag:com:3})} \\[0.5cm]
$\phi^{(\ast)}$ charged scalar  : 2  ~~~~~~ & ~~~~~~
$\phi_1^\prime $ neutral scalar : ~ 1   \\
$A_\mu$ massless vector : 2  ~~~~~~ & ~~~~~~ 
$A_\mu^\prime$ massive vector : 3 \\[-0.3cm]
~~~~~~~~~~~~~~~~~~~~~~~~~~ \underline{~~~~~~~~~} & 
~~~~~~~~~~~~~~~~~~~~~~~~~~~~~~~~~~~~~~~ \underline{~~~~~~~~~} \\
~~~~~~~~~~~~~~~~~~~~~~~~~~ 4 & ~~~~~~~~~~~~~~~~~~~~~~~~~~~~~~~~~~~~~~~ 4
\end{tabular}
\end{center}
\vskip 0.5cm

As we can see, the corresponding degree of freedom of the Goldstone
boson was absorbed by the vector boson that acquires mass. The
Goldstone turned into the longitudinal degree of freedom of the vector
boson.


\newpage
\subsection{The Non--Abelian Case} \label{hig:nonabe} \indent

It is straightforward to generalize the last section's  results for a
non--Abelian group $G$ of dimension $N_G$, and generators $T^a$. In
this case, we introduce $N_G$ gauge bosons, such that the covariant
derivative is written as
\[
\del_\mu \longrightarrow D_\mu = \del_\mu - i g T^a B^a_\mu \; .
\]
After the spontaneous symmetry breaking, a {\it sub--group $g$} of
dimension $n_g$ remains as a symmetry of the vacuum, that is,
\[
T^a_{ij} \; \phi^0_j = 0 \;\; , \;\;\;\; \mbox{for} \;\;\;\; a =
1, \cdots , n_g \; .
\]

We would expect the appearance of $(N_G - n_g)$ massless Goldstone bosons.
Like in (\ref{par:phi}), we parametrize the original scalar field as
\[
\phi = (\tilde{\phi} + v) \; 
\exp\left(i \frac{\phi_{\text{GB}}^a T^a}{v} \right) \; ,
\]
where $T^a$ are the $(N_G - n_g)$ broken generators that do {\it not}
annihilate the vacuum.

Choose the gauge parameter $\alpha^a(x)$ in order to to eliminate
$\phi_{\text{GB}}^a$. This will give rise to  $(N_G - n_g)$ massive
gauge bosons. Counting the total number of degrees of freedom we
obtain $N_\phi + 2 N_G$, both before and after the spontaneous
symmetry breaking: 

\vskip 1.2cm
\begin{center}
\begin{tabular}{cc}
Before SSB & After SSB \\
& \\
$\phi$ massless scalar : $N_\phi$  & 
$\tilde{\phi}$ massive scalar : $N_\phi - (N_G - n_g)$   \\
$B^a_\mu$ massless vector : 2 $N_G$  & 
$\tilde{B}^a_\mu$ massive vector : 3 $(N_G - n_g)$ \\
 &  $B^a_\mu$ massless vector : 2 $n_g$ 
\end{tabular}
\end{center}

\newpage
\chapter{The Standard Model} \label{sm}

\section{Constructing the Model} 

\subsection{General Principles to Construct Gauge Theories}
\indent

Based on what we have learned from the previous sections, we can
establish some quite general principles to construct a gauge theory.
The recipe is as follows,

\begin{itemize}
\item Choose the gauge group $G$ with $N_G$ generators;

\item Add $N_G$ vector fields (gauge bosons) in a specific
representation of the gauge group;

\item Choose the representation, in general the fundamental
representation, for the matter fields (elementary particles);

\item Add scalar fields to give mass to (some) vector bosons;

\item Define the covariant derivative and write the most general
renormalizable Lagrangian, invariant under $G$, which couples all
these fields;

\item Shift the scalar fields in such a way that the minimum of the
potential is at zero;

\item Apply the usual techniques of quantum field theory to verify
the renormalizability and to make predictions;

\item Check with Nature if the model has anything to do with reality;

\item If not, restart from the very beginning!
\end{itemize}

In fact, there were several attempts to construct a gauge theory for
the (electro)weak interaction. In 1957, Schwinger \cite{Schwinger:57}
suggested a model based on the group $O(3)$ with a triplet gauge
fields  $(V^+, V^-, V^0)$. The charged gauge bosons were associated
to weak bosons and the neutral $V^0$ was identified with the photon.
This model was proposed before the structure $V-A$ of the weak
currents have been established \cite{Feynman:58,Marshak:58,Sakurai:58}. 

The first attempt to incorporate the $V-A$ structure in a gauge
theory for the weak interactions was made by Bludman
\cite{Bludman:58} in 1958. His model, based on the $SU(2)$ weak
isospin group, also required three vector bosons. However in this
case the neutral gauge boson was associated to a new massive vector
boson that was responsible for weak interactions without exchange of
charge (neutral currents). The  hypothesis of a neutral vector boson
exchanged in weak interaction was also suggested independently by
Leite Lopes \cite{Lopes:58} in the same year. This kind of process
was observed experimentally for the first time in 1973 at the CERN
neutrino experiment \cite{Hasert:73a}. 

Glashow \cite{Glashow:61} in 1961 noticed that in order to
accommodate both weak and electromagnetic interactions we should go
beyond the $SU(2)$ isospin structure. He suggested the gauge group
$SU(2) \otimes U(1)$, where the $U(1)$ was associated to the leptonic
hypercharge ($Y$) that is related to the weak isospin ($T$) and the
electric charge through the analogous of the Gell-Mann--Nishijima
formula ($Q = T_3 + Y/2$). The theory now requires four gauge bosons:
a triplet ($W^1, W^2, W^3$) associated to the generators of $SU(2)$
and a neutral field ($B$) related to $U(1)$. The charged weak bosons
appear as a linear combination of $W^1$ and $W^2$, while the photon
and a neutral weak boson $Z^0$ are both given by a mixture of $W^3$
and $B$. A similar model was proposed by Salam and Ward
\cite{Salam:64} in 1964. 

The mass terms for $W^\pm$ and $Z^0$ were put ``by hand''. However,
as we have seen, this procedure breaks explicitly the gauge
invariance of the theory. In 1967, Weinberg \cite{Weinberg:67} and
independently Salam \cite{Salam:68} in 1968, employed the idea of
spontaneous symmetry breaking and the Higgs mechanism to give mass to
the weak bosons and, at the same time, to preserve the gauge
invariance,  making the theory renormalizable as shown later by 't
Hooft \cite{Hooft:71}. The Glashow--Weinberg--Salam model is known,
at the present time, as the {\it Standard Model of Electroweak
Interactions}, reflecting its impressive success.  


\subsection{Right-- and Left-- Handed Fermions} \indent

Before the introduction of the Standard Model, let us make an interlude
and discuss some properties of the fermionic helicity states. At high
energies ({\it i.e.} for $E \gg m$), the Dirac spinors 
\[ 
u(p,s) \;\; , \;\;\;\; \mbox{and} \;\;\;\;\; 
v(p,s) \equiv C \, \bar{u}^T (p,s) = i
\, \gamma_2 \,  u^\ast(p,s) \; ,
\]
are eigenstates of the $\gc$ matrix. 

The helicity $+1/2$ (right--handed, $R$) and helicity $-1/2$
(left--handed, $L$) states satisfy 
\[
u \, _{\stackrel{\text{R}}{\text{L}}} = \frac{1}{2} \left(1 \pm \gc \right) u 
\;\;\;  \mbox{and} \;\;\; 
v \, _{\stackrel{\text{R}}{\text{L}}} = \frac{1}{2} \left(1 \mp \gc \right) v 
\; .
\]
It is convenient to define the {\it helicity projectors}:
\begin{equation}
\fbox{$ \; \displaystyle{L \equiv \frac{1}{2} \left(1 - \gc \right)} \;
$} 
\;\;\;\; ,   \;\;\;\;\;\; 
\fbox{$ \; \displaystyle{R \equiv \frac{1}{2} \left(1 + \gc \right)}
\; $} \; ,
\label{proj}
\end{equation}
which satisfy the usual properties of projection operators,
\begin{eqnarray*}
L + R &=& 1 \; ,\\
R \, L = L \, R & = & 0 \; ,\\
L^2 & = & L \; ,\\
R^2 & = & R \; .
\end{eqnarray*}

For the conjugate spinors we have,
\begin{eqnarray*}
\bar{\psi}_L &=& (L \psi)^\dagger \gamma_0 = \psi^\dagger L^\dagger
\gamma_0 = \psi^\dagger L \gamma_0 = \psi^\dagger \gamma_0  R =
\bar{\psi} R \\
\bar{\psi}_R &=& \bar{\psi} L \; .
\end{eqnarray*}

Let us make some general remarks. First of all, we should notice that
fermion mass term mixes right-- and left--handed fermion components,  
\begin{equation}
\bar{\psi} \psi = \bar{\psi}_R \psi_L + \bar{\psi}_L \psi_R \; .
\label{m}
\end{equation}
On the other hand, the electromagnetic (vector) current, does not mix
those components, {\it i.e.}
\begin{equation}
\bar{\psi} \g^\mu \psi = \bar{\psi}_R \g^\mu \psi_R + \bar{\psi}_L
\g^\mu \psi_L  \; .
\label{v:cur}
\end{equation}
Finally, the $(V-A)$  fermionic weak current can be written in terms of the
helicity states as,
\begin{equation}
\bar{\psi}_L \g^\mu \psi_L = \bar{\psi} R \g^\mu L \psi =  
\bar{\psi} \g^\mu L^2 \psi =   \bar{\psi} \g^\mu L \psi = 
\frac{1}{2} \; \bar{\psi} \g^\mu (1 - \gc) \psi \; ,
\label{va}
\end{equation}
what shows that only left--handed fermions play a r\^ole in weak
interactions. 


\subsection{Choosing the gauge group} \indent

Let us investigate which gauge group would be able to unify the
electromagnetic and weak interactions. We start with the charged weak
current for leptons. Since electron--type and muon--type lepton
numbers are separately conserved, they must form separate
representations of the gauge group. Therefore, we refer as $\ell$ any
lepton flavor ($\ell = e,\, \mu,\, \tau$), and the final Lagrangian
will be given by a sum over all these flavors. 

From Eq.\ (\ref{va}), we see that the weak current (\ref{wea:cur}),
for a generic lepton $\ell$, is given by,
\begin{equation}
J_\mu^+ = \bar{\ell} \g_\mu (1 - \gc) \nu = 2 \; \bar{\ell}_L \g_\mu
\nu_L \; .
\label{wea:cur:2}
\end{equation}

If we introduce the left--handed isospin doublet ($T = 1/2$),
\begin{equation}
\ld \equiv 	\left( \ba{c}
	      		\nu\\
	      		\ell
              \ea \right)_L = 
	\left( \ba{c}
	      		L \, \nu\\
	      		L \, \ell
              \ea \right) = 
	\left( \ba{c}
	      		\nu_L\\
	      		\ell_L
              \ea \right) \; ,
\label{l}
\end{equation}
where the $T_3 = + 1/2$ and $T_3 = - 1/2$ components are the
left--handed parts of the neutrino and of the charged lepton
respectively. Since, there is no right--handed component for the
neutrino \footnote{At this moment, we consider that the  neutrinos
are massless. The possible mass term for the neutrinos will be 
discussed later, Sec.\ \ref{sm:lag}.}, the right--handed part of the
charged lepton is accommodated in a weak isospin singlet ($T =0$)
\begin{equation}
\rs \equiv R \, \ell = \ell_R  \; .
\label{r}
\end{equation}

The charged weak current (\ref{wea:cur:2}) can be written in terms of
leptonic isospin currents: 
\[
J_\mu^i =  \bar{\ld} \; \g_\mu \; \frac{\tau^i}{2} \; \ld \; ,
\]
where $\tau^i$ are the Pauli matrices. In a explicit form,
\begin{eqnarray*}
J_\mu^1 &=& \frac{1}{2} ( \bar{\nu}_L \;\; \bar{\ell}_L) \; \g_\mu 
\left( \ba{cc}
 0 & 1 \\
 1 & 0
              \ea \right) 
\left( \ba{c}
	      		\nu_L\\
	      		\ell_L
              \ea \right) 
= \frac{1}{2} \left(\bar{\ell}_L \g_\mu \nu_L + 
                    \bar{\nu}_L \g_\mu \ell_L  
                     \right)  \; , \\
J_\mu^2 &=& \frac{1}{2} ( \bar{\nu}_L \;\; \bar{\ell}_L) \; \g_\mu 
\left( \ba{cc}
 0 & -i \\
 i & 0
              \ea \right) 
\left( \ba{c}
	      		\nu_L\\
	      		\ell_L
              \ea \right) 
= \frac{i}{2} \left( \bar{\ell}_L \g_\mu \nu_L - 
                     \bar{\nu}_L \g_\mu \ell_L   \right)  \; , \\
J_\mu^3 &=& \frac{1}{2} ( \bar{\nu}_L \;\; \bar{\ell}_L) \; \g_\mu 
\left( \ba{cc}
 1 & 0 \\
 0 & -1
              \ea \right) 
\left( \ba{c}
	      		\nu_L\\
	      		\ell_L
              \ea \right) = 
\frac{1}{2} \left( \bar{\nu}_L \g_\mu \nu_L - 
                   \bar{\ell}_L \g_\mu \ell_L  \right)  \; .\\
\end{eqnarray*}


Therefore, the weak charged current (\ref{wea:cur:2}), that couples
with intermediate vector boson $W^-_\mu$, can be written in terms of
$J^1$ and $J^2$ as,
\[
J_\mu^+ = 2 \;  \left( J_\mu^1 - i J_\mu^2\right) \; .
\]

In order to accommodate the third (neutral) current $J^3$, we can
define the {\it hypercharge current} by,
\[
J_\mu^Y \equiv - \left(  \bar{\ld} \; \g_\mu \; \ld  + 
               2 \; \bar{\rs} \; \g_\mu \; \rs \right) = 
-  \left( \bar{\nu}_L \g_\mu \nu_L +  
          \bar{\ell}_L \g_\mu \ell_L + 
          2 \; \bar{\ell}_R \g_\mu \ell_R  \right) \; .
\]

The {\it electromagnetic current} can be written as
\[
J_\mu^{\text{em}} = - \;  \bar{\ell} \; \g_\mu \; \ell  
= -  \left(\bar{\ell}_L \g_\mu \ell_L + \bar{\ell}_R \g_\mu \ell_R \right)
= J_\mu^{3} + \frac{1}{2} J_\mu^Y \; .
\]

We should notice that neither $T_3$ nor $Q$ commute with $T_{1,2}$.
However, the `charges' associated to the currents $J^i$ and $J^Y$,  
\[
T^i = \int d^3 x \; J_0^i \;\;\;\;\;\; \mbox{and} \;\;\;\;\;\;
Y  = \int d^3 x \; J_0^Y \; ,
\]
satisfy the algebra of the $SU(2) \otimes U(1)$ group:
\[
[T^i , T^j] = i \; \epsilon^{ijk} T^k \;\; , \;\;\;\; \mbox{and}
\;\;\;\;
[T^i, Y] = 0 \; ,
\]
and the Gell-Mann--Nishijima relation between $Q$ and $T_3$
emerges in a natural way,
\begin{equation}
\fbox{$ \; \displaystyle{Q = T_3 + \frac{1}{2} Y} \; $} \; .
\label{gell:nish}
\end{equation}
With the aid of (\ref{gell:nish}) we can define the weak hypercharge
of the doublet ($Y_{\ld} = -1$) and of the fermion singlet ($Y_{\rs}
= -2$).

Let us follow our previous recipe for building a general gauge
theory. We have just chosen the candidate for the gauge group,
\[
\fbox{$ \; \displaystyle{SU(2)_L \otimes U(1)_Y} \;  $} \; .
\]
The next step is to introduce {\it gauge fields} corresponding to
each generator, that is,
\begin{eqnarray*}
SU(2)_L \;\; & \longrightarrow & \;\; 
W^1_\mu \; , \;\; W^2_\mu \; , \;\; W^3_\mu \; ,\\
U(1)_Y  \;\; & \longrightarrow & \;\; B_\mu \; .
\end{eqnarray*}

Defining the {\it strength tensors} for the gauge fields according to
(\ref{abe:f}) and (\ref{nonab:f}),
\begin{eqnarray*}
W_{\mu\nu}^i & \equiv & \del_\mu W^i_\nu - \del_\nu W^i_\mu + g \,
\epsilon^{ijk} \, W_\mu^j W_\nu^k \; , \\
B_{\mu\nu} & \equiv & \del_\mu B_\nu - \del_\nu B_\mu \; ,
\end{eqnarray*}
we can write the free Lagrangian for the gauge fields following the
results (\ref{abe:l}) and (\ref{nonab:l}),
\begin{equation}
\lag_{\text{gauge}} = - \frac{1}{4} W_{\mu\nu}^i W^{i \;\; \mu\nu} - 
                        \frac{1}{4} B_{\mu\nu} B^{\mu\nu} \; .
\label{l:gau}
\end{equation}

For the leptons, we write the free Lagrangian,
\begin{eqnarray}
\lag_{\text{leptons}} &=& \bar{\rs} \; i \dels \; \rs + 
                           \bar{\ld} \; i \dels \; \ld \nonumber \\
			&=& \bar{\ell}_R \; i \dels \; \ell_R + 
			    \bar{\ell}_L \; i \dels \; \ell_L +
			    \bar{\nu}_L \; i \dels \; \nu_L  \nonumber \\
	&=& \bar{\ell} \; i \dels \; \ell + \bar{\nu} \; i \dels \;
	\nu \; .
\label{free:fer}
\end{eqnarray}	
Remember that a mass term for the fermions  (\ref{m}) mixes the
right-- and left--components and would break the gauge invariance of
the theory from the very beginning. 

The next step is to introduce the fermion--gauge boson coupling via
the {\it covariant derivative}, {\it i.e.}
\begin{equation}
\ld \;\; : \;\; \;\; \del_\mu + i \; \frac{g}{2} \; { \tau^i} \; W^i_\mu 
+ i \; \frac{g'}{2} \; { Y} B_\mu \; ,
\label{cov:l}
\end{equation}
\begin{equation}
\rs \;\; : \;\; \;\; \del_\mu + i \; \frac{g'}{2}\; { Y} \; B_\mu \; ,
\label{cov:r}
\end{equation}
where $g$ and $g'$ are the coupling constant associated to the groups
$SU(2)_L$ and $U(1)_Y$ respectively, and
\begin{equation}
Y_{\ld_\ell} = -1 \;\;\; , \;\;\;\;\;\; Y_{\rs_\ell} = -2 \; .
\label{hyp:lep}
\end{equation}

Therefore, the fermion Lagrangian (\ref{free:fer}) becomes
\begin{eqnarray}
\lag_{\text{leptons}} & \longrightarrow & \lag_{\text{leptons}} 
 + \;
\bar{\ld} \; i \g^\mu \left( i\frac{g}{2} \tau^i W^i_\mu 
+ i \frac{g'}{2} Y B_\mu \right ) \ld \nonumber \\
&& \; + \; \bar{\rs} \; i \g^\mu \left(i \frac{g'}{2} Y B_\mu \right) \rs
\; .
\label{free:fer1}
\end{eqnarray}

Let us first pick up just the ``left'' piece of (\ref{free:fer1}),
\[
\lag_{\text{leptons}}^{\ld} = - g \; \bar{\ld} \; \g^\mu
\left(\frac{\tau^1}{2} W^1_\mu + \frac{\tau^2}{2} W^2_\mu \right) \ld
- g \; \bar{\ld} \; \g^\mu \frac{\tau^3}{2} \ld \; W^3_\mu
- \frac{g'}{2}  Y \bar{\ld} \; \g^\mu \ld \; B_\mu \; .
\]

The first term is {\it charged} and can be written as
\[
\lag_{\text{leptons}}^{\ld (\pm)} = 
- \frac{g}{2} \; \bar{\ld} \; \g^\mu 
\left( \ba{cc}
0 			& W^1_\mu - i W^2_\mu \\
W^1_\mu + i W^2_\mu 	& 0
\ea \right) \ld  \; .
\]

This suggests the definition of the {\it charged gauge bosons} as
\begin{equation}
\fbox{$ \; \displaystyle{W^\pm_\mu = 
\frac{1}{\sqrt{2}} (W_\mu^1 \mp W_\mu^2)} \; $} \; ,
\label{w}
\end{equation}
in such a way that
\begin{equation}
\lag_{\text{leptons}}^{\ld (\pm)} = - { \frac{g}{2\sqrt{2}}} 
\left[	\bar{\nu}  \g^\mu (1 - \gc) \ell \;  W^+_\mu +  
	\bar{\ell} \g^\mu (1 - \gc) \nu \; W^-_\mu \right] \; ,
\label{l:w2}
\end{equation}
reproduces exactly the $(V-A)$ structure of the weak charged current . 

When we compare the Lagrangian (\ref{l:w2}) with (\ref{l:w}) and take
into account the result from low--energy phenomenology (\ref{gw}) 
we see that $G_W = g/2\sqrt{2}$ and we obtain the
relation 
\begin{equation}
\fbox{$ \; \displaystyle{\frac{g}{2\sqrt{2}} = 
\left(\frac{M_W^2 G_F}{\sqrt{2}} \right)^{1/2}} \; $} \; .
\label{g:rel}
\end{equation}


Now let us treat the neutral piece of
$\lag_{\text{leptons}}$ (\ref{free:fer1}) that contains both left and
right fermion components,
\begin{eqnarray}
\lag_{\text{leptons}}^{(\ld + \rs) (0)} &=& 
- g \; { \bar{\ld} \; \left( \g^\mu \frac{\tau^3}{2} \right)
\ld} \; W^3_\mu
- \frac{g'}{2} \; { \left( \bar{\ld} \g^\mu Y \ld  + 
                      \bar{\rs} \g^\mu Y \rs \right) } B_\mu 
                      \nonumber \\
&=& - g \; { J_3^\mu} \; W^3_\mu - 
\frac{g'}{2} \; { J_Y^\mu} \; B_\mu \; ,
\label{l:0}
\end{eqnarray}
where the currents $J_3$ and $J_Y$ have been defined before, 
\begin{eqnarray*}
J_3^\mu &=& \frac{1}{2} (\bar{\nu}_L \g^\mu \nu_L - \bar{\ell}_L \g^\mu \ell_L)
\\
J_Y^\mu &=& - \left(\bar{\nu}_L \g^\mu \nu_L + \bar{\ell}_L \g^\mu \ell_L
+ 2 \bar{\ell}_R \g^\mu \ell_R \right) \; .
\end{eqnarray*}

Note that the `charges' respect the Gell-Mann--Nishijima relation
(\ref{gell:nish}) and currents satisfy,
\[
J_{\text{em}} = J_3 + \frac{1}{2} J_Y  \; .
\]

In order to obtain the right combination of fields that couples to the
electromagnetic current, let us make the rotation in the neutral
fields, defining the new fields $A$ and $Z$ by,
\begin{equation}
\left( \ba{c}
{ A_\mu} \\
{ Z_\mu}
\ea \right) =
\left( \ba{cc}
\cos \theta_W & \sin\theta_W \\
-\sin\theta_W & \cos \theta_W
\ea \right) 
\left( \ba{c}
B_\mu   \\
W^3_\mu
\ea \right) \; ,
\label{az}
\end{equation}
or,
\begin{eqnarray*}
W^3_\mu &=& \sin\theta_W A_\mu + \cos \theta_W Z_\mu \; , \\
B_\mu   &=& \cos \theta_W A_\mu - \sin\theta_W Z_\mu  \; , 
\end{eqnarray*}
where $\theta_W$ is called the Weinberg angle and the relation with
the $SU(2)$ and $U(1)$ coupling constants hold,
\begin{equation}
\fbox{$ \; \displaystyle{\sin\theta_W = 
\frac{g'}{\sqrt{g^2 + {g'}^2}}} \; $} \;\;\;\;\;
\fbox{$ \; \displaystyle{\cos\theta_W = 
\frac{g}{\sqrt{g^2 + {g'}^2}}}  \; $} \; .
\label{theta}
\end{equation}

In terms of the new fields, the neutral part of the fermion
Lagrangian (\ref{l:0}) becomes 
\begin{eqnarray}
\lag_{\text{leptons}}^{(\ld + \rs) (0)} &=& - ( { g \sin\theta_W}
J_3^\mu + 
\frac{1}{2} { g' \cos \theta_W}  J_Y^\mu) A_\mu
\nonumber \\
&& + (- g \cos\theta_W J_3^\mu + \frac{1}{2} g' \sin \theta_W  J_Y^\mu) Z_\mu
\nonumber \\
&=& - { g \sin\theta_W} \; { (\bar{\ell} \g^\mu \ell)}
\; A_\mu \nonumber \\
&& - \frac{g}{2 \cos\theta_W} \sum_{\psi_i = \nu, \ell} 
\bar{\psi_i} \g^\mu (g_V^i - g_A^i \gc) \psi_i Z_\mu \; ,
\label{neu:cur} 
\end{eqnarray}
and we easily identify the electromagnetic current coupled to the
photon field $A_\mu$ and the {\it electromagnetic charge},
\begin{equation}
\fbox{$ \; \displaystyle{e = g \sin\theta_W = g' \cos\theta_W} \; $}
\; .
\label{e}
\end{equation}

The Standard Model introduces a new ingredient, weak interactions
without change of charge, and make a specific prediction for the
vector ($V$) and axial ($A$) couplings of the $Z$ to the fermions,
\begin{equation}
\fbox{$ \; \displaystyle{g_V^i \equiv T_3^i - 2 Q_i \sin^2\theta_W} \;
$} \; ,
\label{gv}
\end{equation}
\begin{equation}
\fbox{$ \; \displaystyle{g_A^i \equiv T_3^i} \; $} \; .
\label{ga}
\end{equation}

This was a very successful prediction of the Standard Model since at
that time we had no hint about this new kind of weak interaction. 
The experimental confirmation of the existence of weak neutral
currents occurred more than five years after the model was proposed
\cite{Hasert:73a}.

Up to now we have in the theory:   
\begin{itemize}
\item 4 massless gauge fields $W^i_\mu$, $B_\mu$ or equivalently,
$W^\pm_\mu$, $Z_\mu$, and $A_\mu$;

\item 2 massless fermions: $\nu$, $\ell$.
\end{itemize}

The next step will be to add scalar fields in order to break
spontaneously the symmetry and use the Higgs mechanism to give mass
to the three weak intermediate vector bosons, making sure that the
photon remains massless.


\subsection{The Higgs Mechanism and the $W$ and $Z$ mass} \indent

In order to apply the Higgs mechanism to give mass to $W^\pm$ and
$Z^0$, let us introduce the scalar doublet
\begin{equation}
\Phi \equiv 	\left( \ba{c}
	      		\phi^+\\
	      		\phi^0
              \ea \right) \; .
\label{phi}
\end{equation}
From the relation (\ref{gell:nish}), we verify that the hypercharge of
the Higgs doublet is $Y = 1$. We introduce the Lagrangian
\[
\lag_{\text{scalar}} = \del_\mu \Phi^\dagger \; \del^\mu \Phi -
V(\Phi^\dagger\Phi) \; ,
\]
where the potential is given by
\begin{equation}
V(\Phi^\dagger\Phi) = \mu^2 \; \Phi^\dagger\Phi + 
\lambda \; (\Phi^\dagger\Phi)^2 \; .
\label{hig:pot}
\end{equation}

In order to maintain the gauge invariance under the $SU(2)_L \otimes
U(1)_Y$, we should introduce the covariant derivative 
\[
\del_\mu \to D_\mu = \del_\mu + i \; g \frac{\tau^i}{2} W^i_\mu 
+ i \; \frac{g'}{2} Y B_\mu \; .
\]

We can choose the {vacuum expectation value} of the Higgs field as,
\[
<\Phi>_0 = \left( \ba{c}
	      		0\\
	      		v/\sqrt{2}
              \ea \right) \; ,
\]
where
\begin{equation}
\fbox{$ \; \displaystyle{
v = \sqrt{ - \frac{\mu^2}{\lambda} }} \; $} \; .
\label{v}
\end{equation}

Since we want to preserve the exact electromagnetic symmetry to
maintain the electric charged conserved, we must break the original
symmetry group as
\[
SU(2)_L \otimes U(1)_Y  \to U(1)_{\text{em}} \; ,
\]
{\it i.e.} after the spontaneous symmetry breaking, the sub--group $
U(1)_{\text{em}}$, of dimension $1$, should remain as a symmetry of
the vacuum.

In this case the corresponding gauge boson, the photon, will remain
massless, according to results of section \ref{hig:nonabe}. We can
verify that our choice let indeed the vacuum invariant under
$U(1)_{\text{em}}$. This invariance requires that 
\[
e^{i \alpha Q} <\Phi>_0 \; \simeq \; \left(1 + i \; \alpha \; Q \right) 
<\Phi>_0 \; = \; <\Phi>_0 \; ,
\]
or, the operator $Q$ annihilates the vacuum, $Q  <\Phi>_0 \; = \; 0$.
This is exactly what happens: the electric charge of the vacuum is
zero,
\begin{eqnarray*}
Q  <\Phi>_0 &=& \left( T_3 + \frac{1}{2} Y \right) <\Phi>_0 
\\
&=&  \frac{1}{2}\left[ \left( \ba{cc}
	      		1 & 0 \\
	      		0 & -1
              \ea \right) + \left( \ba{cc}
	      		1 & 0 \\
	      		0 & 1
              \ea \right) \right] \left( \ba{c}
	      		0\\
	      		v/\sqrt{2}
              \ea \right) = 0 \; .
\end{eqnarray*}

The other gauge bosons, corresponding to the {\it broken generators}
$T_1$, $T_2$, and $(T_3 - Y/2) = 2 T_3 - Q$ should acquire mass. In
order to make this explicit, let us parametrize the Higgs doublet
{\it c.f.} (\ref{par:phi}),
\begin{eqnarray*}
\Phi &\equiv& \exp\left( i \frac{\tau^i}{2} \frac{\chi_i}{v} \right)
        	\left( \ba{c}
	      		0\\
	      		(v + H)/\sqrt{2}
              \ea \right) 
\\
&\simeq& <\Phi>_0 + \frac{1}{2\sqrt{2}} 
			\left( \ba{c}
	      		\chi_2 + i \chi_1\\
	      		2H - i \chi_3
                         \ea \right) =
\frac{1}{\sqrt{2}} 
			\left( \ba{c}
	      		i\sqrt{2} \omega^+\\
	      		v + H - i z^0
                         \ea \right) \; .
\end{eqnarray*}
where {$\omega^\pm$} and {$z^0$} are the {Goldstone
bosons}.

Now, if we make a $SU(2)_L$ gauge transformation with {$\alpha_i =
\chi_i/v$} (unitary gauge) the fields become
\begin{equation}
\Phi  \to \Phi' = 
\exp\left(- i \frac{\tau^i}{2} \frac{\chi_i}{v} \right) \Phi
= { \frac{(v + H)}{\sqrt{2}} 
        	\left( \ba{c}
	      		0\\
	      		1
              \ea \right) } \; .
\label{hig:dou}
\end{equation}
and the scalar Lagrangian can be written in terms of these new field
as
\begin{eqnarray}
\lag_{\text{scalar}} &=& \left|
\left(\del_\mu + i g \frac{\tau^i}{2} W^i_\mu + i \frac{g'}{2} Y B_\mu 
\right) \frac{(v + H)}{\sqrt{2}} 	\left( \ba{c}
	      		0\\
	      		1
              \ea \right)  \right|^2 
\nonumber \\
&& - \,  \mu^2 \, \frac{(v + H)^2}{2} - \lambda \, \frac{(v + H)^4}{4}
\; .
\label{l:sca}
\end{eqnarray}

In terms of the physical fields $W^\pm$ (\ref{w}) and  $Z^0$
(\ref{az}) the first term of (\ref{l:sca}), that contain the
vector bosons, is
\begin{eqnarray}
\left|
		\left( \ba{c}
	      		0\\
	      		\del_\mu H/\sqrt{2}
              \ea \right) + 
i \frac{g}{2} (v + H)	\left( \ba{c}
	      		W_\mu^+\\
	      		(-1/\sqrt{2}c_W) Z_\mu
              \ea \right) \right|^2 
\nonumber \\
= \frac{1}{2} \del_\mu H \del^\mu H +
\frac{g^2}{4} (v + H)^2 \left(W_\mu^+ W^{- \; \mu} + 
\frac{1}{2 c_W^2} Z_\mu Z^\mu \right) \; ,
\label{l:sca:1}
\end{eqnarray}
where we defined $c_W \equiv \cos\theta_W$. 

The quadratic terms in the vector fields, are,
\[
\frac{g^2 v^2}{4} W_\mu^+ W^{- \; \mu}  + 
\frac{g^2 v^2}{8\cos^2\theta_W} Z_\mu Z^\mu \; ,
\]
When compared with the usual mass terms for a charged and neutral
vector bosons,
\[
M_W^2 W_\mu^+ W^{- \; \mu}  + \frac{1}{2} M_Z^2 Z_\mu Z^\mu \; ,
\]
and we can easily identify
\begin{equation}
\fbox{$ \; \displaystyle{M_W = \frac{g v}{2}} \; $} \;\;\;\;\; 
\fbox{$ \; \displaystyle{M_Z = \frac{g v}{2 c_W} = \frac{M_W}{c_W}} \;
$} \; .
\label{mw:mz}
\end{equation}
We can see from (\ref{l:sca:1}) that no quadratic term in $A_\mu$
appears, and therefore, the photon remains massless, as we could
expect since the $U(1)_{\text{em}}$ remains as a symmetry of the
theory. 

Taking into account the low--energy phenomenology via the relation
(\ref{g:rel}), we obtain for the vacuum expectation value
\begin{equation}
\fbox{$ \; \displaystyle{v = \left( \sqrt{2} G_F\right )^{1/2} 
\simeq 246}$ GeV \; } \; ,
\label{v:num}
\end{equation}
and the Standard Model predictions for the $W$ and $Z$ masses are 
\[
M_W^2 = \frac{e^2}{4 s_W^2} v^2 =  \frac{\pi \alpha}{s_W^2} v^2
\simeq \left(\frac{37.2}{s_W} \; \mbox{GeV} \right)^2 
\sim (80 \; \mbox{GeV})^2 \; ,
\]
\[
M_Z^2 \simeq \left(\frac{37.2}{s_W c_W} \; \mbox{GeV}\right)^2 \sim
(90 \; \mbox{GeV})^2 \; ,
\]
where we assumed a experimental value for $s_W^2 \equiv \sin^2\theta_W
\sim 0.22$.

We can learn from (\ref{l:sca}) that one scalar boson, out of the
four degrees of freedom introduced in (\ref{phi}), is remnant of the
symmetry breaking. The search for the so called Higgs boson, remains
as one of the major challenges of the experimental high energy
physics, and will be discussed later in this course (see Sec.\
\ref{higgs:bos}).

The second term of (\ref{l:sca}) gives rise to terms involving
exclusively the scalar field $H$, namely,
\begin{equation}
-\frac{1}{2} (- 2 \mu^2) H^2 + \frac{1}{4} \mu^2 v^2
\left(\frac{4}{v^3} H^3 + \frac{1}{v^4} H^4 - 1 \right) \; .
\label{h:self}
\end{equation}

In (\ref{h:self}) we can also identify the Higgs boson mass term with
\begin{equation}
\fbox{$  \; \displaystyle{M_H = \sqrt{- 2 \mu^2}} \; $} \; ,
\label{mh}
\end{equation}
and the self--interactions of the $H$ field. In spite of predicting
the existence of the Higgs boson, the Standard Model does not give a
hint on the value of its mass since $\mu^2$ is {\it a priori}
unknown.


\section{Some General Remarks} \label{remarks} \indent

Let us  address some general features of the Standard Model:

\subsection{On the mass matrix of the neutral bosons} \indent

In order to have a different view of the rotation (\ref{az}) we
analyze the mass term for $W^3_\mu$ and $B_\mu$ in (\ref{l:sca}).
It can be written as
\begin{eqnarray*}
\lag^{W^3-B}_{\text{scalar}} &=& \frac{v^2}{2} \left|
\left(g \frac{\tau^3}{2} W^3_\mu + \frac{g'}{2} Y B_\mu 
\right)	\left( \ba{c}
	           0\\
	           1
              \ea \right)  \right|^2 
\\
&=& \frac{v^2}{8} \left[ 
\left( B_\mu  \;\; W^3_\mu \right)
			\left( \ba{cc}
	      		g^{\prime \, 2}	& - g g^\prime \\
	          	- g g^\prime 	&  g^2
              		\ea \right)	
\left( \ba{c}
B^\mu \\
W^{3 \, \mu}
\ea \right) \right] \; . 
\end{eqnarray*}

The mass matrix is not diagonal and has two eigenvalues, namely, 
\[
{ 0} \;\;\;\;  \mbox{and} \;\;\;\;
{ \left(\frac{1}{2}\right) 
\frac{(g^2  + g^{\prime \, 2})v^2}{4} =
\frac{1}{2} M_Z^2 } \; ,
\]
which correspond exactly to the photon ($M_A =0$) and $Z$ mass
(\ref{mw:mz}). 

We obtain a better understanding of the meaning of the Weinberg
angle rotation by noticing that the same rotation matrix used to
define the physical fields in (\ref{az}),
\[
R_W =
\left( \ba{cc}
\cos \theta_W & \sin\theta_W \\
-\sin\theta_W & \cos \theta_W
\ea \right) \; ,
\]
is the one that diagonalizes the mass matrix for the neutral gauge
bosons, {\it i.e.}
\[
R_W \; \frac{v^2}{4} \left( \ba{cc}
	      		g^{\prime \, 2}	& - g g^\prime \\
	          	- g g^\prime 	&  g^2
              		\ea \right) R_W^T = 
\left( \ba{cc}
	      	     0	& 0 \\
	             0 	& M_Z^2
              		\ea \right) \; .
\]


\subsection{On the $\rho$ Parameter}\indent

We can define a dimensionless parameter $\rho$ by:
\[
\rho = \frac{M_W^2}{\cos^2 \theta_W M_Z^2} \; ,
\]
that represents the relative strength of the neutral and charged
effective Lagrangians 
($J^{0 \; \mu} J^0_\mu/J^{+ \; \mu} J^-_\mu$),
\[
\rho = \frac{g^2}{8 \cos^2 \theta_W M_Z^2}\left/ \frac{g^2}{8
M_W^2}\right. \; .
\]

In the Standard Model, at tree level, the $\rho$ parameter is 1. This
is not a general consequence of the gauge invariance of the model, but
it is, in fact, a successful prediction of the model.

In a model with an arbitrary number of Higgs multiplets $\phi_i$
with isospin $T_i$ and third component $T_i^3$, and vacuum expectation
value $v_i$, the $\rho$ parameter is given by
\[
\rho = 
\frac{\sum_i \left[ T_i (T_i + 1) - (T_{3 \; i})^2 \right] v_i^2}{2 
\sum_i (T_{3 \; i})^2 v_i^2} \; ,
\]
which is 1 for an arbitrary number of doublets.

Therefore, $\rho$ represents a good test for the isospin structure of
the Higgs sector. As we will see later, it is also sensitive to
radiative corrections.


\subsection{On the Gauge Fixing Term} \indent 

The unitary gauge chosen in (\ref{hig:dou}) has the great
advantage of making the physical spectrum clear: the $W^\pm$ and $Z^0$
become massive and no massless Goldstone boson appears in the
spectrum. 

In this gauge the vector boson ($V$) propagator is given by 
\[
P_{\mu\nu}^U (V) = \frac{-i}{q^2 - M_V^2} \left(g_{\mu\nu} -
{ \frac{q_{\mu} q_{\nu}}{M_V^2} } \right) \; .
\]

Notice that $P_{\mu\nu}^U$ does {\it not} go like $\sim 1/q^2$ as $q
\to \infty$ due to the term proportional to $q_{\mu} q_{\nu}$. This
feature has some very unpleasant consequences. First of all there are
very complicated cancellations in the invariant amplitudes involving
the vector boson propagation at high energies. More dramatic is the
fact that it is very hard to prove the renormalizability of the
theory since it makes use of power counting analysis in the loop
diagrams. 

A way out to this problem \cite{Hooft:71,Fujikawa:72} is to add a {\it
gauge--fixing} term to the original Lagrangian,
\[
\lag_{\text{gf}} = -\frac{1}{2} 
\left(2 G_W^+ G_W^- + G_Z^{\; 2} + G_A^{\; 2} \right) \; ,
\]
with
\begin{eqnarray*}
 G_W^\pm &=& \frac{1}{\sqrt{\xi_W}} \left(\del^\mu W^\pm_\mu \mp i
\xi_W M_W \omega^\pm  \right) \; , \\
 G_Z &=& \frac{1}{\sqrt{\xi_Z}} \left(\del^\mu Z_\mu - 
\xi_Z M_Z z  \right) \; , \\
G_A &=& \frac{1}{\sqrt{\xi_A}} \del^\mu A_\mu \; ,
\end{eqnarray*}
where $ \omega^\pm$ and $z$ are the Goldstone bosons. This is called
the R$_\xi$ gauge.

Notice, for instance, that, 
\begin{eqnarray*}
- \frac{1}{2} G_Z^{\; 2} &=& -  \frac{1}{2 \xi_Z} (\del_\mu Z^\mu -
\xi_Z M_Z z)^2 \\
&=& \frac{1}{2} Z_\mu \left( \frac{1}{\xi_Z} \del^\mu \del^\nu
\right)Z_\nu
- \frac{1}{2} \; \xi_Z \; M_Z^2 \; z^2 + M_Z \; z \; \del^\mu Z_\mu \; ,
\end{eqnarray*}
where the last term that mixes the Goldstone ($z$) and the vector
boson ($\del^\mu Z_\mu$) is canceled by an identical term that comes
from the scalar Lagrangian [see Eq.\ (\ref{lag:com:2})]. 

In the R$_\xi$ gauge the vector boson propagators is
\begin{equation}
P_{\mu\nu}^{R_\xi} (V) = \frac{-i}{q^2 - M_V^2} \left[ g_{\mu\nu} -
(1 - \xi_V) \frac{q_{\mu} q_{\nu}}{q^2 - \xi_V M_V^2} \right] \; .
\label{rx}
\end{equation}

In this gauge the Goldstone bosons, with mass $\sqrt{\xi_V} M_V$, remain in
the spectrum and their propagators are given by,
\[
P^{R_\xi} (GB) = \frac{i}{q^2 - \xi_V M_V^2} \; .
\]
and the physical Higgs propagator remains the same.

In the limit of $\xi_V \to \infty$ the Goldstone bosons disappear and
the unitary gauge is recovered. Other gauge choices like Landau gauge
($\xi_V \to 0$) and  Feynman gauge ($\xi_V \to 1$) are contained in
(\ref{rx}). Therefore, all physical processes should not depend on the
parameter $\xi_V$.


\subsection{On the Measurement of $\sin^2\theta_W$ at Low Energies}
\indent

The value of the Weinberg angle is not predicted by the Standard
Model and should be extract from the experimental data. Once we have
measured $\theta_W$ (and of course, $e$) the value of the $SU(2)_L$
and $U(1)_Y$ coupling constants are determined via (\ref{e}).

At low energies the value of $\sin^2\theta_W$ can be obtained from
different reactions. For instance:

$\bullet$ The cross section for elastic neutrino--lepton
scattering
\[
\stackrel{\mbox{$\nu_\mu$}}{\mbox{$\bar{\nu}_\mu$}} + \; e \; \to \;\; 
\stackrel{\mbox{$\nu_\mu$}}{\mbox{$\bar{\nu}_\mu$}} + \; e \; ,
\]
which involve a $t$--channel $Z^0$ exchange is given by 
\[
\sigma = \frac{G_F^2 M_e E_\nu}{2 \pi}
\left[ \left(g_V^e \pm  g_A^e \right)^2 + 
\frac{1}{3}   \left( g_V^e \mp  g_A^e \right)^2 \right] \; .
\]
The vector and axial couplings of the electron to the $Z$ are
given by (\ref{gv}) and (\ref{ga}),
\[
g_V^e = -\frac{1}{2} + 2 \; { \sin^2\theta_W}
\;\; , \;\;\;\; 
g_A^e = -\frac{1}{2} \; , 
\]
and depend on the $\sin^2\theta_W$. For  $\nu_e$ reaction we should
make the substitution $g_{V,A}^e \to (g_{V,A}^e + 1)$ since in this
case there is also a $W$ exchange contribution. When the ratio $\sigma
(\nu_\mu e)/\sigma (\bar{\nu}_\mu e)$ is measured the systematic
uncertainties cancel out and yields $\sin^2\theta_W = 0.221 \pm 0.008$
\cite{pdg:98}. 

$\bullet$ Deep inelastic neutrino scattering from isoscalar targets
($N$). The ratio between the neutral ($NC$) and charged ($CC$) current
cross sections
\[
R_{\nu (\bar{\nu})} \equiv \frac{\sigma^{\text{NC}}[\nu (\bar{\nu}) N]}
                    {\sigma^{\text{CC}}[\nu (\bar{\nu}) N]}
                              \; ,
\]
depends on the $\sin^2\theta_W$ as
\[ 
R_{\nu ({ \bar{\nu}})} \simeq \frac{1}{2} - 
{ \sin^2\theta_W} + 
\frac{5}{9} [1 + r { (1/r)}] \; { \sin^4\theta_W} \; ,
\]
with $r \equiv \sigma^{\text{CC}}(\bar{\nu} N)/
\sigma^{\text{CC}}(\nu N)\simeq 0.44$. The measurement of these
reactions yields $\sin^2\theta_W = 0.226 \pm 0.004$ \cite{pdg:98}. 

$\bullet$ {Atomic parity violation}. The $Z^0$ mediated
electron--nucleus interaction in cesium, thallium, lead and 
bismuth can be described by the interaction Hamiltonian,
\[
{\cal H} = \frac{G_F}{2\sqrt{2}} \; Q_W \; \gc \;
\rho_{\, _{\text{nuc}}} \; ,
\]
with $Q_W$ being the ``weak charge'' that depends on the Weinberg
angle,  
\[
Q_W \simeq Z (1 - 4 \; { \sin^2\theta_W}) - N \; ,
\]
where $Z(N)$ is the number of protons (neutrons). This measurement
furnishes  $\sin^2\theta_W = 0.220 \pm 0.003$ \cite{pdg:98}. 

Nevertheless, the most precise measurements of the Weinberg angle are 
obtained at high energies, for instance in electron--positron
collisions at the $Z$ pole (see section \ref{z:phy}).  


\subsection{On the Lepton Mass} \indent

Note that the charged lepton is still massless, since
\[
M_\ell \; \bar{\ell} \; \ell = M_\ell \; (\bar{\ell}_R \; \ell_L + 
\bar{\ell}_L \; \ell_R)  \; ,
\]
mixes $L$ and $R$ components and breaks gauge invariance.  A way to
give mass in a gauge invariant way is via the Yukawa coupling of the
leptons with the Higgs field (\ref{hig:dou}), that is,
\begin{eqnarray}
\lag_{\text{yuk}}^\ell &=& - G_\ell 
\left[ \bar{\rs} \; \left( \Phi^\dagger \; \ld \right) +
 \left( \bar{\ld} \; \Phi \right) \; \rs \right] 
\nonumber \\
&=&  - G_\ell \frac{(v + H)}{\sqrt{2}} 
\left[ \bar{\ell}_R \; (0  \;\;\; 1) \left( \ba{c}
	      				\nu_L\\
	      				\ell_L
              				\ea \right) +
(\bar{\nu_L} \;\;\; \bar{\ell_L}) \left( \ba{c}
	      				0\\
	      				1
              				\ea \right) \ell_R \right]
\nonumber \\
&=& - \frac{G_\ell \; v}{\sqrt{2}} \; \bar{\ell} \; \ell  - 
      \frac{G_\ell}{\sqrt{2}} \; \bar{\ell} \; \ell \; H \; .
\label{yuk:lep}
\end{eqnarray}

Thus, we can identify the charged lepton mass,
\begin{equation}
\fbox{$ \; \displaystyle{M_\ell = \frac{G_\ell \; v}{\sqrt{2}}} \; $}
\; .
\label{ml}
\end{equation}

We notice that this procedure is able to generate a mass term for the
fermion in a gauge invariant way. However, it does not specify the
value of the mass since the Yukawa constant $G_\ell$ introduced in
(\ref{yuk:lep}) is arbitrary.

As a consequence, we obtain the Higgs--lepton coupling with strength,
\begin{equation}
\fbox{$ \; \displaystyle{C_{\bar{\ell}\ell H} = \frac{M_\ell}{v}} \; $} \; ,
\label{hl}
\end{equation}
which is a precise prediction of the Standard Model that should also
be checked experimentally.


\subsection{On the Cross Sections $e^+ e^- \to W^+ W^-$} \indent

A very interesting example on how the Standard Model is able to
improve the unitarity behavior of the cross sections is provided by
the $e^+ e^- \to W^+ W^-$ processes, which is presented in Fig.\ 5.

The first two diagrams are the $t$--channel neutrino exchange,
similar to the contribution of Fig.\ 1, and the $s$--channel photon
exchange.   Both of them are present in any theory containing charged
intermediate vector boson. However, the Standard Model introduces two
new contributions: the neutral current contribution ($Z$ exchange)
and the Higgs boson exchange ($H$).

The leading $p$--wave  divergence of the neutrino diagram, which is
proportional to $s$, is analogous to the one found in the reaction
$\nu \bar{\nu} \to W^+ W^-$. However, in this case it  is exactly 
canceled by the sum of the contributions of the photon ($A$) and the
$Z$. This delicate canceling is a direct consequence of the gauge
structure of the theory \cite{Llwellyn:73}.

\includegraphics[scale=0.8]{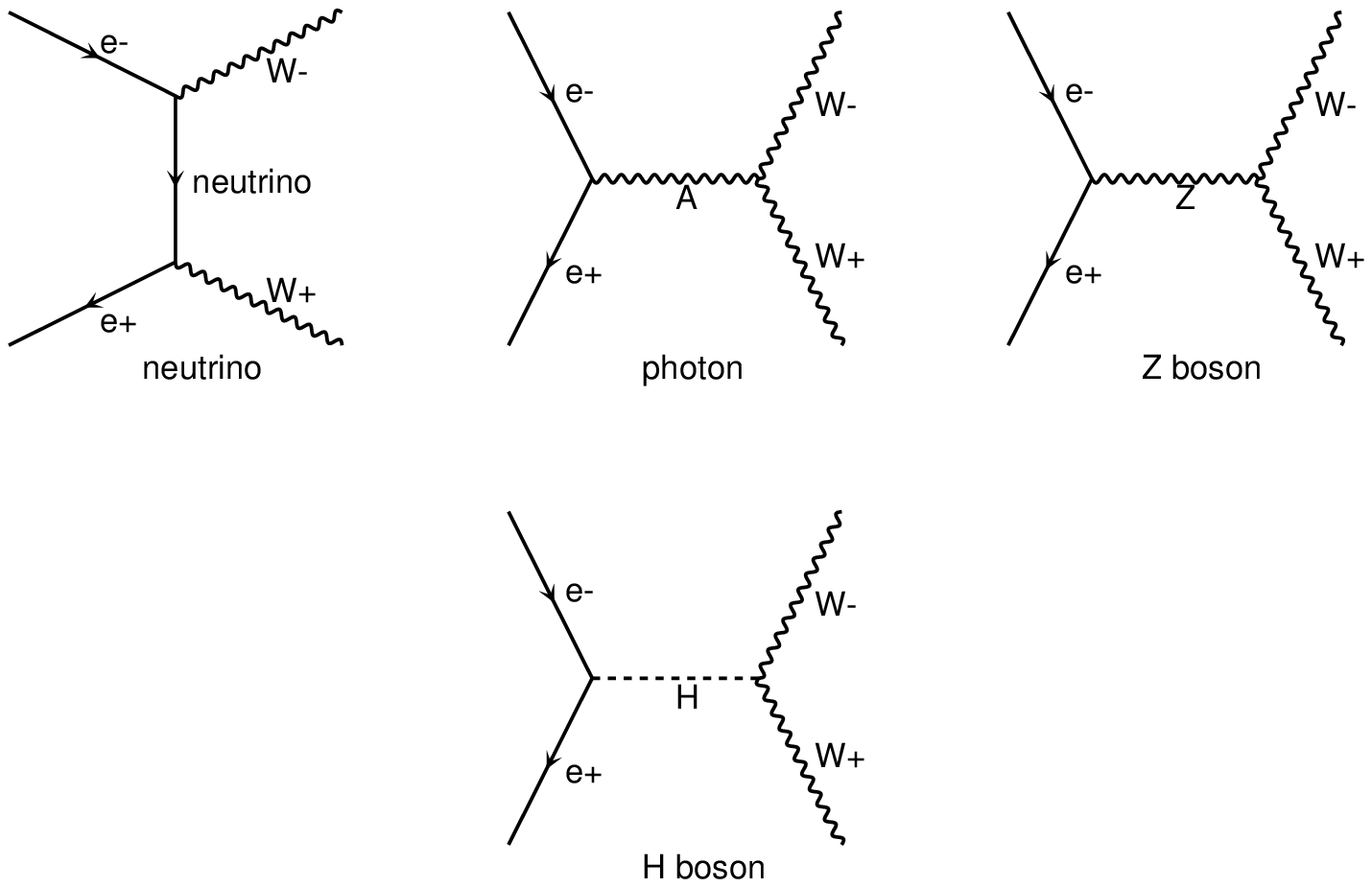}
\vskip 0.3cm
\begin{minipage}[h]{12cm}
\begin{center}
{\it Fig.\ 5: Feynman diagram for the reaction $e^+ e^- \to W^+ W^-$.}
\end{center}
\end{minipage}

\vskip 1cm
However, the $s$--wave scattering amplitude is  proportional to ($
m_f \, \sqrt{s}$) and, therefore, is also divergent at high energies.
This remaining divergence is canceled by the Higgs exchange diagram.
Therefore, the existence of a scalar boson, that gives rise to a
$s$--wave contribution and couples proportionally to the fermion mass,
is an essential ingredient of the theory. In Quigg's words
\cite{Quigg:83},

\begin{center}
\begin{minipage}[h]{12cm}
{\it ``If the Higgs boson did not exist, we should have to invent
something very much like it.''}
\end{minipage}
\end{center}

\newpage
\section{Introducing the Quarks} \indent 

In order to introduce the strong interacting particles in the
Standard Model we shall first examine what happens with the hadronic
neutral current when the Cabibbo angle (\ref{cab:ang}) is taken into
account.
We can write the hadronic neutral current in terms of the quarks $u$
and $d^\prime$,
\begin{eqnarray*}
J_\mu^H (0) &=& \bar{u} \g_\mu (1 - \gc) u + 
\bar{d} {}^\prime \g_\mu (1 - \gc) d^\prime
\\
&=&\bar{u} \g_\mu (1 - \gc) u + \cos^2\theta_C \; \bar{d} \g_\mu (1 - \gc) d
+ \sin^2\theta_C  \; \bar{s} \g_\mu (1 - \gc) s 
\\
&+& \cos\theta_C \sin\theta_C \; { 
\left[\bar{d} \g_\mu (1 - \gc) s + \bar{s} \g_\mu (1 - \gc) d \right]}
\; .
\end{eqnarray*}
We should notice that the last term generates flavor changing neutral
currents (FCNC), {\it i.e.} transitions like $d + \bar{s}
\leftrightarrow \bar{d} + s$, with the same strength of the usual
weak interaction. However, the observed FCNC processes are extremely
small. For instance, the branching ratio of charged kaons decaying
via charged current is,
\[
BR\left( K^+_{u\bar{s}} \to W^+ \to \mu^+ \nu\right) \simeq 63.5\% \; ,
\]
while process involving FCNC are very small \cite{pdg:98}:
\begin{eqnarray*}
BR\left(K^+_{u\bar{s}} \to \pi^+_{u\bar{d}} \nu \bar{\nu}
\right)  & \simeq &  4.2 \times 10^{-10} \; , \\
BR\left(K^L_{d\bar{s}} \to \mu^+ \mu^-\right) & \simeq & 7.2 
\times 10^{-9} \; .
\end{eqnarray*}

In 1970, Glashow, Iliopoulos, and Maiani proposed the GIM mechanism
\cite{Glashow:70}. They consider a fourth quark flavor, the charm
($c$), already introduced by Bjorken and Glashow in 1963. This extra
quark completes the symmetry between quarks ($u$, $d$, $c$, and $s$)
and leptons ($\nu_e$, $e$, $\nu_\mu$, and $\mu$) and suggests the
introduction of the weak doublets
\begin{eqnarray}
\ld_U & \equiv & \left( \ba{c}
	      		u\\
	      	 	d^\prime
              \ea \right)_L  = 
              \left( \ba{c}
              u\\
              \cos\theta_C \; d + \sin\theta_C \; s
                  \ea \right)_L 
              \; , \nonumber \\
\ld_C & \equiv & \left( \ba{c}
	      		c\\
	      		s^\prime
              \ea \right)_L  = 
              \left( \ba{c}
              c\\
               -\sin\theta_C \; d +  \cos\theta_C \; s
                  \ea \right)_L \; .
\label{dou:qua}
\end{eqnarray}
and the right--handed quark singlets,
\begin{equation}
\rs_U \;\;\; , \;\;\;\;\; \rs_D \;\;\; , \;\;\;\;\;  
\rs_S \;\;\; , \;\;\;\;\; \rs_C  \; .
\label{sin:qua}
\end{equation}
Notice that now all particles, {\it i.e.} the $T_3 = \pm 1/2$
fields,  have also the right components to enable a mass term for
them.

In order to introduce the quarks in the Standard Model, we should
start, just like in the leptonic case (\ref{free:fer}), from the free
massless Dirac Lagrangian for the quarks,
\begin{eqnarray}
\lag_{\text{quarks}} &=& \bar{\ld}_U \; i \dels \; \ld_U 
+ \bar{\ld}_C \; i \dels \; \ld_C \nonumber \\
&& + \bar{\rs}_U \; i \dels \; \rs_U + \cdots + 
\bar{\rs}_C \; i \dels \; \rs_C \; .
\label{l:qua}
\end{eqnarray}

We should now introduce the gauge bosons interaction via the
covariant derivatives (\ref{cov:l}) with the quark hypercharges
determined by the Gell-Mann--Ni\-shi\-ji\-ma relation
(\ref{gell:nish}), in such a way that the up--type quark charge is
$+2/3$ and the down--type $-1/3$,
\begin{equation}
Y_{\ld_Q} =   \frac{1}{3}  \;\;\; , \;\;\;\;\;
Y_{\rs_U} =   \frac{4}{3}  \;\;\; , \;\;\;\;\;
Y_{\rs_D} = - \frac{2}{3} \; .
\label{hyp:qua}
\end{equation}

Therefore, the charged weak couplings quark--gauge bosons, is given
by,
\begin{equation}
\lag_{\text{quarks}}^{(\pm)} = \frac{g}{2\sqrt{2}} 
\left[ \bar{u} \g^\mu (1 - \gc) d^\prime + 
\bar{c} \g^\mu (1 - \gc) s^\prime \right]
W_\mu^+ + \mbox{h.c.} \; .
\label{cha:qua}
\end{equation}

On the other hand, the neutral current receives a new contribution
proportional to  \[ \bar{c} \g_\mu (1 - \gc) c +  \bar{s} {}^\prime
\g_\mu (1 - \gc) s^\prime \] and becomes diagonal in the quarks
flavors, since the inconvenient  terms of $J_\mu^H (0)$ cancels out, 
avoiding the phenomenological problem with the FCNC.  For instance,
for the process $K^L \to \mu^+ \mu^-$, the GIM mechanism introduces a
new box contribution containing the $c$--quark that cancels most of
the $u$--box contribution and gives a result in agreement with 
experiment \cite{Gaillard:74}.

Finally, the neutral current interaction of the quarks become,
\begin{equation}
\lag_{\text{quarks}}^{(0)}  =  - \frac{g}{2c_W} 
 \sum_{\psi_q = u, \cdots, c} 
\bar{\psi_q} \g^\mu (g_V^q - g_A^q \gc) \psi_q \; Z_\mu \; ,
\label{neu:qua}
\end{equation}
with the vector and axial couplings for the quarks given by (\ref{gv})
and  (\ref{ga}), for $i=q$.


\subsection{On Anomaly Cancellation} \indent

In field theory, some loop corrections can violate a classical local
conservation law, derived from gauge invariance via Noether's theorem
The so--called anomaly is a disaster since it breaks Ward--Takahashi
identities and invalidate the proofs of renormalizability. The
vanishing of the anomalies is so important that have been used as a
guide for constructing realistic theories. 

Let us consider a generic theory with Lagrangian
\[
\lag_{\text{int}} = - g \left( \bar{R} \; \g^\mu \;  T^a_+ \;  R + 
       \bar{L} \; \g^\mu \;  T^a_- \; L \right) \; {\cal V}^a_\mu \; ,
\]
where $T^a_\pm$  are the generators in the right ($+$) and left
($-$)  representation of the matter fields, and   ${\cal V}^a_\mu$
are the gauge bosons. This theory will be anomaly free if 
\[
{\cal A}^{abc} = {\cal A}^{abc}_+ - {\cal A}^{abc}_- = 0 \; ,
\]
where ${\cal A}^{abc}_{\pm}$ is given by the following trace of
generators
\[
{\cal A}^{abc}_{\pm} \equiv 
\mbox{Tr} \left[ \{ T^a_{\pm} , T^b_{\pm}\} T^c_{\pm} \right] \; .
\]

In a $V-A$ gauge theory like the Standard Model, the only possible
anomalies come from $VVA$ triangle loops, {\it i.e.} loops with  two
vectors and one axial vertex and are proportional to:
\begin{eqnarray*}
{SU(2)^2 U(1)} \; & : & \;\;\;
\mbox{Tr} \left[ \{ \tau^a  , \tau^b \} Y \right] = 
\mbox{Tr} \left[ \{ \tau^a  , \tau^b \} \right] \mbox{Tr} \left[ Y \right] 
\propto \sum_{doub.} Y \\
{U(1)^3} \; & : &  \;\;\; 
\mbox{Tr} \left[ Y^3 \right] \propto \sum_{ferm.} Y^3 \; .
\end{eqnarray*}

Remembering the value of the hypercharge of the leptons
(\ref{hyp:lep}) and quarks (\ref{hyp:qua}), we can write for the
${SU(2)^2 U(1)}$ case,
\[
{\cal A}^{abc} \propto - \sum_{doub.} Y 
= - \left[ - 1 + { 3} \left(\frac{1}{3} \right) \right] = 0 \; ,
\]
and for the ${U(1)^3}$ case,
\begin{eqnarray*}
{\cal A}^{abc} \propto \sum_{ferm} Y^3_+ -  Y^3_- &=& 
\left\{ (-2)^3 + { 3} \left[ \left(
\frac{4}{3} \right)^3 + \left(\frac{- 2}{3}\right)^3 \right] \right\} 
\\
& - & \left\{ (-1)^3 + (-1)^3 +
{ 3} \left[ \left( \frac{1}{3} \right)^3 +  
\left( \frac{1}{3} \right)^3 \right] \right\} = 0 \; .
\end{eqnarray*}
where the 3 colors of the quarks were taken into account. 

This shows that the Standard Model is free from anomalies if the
fermions appears in complete multiplets, with the general structure: 
\[
\left\{	\left( \ba{c}
	      		 \nu_e    \\
	      		   e  
                        \ea \right)_L   ,  \;\; e_R  \; ,  \;\;
 	\left( \ba{c}
	      	  u  \\
	      	  d  
              \ea \right)_L   ,  \;\;  u_R \; ,  \;\; d_R 
\right\} \; ,
\]
that should be repeated always respecting this same structure:
\[
\left\{	\left( \ba{c}
	      		 \nu_\mu    \\
	      		 \mu  
                        \ea \right)_L   ,  \;\; \mu_R  \; ,  \;\;
 	\left( \ba{c}
	      	  c  \\
	      	  s  
              \ea \right)_L   ,  \;\;  c_R \; ,  \;\; s_R 
\right\} \; ,
\]

The discovery of the $\tau$ lepton in 1975 \cite{Perl:75}, and
of a fifth quark flavor, the $b$ \cite{Herb:77}, two years later, were
the evidence for a third fermion generation,
\[
\left\{	\left( \ba{c}
	      		 \nu_\tau    \\
	      		 \tau
                        \ea \right)_L   ,  \;\; \tau_R  \; ,  \;\;
 	\left( \ba{c}
	      	  t  \\
	      	  b  
              \ea \right)_L   ,  \;\;  t_R \; ,  \;\; b_R 
\right\} \; .
\]

The existence of complete {\it generations}, with no missing
partner,  is essential for the vanishing of anomalies. This was a
compelling  theoretical argument in favor of the existence of a top
quark before its discovery in 1995 \cite{Abe:95,Abachi:95}.


\subsection{The Quark Masses} \indent

In order to generate mass for both the up ($U_i = u$, $c$, and $t$)
and down  ($D_i = d$, $s$, and $b$) quarks, we need a $Y = -1$ Higgs
doublet. Defining the conjugate doublet Higgs as,
\begin{equation}
\tilde{\Phi} = i \; \sigma_2 \; \Phi^* = \left( \ba{c}
	      		\phi^{0^*} \\
	      	 	- \phi^-
              \ea \right) \; ,
\label{conj:hig}
\end{equation}
we can write the Yukawa Lagrangian for three generations of quarks
as, 
\begin{equation}
\lag_{\text{yuk}}^q = - \sum_{i,j = 1}^3 
\left[ G^U_{ij} \; \bar{\rs}_{U_i} \left( \tilde{\Phi}^\dagger \ld_j \right) +
G^D_{ij}\;  \bar{\rs}_{D_i} \left( \Phi^\dagger \ld_j \right) \right] +
\mbox{h.c.} \; .
\label{yuk:qua}
\end{equation}

From the vacuum expectation values of $\Phi$ and $\tilde{\Phi}$
doublets, we obtain the mass terms for the up
\[
\overline{(u^\prime, \; c^\prime, \; t^\prime)_R} \; {\cal M}^U \; \left( \ba{c}
	      						u^\prime\\
	      	 					c^\prime\\
							t^\prime
             					 \ea \right)_L +
\; \mbox{h.c.} \; ,
\]
and down quarks
\[
\overline{(d^\prime, \; s^\prime, \; b^\prime)_R} \; {\cal M}^D \; \left( \ba{c}
	      						d^\prime\\
	      	 					s^\prime\\
							b^\prime
             					 \ea \right)_L +
\; \mbox{h.c.} \; ,
\]
with the non--diagonal matrices ${\cal M}^{U(D)}_{ij} = (v/\sqrt{2})
\, G^{U(D)}_{ij}$.

The weak eigenstates ($q^\prime$) are linear superposition of the
mass eigenstates ($q$) given by the unitary transformations:
\[
\left( \ba{c}
{ u^\prime}\\
{ c^\prime}\\
{ t^\prime}
\ea \right)_{L,R} = U_{L,R} \left( \ba{c}
{ u} \\
{ c} \\
{ t} 
\ea \right)_{L,R} \;\; , \;\;\;\;\;\;\;
\left( \ba{c}
{ d^\prime}\\
{ s^\prime}\\
{ b^\prime}
\ea \right)_{L,R} = D_{L,R} \left( \ba{c}
{ d} \\
{ s} \\
{ b} 
\ea \right)_{L,R} \; , 
\]
where $U(D)_{L,R}$ are unitary matrices to preserve the form of the
kinetic terms of the quarks (\ref{l:qua}). These matrices diagonalize the mass
matrices, {\it i.e.},
\begin{eqnarray*}
U_R^{-1} {\cal M}^U U_L &=& 
\left( \ba{ccc}
m_u 	& 0 	& 0\\
0   	& m_c 	& 0\\
0	& 0	& m_t 
\ea \right)
\\
D_R^{-1} {\cal M}^D D_L &=& 
\left( \ba{ccc}
m_d 	& 0 	& 0\\
0   	& m_s 	& 0\\
0	& 0	& m_b 
\ea \right) \; .
\end{eqnarray*}

The $(V-A)$ charged weak current (\ref{cha:qua}), for three
generations, will be proportional to
\[
\overline{(u^\prime, \; c^\prime, \; t^\prime)_L} \; \g_\mu \; \left( \ba{c}
	      						d^\prime\\
	      	 					s^\prime\\
							b^\prime
             					 \ea \right)_L =
\overline{(u, \; c, \; t)_L} \;\; (U^\dagger_L D_L) \; \g_\mu \; 
						\left( \ba{c}
	      						d\\
	      	 					s\\
							b
             					 \ea \right)_L \; ,
\]
with the generation mixing of the mass eigenstates ($q$) described by: 
\[
V \equiv (U^\dagger_L \; D_L) \; .
\]

On the other hand, for the neutral current of the quarks
(\ref{neu:qua}), now becomes,
\[
\overline{(u^\prime, \; c^\prime, \; t^\prime)_L} \; \g_\mu \; \left( \ba{c}
	      						u^\prime\\
	      	 					c^\prime\\
							t^\prime
             					 \ea \right)_L =
\overline{(u, \; c, \; t)_L} \;\; (U^\dagger_L U_L) \; \g_\mu \; 
						\left( \ba{c}
	      						u\\
	      	 					c\\
							t
             					 \ea \right)_L \; .
\]
We can notice that there is no mixing in the neutral sector (FCNC) since 
the matrix $U_{L}$ is unitary: $U^\dagger_{L} U_{L} = 1$. 


The quark mixing, by convention, is restricted to the down quarks,
that is  with $T_3^q = -1/2$,
\[
\left( \ba{c}
d^\prime\\
s^\prime\\
b^\prime
\ea \right)_L  = V \; \left( \ba{c}
	      			d\\
	      	 		s\\
				b
             			 \ea \right)_L \; .
\]

$V$ is the Cabibbo--Kobayashi--Maskawa matrix
\cite{Cabibbo:63,Kobayashi:73}, that can be parame\-tri\-zed as
\[
V = R_1 (\theta_{23}) R_2(\theta_{13}, \delta_{13}) R_3(\theta_{12})
\; ,
\]
where $R_i(\theta_{jk})$ are rotation matrices around the axis $i$, 
the angle $\theta_{jk}$ describes the mixing of the generations $j$
and $k$ and $\delta_{13}$ is a phase. 

We should notice that, for three generations, it is not always
possible to choose the $V$ matrix to be real, that is $\delta_{13}
=0$, and therefore the weak interaction can violate $CP$ and $T$
\footnote{The violation of $CP$ can also occur in the interaction of
scalar bosons, when we have two or more scalar doublets. For a review
see Ref.\ \cite{Winstein:93}.}.

The Cabibbo--Kobayashi--Maskawa matrix can be written as
\[
V = \left( \ba{ccc}
 c_{12}c_{13}						&  
 s_{12}c_{13}						&   
 s_{13} \; e^{-i\delta_{13}}				\\
-s_{12}c_{23} - c_{12}s_{23}s_{13} \; e^{i\delta_{13}} 	& 
 c_{12}c_{23} - s_{12}s_{23}s_{13} \; e^{i\delta_{13}}	& 
 s_{23}c_{13}						\\
 s_{12}s_{23} - c_{12}c_{23}s_{13} \; e^{i\delta_{13}} 	&
-c_{12}s_{23} - s_{12}c_{23}s_{13} \; e^{i\delta_{13}}	&  
 c_{23}c_{13}	 
\ea \right) \; ,
\]
where $s_{ij}(c_{ij}) \equiv \sin (\cos) \theta_ij$. Notice that, in
the limit of $\theta_{23} = \theta_{13} \to 0$, we associate
$\theta_{12} \to \theta_{C}$, the Cabibbo angle (\ref{cab:ang}), and
\[
V \to 
\left( \ba{ccc}
 c_{12}		&  
 s_{12}		&   
 0		\\
-s_{12}  	& 
 c_{12}  	& 
 0		\\
 0		&
 0 		&  
 1	 
\ea \right) \; .
\]

Using unitarity constraints and assuming only three generations the
experimental value for the elements of the matrix $V$, with 90\% of
C.L., can be extract from weak quark decays and from deep inelastic
neutrino scattering \cite{pdg:98},
\[
V = \left( \ba{ccc}
 0.9742-0.9757	&  
 0.219-0226	&   
 0.002-0.005	\\
 0.219-0.225	& 
 0.9734-09749	& 
 0.037-0.043	\\
 0.004-0.014	&
 0.035-0.043	&  
 0.9990-0.9993	 
\ea \right) \; .
\]


\section{The Standard Model Lagrangian} \label{sm:lag} \indent

We end this chapter giving a birds' eye view of the Standard Model,
putting all terms together and writing the whole Lagrangian in a 
sche\-ma\-tic way.

\subsubsection{Gauge--boson + Scalar} \indent

The gauge--boson (\ref{l:gau}) and the scalar (\ref{l:sca})
Lagrangians give rise to the free Lagrangian for the photon, $W$,
$Z$, and the Higgs boson. Besides that, they generate triple and
quartic couplings among the vector bosons and also couplings involving
the Higgs boson:  
\begin{eqnarray}
&&\lag_{\text{gauge}} + \lag_{\text{scalar}} =
\label{gau:sca}
\\
&-& \frac{1}{4} F_{\mu\nu} F^{\mu\nu} 
- \frac{1}{2} W^+_{\mu\nu} W^{-\;\mu\nu} + M_W^2 W^+_{\mu} W^{-\;\mu}
\nonumber \\
&-& \frac{1}{4} Z_{\mu\nu} Z^{\;\mu\nu} + M_Z^2 Z_{\mu} Z^{\mu}
+ \frac{1}{2} \del_\mu H  \del^\mu H - \frac{1}{2} M_H^2 H^2 
\nonumber \\ 
&+& \fbox{$W^+W^-A$} + \fbox{$W^+W^-Z$} 
\nonumber \\  
&+& \fbox{$W^+W^-AA$}+\fbox{$W^+W^-ZZ$}+\fbox{$W^+W^-AZ$}+\fbox{$W^+W^-W^+W^-$} 
\nonumber \\  
&+&\fbox{$HHH$} + \fbox{$HHHH$} 
\nonumber \\ 
&+&\fbox{$W^+W^-H$} + \fbox{$W^+W^-HH$} + \fbox{$ZZH$} + \fbox{$ZZHH$}
\; .
\nonumber 
\end{eqnarray}

The vector--boson self--couplings that appear in (\ref{gau:sca}) are
strictly constrained by the $SU(2)_L \otimes U(1)_Y$ gauge invariance
and any small deviation from the Standard Model predictions would
destroy, for instance, the precise cancellation of the high--energy
behavior between the various diagrams, giving rise to an anomalous
growth of the cross section with energy. Therefore, the careful study
of the vector--boson self--interactions is a important test of the
Standard Model (see M.\ E.\ Pol, these Proceedings).


\subsubsection{Leptons + Yukawa} \indent

The leptonic (\ref{free:fer1}) and the Yukawa (\ref{yuk:lep})
Lagrangians are responsible for the lepton free Lagrangian and for the
couplings with the gauge bosons: photon (QED interaction), $W$ (charged
weak current) and $Z$ (neutral weak current). The mass terms are
generated by the Yukawa interaction which also gives rise to the coupling
of the massive lepton with the Higgs boson:
\begin{eqnarray}
&&\lag_{\text{leptons}} + \lag_{\text{yuk}}^\ell = 
\label{lep:yuk} \\
&& \sum_{\ell = e, \mu, \tau} \bar{\ell} (i \dels - m_\ell) \ell + 
\sum_{\nu_\ell = \nu_e, \nu_\mu, \nu_\tau} 
\bar{\nu}_\ell (i \dels) \nu_\ell  
\nonumber \\  
&+& \fbox{$\bar{\ell} \, \ell \, A$} 
+ \fbox{$\bar{\nu}_\ell \, \ell \, W^+$} + 
\fbox{$\bar{\ell} \, \nu_\ell \, W^-$} 
+ \fbox{$\bar{\ell} \, \ell \, Z$} + \fbox{$\bar{\nu}_\ell \, \nu_\ell \, Z$} 
\nonumber \\  
&+& \fbox{$\bar{\ell} \, \ell \, H$} \; .\nonumber
\end{eqnarray}

Even if the neutrinos have mass, as seems to suggest the recent 
experimental results \cite{SolarNeutrino:99,AtmosphericNeutrino:99},
their Dirac mass terms could be incorporated in the framework of the
Standard Model without any difficulty. The procedure would be similar
to the one that lead to the quark mass terms, that is, introducing a
right--handed component of the neutrino and a Yukawa coupling with
the conjugate Higgs doublet (\ref{conj:hig}). One may notice,
however, than being electrically  neutral neutrinos may also have a
Majorana mass which violates lepton number. The simultaneous
existence of both type of mass terms, Dirac and Majorana, can be used
to explain the smallness of the  neutrino mass as compared to the
charged leptons via the so called see-saw mechanism \cite{Seesaw:79}.


\subsubsection{Quarks + Yukawa} \indent

The quark Lagrangian (\ref{l:qua}) and the corresponding Yukawa
interaction (\ref{yuk:qua}) give rise to the free Dirac term and to
the electromagnetic and weak interaction of the quarks. A quark--Higgs
coupling is also generated,
\begin{eqnarray}
&&\lag_{\text{quarks}} + \lag_{\text{Yuk}}^q =
\label{qua:yuk} \\
&&\sum_{q = u, \cdots , t}  \bar{q} (i \dels - m_q) q 
\nonumber \\   
&+& \fbox{$\bar{q} \, q \, A$} 
\nonumber \\  
&+& \fbox{$\bar{u} \, d^\prime \, W^+$} + \fbox{$\bar{d}^\prime \, u \, W^-$} + 
\fbox{$\bar{q} \, q \, Z$} 
\nonumber \\  
&+& \fbox{$\bar{q} \, q \, H$} \; . \nonumber
\end{eqnarray}

Besides the propagators and couplings presented above, in a general
R$_\xi$ gauge, we should also take into account the contribution of
the Goldstone bosons and of the ghosts. The Faddeev--Popov ghosts
\cite{Faddeev:67} are important to cancel the contribution of the
unphysical (timelike and longitudinal) degrees of freedom of the gauge
bosons. 

A practical guide to derive the Feynman rules for the vertex and
propagators can be found, for instance, in Ref.\ \cite{Cheng:84},
where the complete set of rules for the Standard Model is presented.


\chapter{ ~ Beyond the Trees} \label{loops}

\section{Radiative Corrections to the Standard Model}\indent

It was shown that the Standard Model is a renormalizable field
theory.  This means that when we go beyond the tree level (Born
approximation) we are still able to make definite predictions for
observables. The general procedure to evaluate these quantities at
the quantum level is to collect and evaluate all the loop diagrams up
to a certain level. Many of these contributions are ultraviolet
divergent and a convenient regularization method ({\it e.g.}
dimensional regularization) should be used to isolate the divergent
parts. These divergences are absorbed in the bare couplings and
masses of the theory. Assuming a  renormalization condition ({\it
e.g.} on--shell or $\overline{\mbox{MS}}$ scheme), we can evaluate
all the counterterms. After all these ingredients are put together we
are able to obtain finite results for S--matrix elements that can be
translated in, for instance, cross sections and decay widths. The
predictions of the Standard Model for several observables are
obtained and can be compared with the experimental results for these
quantities enabling the theory to be falsified (in the Popperian
sense).

The subject of renormalization is very cumbersome  and deserves a
whole course by itself. Here we want to give just the minimum
necessary tools to enable the reader to appreciate the astonishing
agreement of the Standard Model, even at the quantum level, with the
recent experimental results. Very good accounts of the electroweak
radiative corrections can be found elsewhere
\cite{Hollik:94,Barbieri:95,Sirlin:99}.

Let us start considering the Standard Model Lagrangian which is given
by the sum of the contributions (\ref{gau:sca}), (\ref{lep:yuk}), and
(\ref{qua:yuk}). The $\lag_{SM}$ is a function of coupling constants
$g$ and $g^\prime$ and of the vacuum expectation value of the Higgs
field, $v$. The observables can be determined in terms of these
parameters and any possible dependence on other quantities like
$M_H$ or $m_t$ appears just through radiative corrections.

\linha

Therefore, we need three precisely known observables in order
to determine the basic input parameters of the model. A natural choice
will be the most well measured quantities, like, {\it e.g.}:
\begin{itemize}
\item The electromagnetic fine structure constant that can be
extracted, for instance, from the electron $g_e-2$ or from the quantum
Hall effect,
\[
\alpha (0)  = 1/137.0359895 (61)  \; ;
\]
\item The Fermi constant measured from the muon lifetime,
\[
G_F (\mu) = 1.16639 (1)  \times 10^{-5} \; \mbox{GeV}^{-2} \; ;
\]
\item The Z boson mass that was obtained by LEP at the $Z$ pole,
\[
M_Z = 91.1867(21) \; \mbox{GeV} \; .
\]
\end{itemize}
These input parameters can be written, at tree level, in terms of  
just $g$, $g^\prime$ and $v$ as
\begin{eqnarray}
\alpha_0 (0) &=& \frac{g^2 s_W^2}{4\pi} \; ,
\nonumber \\
G_{F_0} &=&  \frac{1}{\sqrt{2} v^2} \; , \\
M_{Z_0}^2 &=& \frac{g^2 v^2}{4 c_W^2}   \; . \nonumber
\label{input}
\end{eqnarray}
where the subscript $0$ indicates that these relations are valid at
tree level, and
\[
s_W^2 \equiv \frac{g^{\prime \; 2}}{g^2 + g^{\prime \; 2}}
\;\; , \;\;\;\; \mbox{and} \;\; 
{c_W^2} \equiv \frac{g^{2}}{g^2 + g^{\prime \; 2}}  \; ,
\]
depend only on $g$ and $g^\prime$:


\subsection{One Loop Calculations} \indent

Let us write the vacuum polarization amplitude (self--energy)
for vector bosons ($a, b = \gamma, W, Z$) as
\[
\Pi^{\mu\nu}_{ab} (q^2) \equiv g^{\mu\nu} \; \Pi_{ab}(q^2) + (q^\mu
q^\nu \;
\mbox{terms}) \; .
\]     

The terms proportional to $q^\mu q^\nu$ can be dropped since these
amplitudes should be plugged in conserved fermion currents, and from the
Dirac equation, they will give rise to terms that goes with the external
fermion mass that can be neglected in the usual experimental situation.

We can summarize the relevant quantities for the loop corrections of
the Standard Model \cite{Barbieri:95}:

\begin{itemize}
\item The vector and axial form factors of the $Z^0$ coupling, at $q^2
= M_Z^2$, which include both the vertex and the fermion self--energy
radiative corrections. From (\ref{neu:cur}) we can write
\[
V_{Zf\bar{f}}^\mu = - i \frac{g}{2 \cos\theta_W} 
         \bar{\psi_f} \g^\mu ( {  g_V^f} - 
                               {  g_A^f} \; \gc) \psi_f \; , 
\]
where (\ref{gv}) and (\ref{ga}) now are given by
\begin{eqnarray}
g_V^f & = & { \sqrt{\rho}} \; \left( T_3^f - 2 \; {\kappa_f} \;
Q_f \sin^2\theta_W \right) \; ,
\nonumber \\
g_A^f & = & { \sqrt{\rho}} \; T_3^f \; .
\label{cor:gvga}
\end{eqnarray}
which define the relative strength of the neutral and charged
currents, $\rho$, and the coefficient of the $\sin^2\theta_W$,
$\kappa_f$. Notice that at tree level, $\rho = \kappa_f =1$.

\item Correction to $\mu$--decay amplitude at $q^2 = 0$, which 
includes the box ($B$), vertex ($V$) and the fermion self--energy
corrections
\[
{\cal M} (\mu) = -i \; { \delta G_{(B,V)} }
\left[ \bar{\psi_e} \g^\mu (1 - \gc) \psi_{\nu_e} \right]
\left[ \bar{\psi_{\nu_\mu}} \g^\mu (1 - \gc) \psi_{\mu} \right] \; .
\]
\end{itemize}

We can write the corrections to the input parameters as,
\begin{eqnarray}
\alpha &=& \alpha_0 + { \delta \alpha}\; , \nonumber \\
M_Z^2 &=& M_{Z_0}^2 + { \delta M_Z^2} \; , \\
G_F &=& G_{F_0} + { \delta G_F}  \; ,    \nonumber 
\label{cor:imp:1}
\end{eqnarray}
where, in terms of the vacuum polarization amplitude, and $\delta
G_{(B,V)}$ the corrections become
\begin{eqnarray}
\frac{\delta \alpha}{\alpha} &=& - \Pi_{\gamma\gamma} (0) -  2
\frac{s_W}{c_W} \frac{\Pi_{\gamma Z} (0)}{M_Z^2} \; , \nonumber \\
\frac{\delta M_Z^2}{M_Z^2} &=& - \frac{\Pi_{ZZ}(M_Z^2)}{M_Z^2} \; ,  \\
\frac{\delta G_F}{G_F} &=& \frac{\Pi_{WW}(0)}{M_W^2}  +
\frac{\delta G_{(B,V)}}{G_F} \; . \nonumber
\label{cor:imp:2}
\end{eqnarray}              


\subsubsection{Correction to the Derived Observables} \indent

From the corrections to the input parameters we can estimate the
radiative corrections to the derived observables. Let us write the
tree level input variables $\alpha_0$, $G_{F_0}$, and $M_{Z_0}$ as
${\cal I}_0^i$. When we compute the radiative correction to the input
parameters ${\cal I}_0^i$, we have
\[
{\cal I}_0^i \;\; \longrightarrow \;\; {\cal I}^i ({\cal I}_0^i) = 
{\cal I}_0^i + \delta {\cal I}^i ({\cal I}_0^i) \; .
\]
The relation for the renormalized input variables, ${\cal I}^i$,  can
be inverted to write $ {\cal I}_0^i = {\cal I}_0^i ({\cal I}^i)$. 

The same holds true for any derived observable (${\cal O}$) or any
$S$--matrix element, that is,
\begin{eqnarray}
{\cal O}[{\cal I}^i_0 ({\cal I}^i)] &=& 
{\cal O}_0({\cal I}^i_0) + \delta {\cal O}({\cal I}^i_0) \nonumber\\
&=& {\cal O}_0({\cal I}^i - \delta {\cal I}^i) + \delta {\cal O}({\cal I}^i 
- \delta {\cal I}^i) \nonumber \\
&=&  {\cal O}_0({\cal I}^i)  - 
\sum_i \frac{\del {\cal O}_0}{\del {\cal I}^i} \, \delta {\cal I}^i 
+ \delta {\cal O}({\cal I}^i)
\nonumber \\
&\equiv&  {\cal O}_0({\cal I}^i) + \Delta {\cal O}({\cal I}^i) \; .
\label{cor:obs}
\end{eqnarray}
At one loop it is enough to renormalize just the input variables
${\cal I}^i$. However, at two loops it is necessary also to
renormalize all other parameters that intervene at one loop level
like $m_t$ and $M_H$.

As an example, let us compute the radiative correction to the $W$
boson mass. At tree level $M_W$ is given by [see (\ref{mw:mz})]
\[
M_{W_0}^2 =  c_{W_0}^2 M_{Z_0}^2 \; .
\]

Writing $c_W^2$ in terms of the input variables, we have
\[
{ c_W^2} = (1 - s_W^2)  = 
\left[1 - \left(\frac{4 \pi \alpha}{g^2}\right) \right] =
\left[1 - 
\left(\frac{\pi \alpha}{\sqrt{2} G_F M_Z^2 { c_W^2}}\right) \right] \; .
\]
Solving for $c_W^2$, we get
\[
M_{W_0}^2 ({\cal I}^i) = \left\{ \frac{1}{2} \left[ 1 + 
\left(1 - \frac{4 \pi \alpha}{\sqrt{2} G_F M_Z^2} \right)^{1/2} \right]
\right\} M_Z^2 \; .
\]
Taking into account the derivatives,
\begin{eqnarray*}
\frac{\del M_{W_0}^2}{\del \alpha} & = & \frac{s_W^2 c_W^2}{s_W^2 - c_W^2}
\frac{M_Z^2}{\alpha}  
\; , \\
\frac{\del M_{W_0}^2}{\del G_F} & = & - \frac{s_W^2 c_W^2}{s_W^2 - c_W^2} 
\frac{M_Z^2}{G_F} 
\; , \\
\frac{\del M_{W_0}^2}{\del M_Z^2} & = & - \frac{c_W^4}{s_W^2 - c_W^2}
\; .
\end{eqnarray*}
We obtain from (\ref{cor:obs}) the $M_W$ correction
\begin{eqnarray*}
M_W^2 &=& M_{W_0}^2 ({\cal I}^i)  - 
\sum_i \frac{\del M_{W_0}^2}{\del {\cal I}^i} \, \delta {\cal I}^i 
+ \delta M_{W}^2 ({\cal I}^i) \\
&=& c_W^2 M_{Z}^2  - \frac{c_W^2 M_Z^2}{s_W^2 - c_W^2}
\left(s_W^2 \frac{\delta \alpha}{\alpha} 
-  s_W^2 \frac{\delta G_F}{G_F}  
- c_W^2 \frac{\delta M_Z^2}{M_Z^2} \right) + \delta M_W^2 \; ,
\end{eqnarray*}
with
\[
\delta M_W^2 = - \Pi_{WW} (M_W^2) \; .
\]


\section{The $Z$ boson Physics} \label{z:phy} 

\subsection{Introduction} \indent

The most important experimental tests of the Standard Model in this
decade was performed at the $Z$ pole. The four LEP
Collaborations (Aleph, Delphi, L3, and Opal)\cite{Lep:99} and the
SLAC SLD Collaboration \cite{Sld:97} studied the reaction,
\[
e^+ e^- \to Z^0 \to f \bar{f} \; .
\]
The main purpose of these experiments was to test the Standard Model
at the level of its quantum corrections and also to try to obtain
some hint on the top quark mass and on the Higgs boson. 

\begin{figure}[ht]
\protect
\epsfxsize=10cm
\begin{center}
\mbox{\centerline{\psfig{file=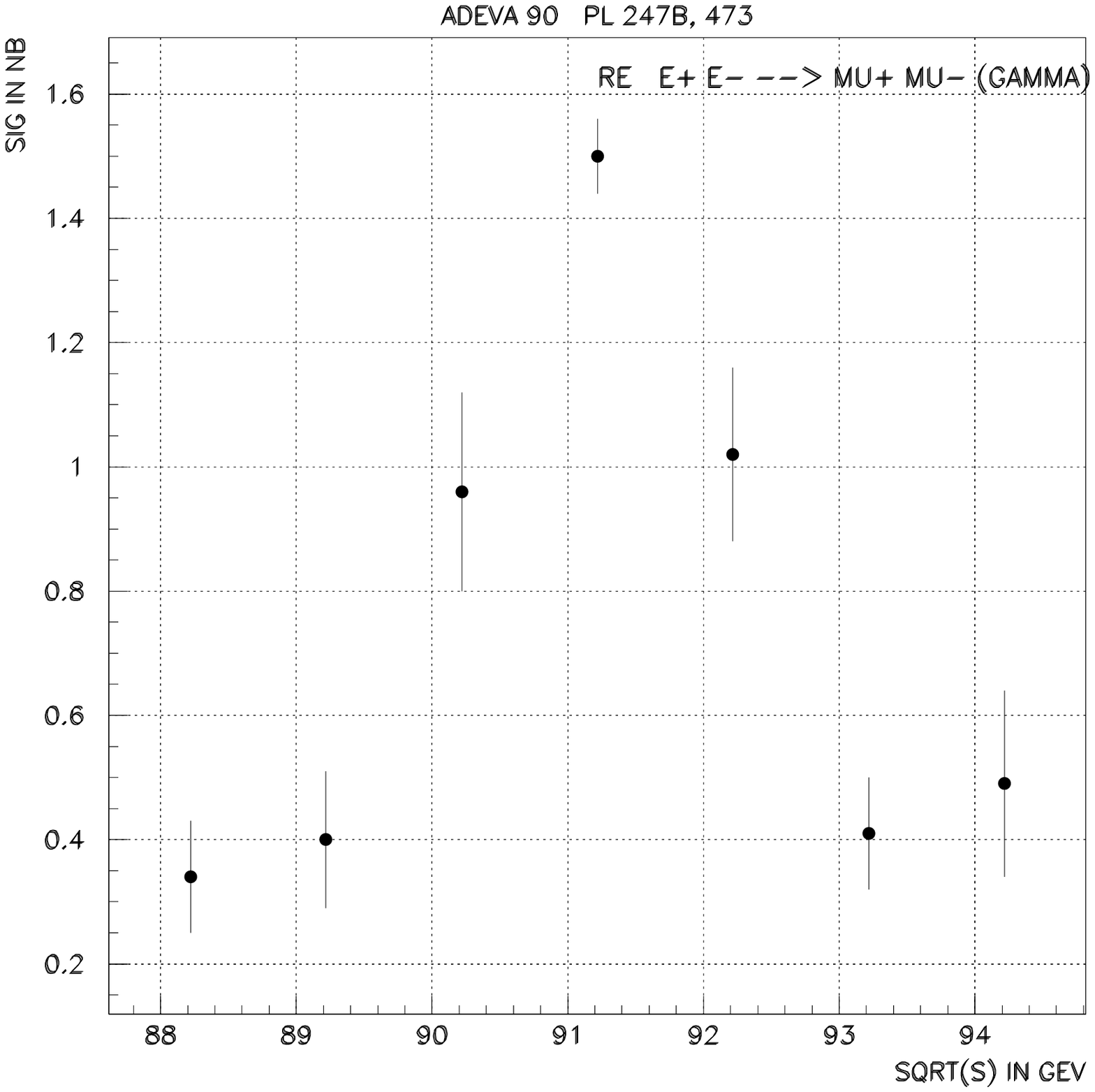,width=0.5\textwidth}}}
\end{center}
\end{figure}
\begin{minipage}[h]{12cm}
\begin{center}
{\it Fig.\ 6: The $Z$ profile measured by the L3 Collab. \cite{Adeva:90}}
\end{center}
\end{minipage}

\vskip 0.5cm
At CERN, after scanning the $Z$ resonance (see Fig.\ 6), data were
collected at the $Z$ peak, and around 17 millions of $Z$'s were
produced and studied.

The shape of the resonance is characterised by the cross section for the
fermion pair ($f\bar{f}$) production at the $Z$ peak,
\[
\sigma^0_{f\bar{f}} = \frac{12 \pi}{M_Z^2} 
		      \frac{\Gamma_e \; \Gamma_f}{\Gamma_Z^2} \; ,
\]
where the position of the peak gives the value  $M_Z = 91186.7 \pm
2.1$ MeV, the full width at half maximum (FWHM) represents the $Z$
width, $\Gamma_Z = 2493.9 \pm 2.4$ MeV, and the height  of the peak
gives the value of the total cross section for $f\bar{f}$
production, $\sigma^0_{f\bar{f}}$.

For the analysis of the $Z$ physics it is necessary to choose the
input parameters at the appropriate scale, $M_Z$.  The
relative uncertainty of the input parameters are:
\begin{center}
\begin{tabular}{ccc}
\hline\hline
&&\\[-0.2cm]
Parameter & Value &   Uncertainty   \\[0.2cm]
\hline
&&\\[-0.2cm]
$m_t$ \cite{Kim:99}	
& $174.3 \pm 5.1$ GeV  &  2.9 \% \\[0.2cm]
$\alpha_s (M_Z^2)$ \cite{Lep:99}	
&  $0.119 \pm 0.002$	&   1.7 \% \\[0.2cm]
$\alpha^{-1}(M_Z^2)$ \cite{Eidelmann:95,Steinhauser:98}  
&  $128.878 \pm 0.090$ 	  &   0.07 \%\\[0.2cm]
$M_Z$ \cite{Lep:99}
& $91186.7 \pm 2.1$ MeV & 0.0023 \% \\[0.2cm]
$G_F (\mu)$ \cite{pdg:98} 
& $(1.16639 \pm 0.00001) \times 10^{-5}$ GeV$^{-2}$ & 
0.00086 \% \\[0.2cm]
\hline\hline
\end{tabular}
\end{center}
\centerline{\it Table I: Relative uncertainty of the input parameters.}

\vskip 0.5cm
\noindent
For the Higgs boson mass we just have available a lower bound of $M_H
> 95.2$ GeV at 95\% of C.L. \cite{Higgs:99}.

Notice that, in spite of the very precise measurement of the
electromagnetic structure constant at low energy, which has a
relative uncertainty of just 0.0000045 \%, its value at $M_Z$ is much
less precise. This uncertainty arises from  the  contribution of
light quarks to the vacuum polarization. The evolution of 
$\alpha$ is given by
\begin{equation}
\alpha(M_Z^2) = \frac{\alpha (0)}{1 - \Delta\alpha} \; ,
\label{alp:mz}
\end{equation}
where
\[
\Delta\alpha = \Delta\alpha_{\text{lep}} + \Delta\alpha_{\text{had}} +
\Delta\alpha_{\text{top}} \; .
\]
The top quark contribution is proportional to $1/m_t^2 \sim 10^{-5}$.
The contributions from leptons ($\ell$) \cite{Steinhauser:98}  and
light quarks ($q$) \cite{Eidelmann:95} are:
\begin{eqnarray}
\Delta\alpha_{\text{lep}} &=&  0.031498 
\; ,
\nonumber \\
\Delta\alpha_{\text{had}} &=& - (0.02804 \pm
0.00065) \; .
\label{del:alp}
\end{eqnarray}
where the error in $\Delta\alpha_{\text{lep}}$ is negligible.
Therefore, the loss of precision comes from
$\Delta\alpha_{\text{had}}$ due to non--perturba\-ti\-ve QCD effects
that are large at low energies and to the imprecision in the light
quark masses.

Other important pure QED corrections are the initial and
final state photon radiation. The initial state radiation is taken
into account by convoluting the cross section with the radiator
function $H(k)$,
\[
\sigma(s) = \int_0^{k_{\text{max}}} dk \; H(k) \sigma_0[s(1-k)] \; ,
\]
where $k_{\text{max}}$ represents a cut in the maximum radiated
energy. The radiator function takes into account virtual and real photon
emissions and includes soft photon resummation \cite{Montagna:97}.

The final state radiation is included by multiplying the bare cross
sections and widths by the QED correction factor, 
\[
\left(1 + \frac{3\alpha Q_f^2}{4\pi}\right)
\simeq ( 1 + 0.002 \; Q_f^2 ) \; .
\]
where $Q_f$ is the fermion charge.


\subsection{The Standard Model Parameters} \indent

We present in the following sections the Standard Model predictions
for some observables. We compare these predictions with the values
measured by the CERN LEP and at the SLAC SLD Collaborations, and 
stress the very impressive agreement between them.

\subsubsection{$Z$ Partial Widths} \indent

The $Z$ width into a fermion pair, at tree level, is given in the
Standard Model by,
\begin{equation}
\Gamma(Z \to f \bar{f} ) = 
C \frac{G_F M_Z^3}{6\sqrt{2}\pi} \left[ (g_A^f)^2 + (g_V^f)^2 \right] \; ,
\label{gam:l}
\end{equation}
where $C$ refers to the fermion color, {\it i.e.},
\[
C = \left\{ \ba{l}
1 \; , \;\; \mbox{for { leptons}} \; ,\\
3 \left[ 1 + \alpha_s(M_Z)/\pi + 1.409 \alpha_s^2 (M_Z)/\pi^2 +
\cdots   \right] \; , \;\; \mbox{for quarks} \; .
 \ea \right.  
\]
where the QCD corrections were included for quarks. At loop level we
should consider the modifications to $g_V^f$ and $g_A^f$
(\ref{cor:gvga}) and the appropriate QED corrections discussed in the
last section. 

The value of the partial width for the different fermion flavors are
\begin{center}
\begin{tabular}{cr}
\hline\hline
&\\[-0.2cm]
~~~ $f \bar{f}$ Pair ~~~  & ~~~~~ Partial Width ~~~ \\[0.2cm]
\hline
&\\[-0.2cm]
$\nu \bar{\nu}$ & ~~~ 167.25 MeV \\ 
$e^+e^-$ 	& ~~~  84.01 MeV \\ 
$u\bar{u}$ 	& ~~~ 300.30 MeV \\ 
$d\bar{d}$ 	& ~~~ 383.10 MeV \\ 
$b\bar{b}$ 	& ~~~ 376.00 MeV \\ [0.2cm]
\hline\hline
\end{tabular}
\end{center}
\centerline{\it Table II: $Z \to f \bar{f}$ partial widths.}

\vskip 0.5cm
The experimental results for the partial widths are \cite{Lep:99},
\begin{eqnarray*}
\Gamma_\ell & \equiv & \Gamma(Z \to \ell^+ \ell^-) =  
83.90 \pm 0.10 \; \mbox{MeV} \; , 
\\ 
\Gamma_{\text{had}}    & \equiv & \Gamma(Z \to \mbox{hadrons}) =
1742.3 \pm 2.3 \; \mbox{MeV} \; , 
\\
\Gamma_Z & \equiv & \Gamma(Z \to \mbox{all}) =  2493.9 \pm 2.4 \; 
\mbox{MeV} \; , 
\\
\Gamma_{\text{inv}} & \equiv & \Gamma_Z - 3 \, 
\Gamma_\ell - \Gamma_{\text{had}} = 500.1 \pm 1.9 \; \mbox{GeV} \; , 
\end{eqnarray*}
where we assume three leptonic channels ($e^+e^-$, $\mu^+\mu^-$, and
$\tau^+\tau^-$), and $\Gamma_{\text{inv}}$ is the invisible $Z$
width. 

\subsubsection{Number of Neutrino Species} \indent

We can extract information on the number of light neutrino species by
supposing that they are responsible for the invisible width, {\it
i.e.} $\Gamma_{\text{inv}} = N_\nu  \Gamma_\nu$. The LEP data
\cite{Lep:99} gives the ratio of the invisible and leptonic $Z$
partial widths, $\Gamma_{\text{inv}}/\Gamma_\ell = 5.961 \pm
0.023$. On the other hand, Standard Model predicts the
$(\Gamma_\nu/\Gamma_\ell)_{\text{SM}} = 1.991 \pm 0.001$, where the
error is associated to the variation of $m_t$ and $M_H$. In the ratio
of these two expressions, $\Gamma_\ell$ cancels out and yields the
number of neutrino species,
\[
N_\nu = 2.994 \pm 0.011 \; ,
\]
where $N_\nu$ represents the total number of neutrino flavors that
are accessible kinematically to the $Z$, that is $M_\nu < M_Z/2$.
This result  indicates that, if the observed pattern of the first
three generations is assumed, then we have only these families of
fermions in nature.  


\subsubsection{Radiative Corrections Beyond QED}\indent

An important question to be asked when comparing the Standard Model
predictions with experimental data is if the effect of pure weak
radiative correction  could indeed be measured.  This question can be
answered by looking, for instance, at the plot of $\sin^2
\theta_{\text{eff}} \times \Gamma_\ell$ (Fig.\ 7), where
$\Gamma_\ell$ is given by (\ref{gam:l}) and, 
\[
{\sin^2 \theta_{\text{eff}}} \equiv \frac{1}{4} 
\left(1 - \frac{g_V^\ell}{g_A^\ell} \right) = 0.23157 \pm 0.00018 \; .
\]

The point at the lower--left corner shows the prediction when only
the QED (photon vacuum polarization) correction is included and the
respective variation for $\alpha(M_Z^2)$ varying by one standard
deviation. The Standard Model prediction, with the full (QED + weak)
radiative correction, is represented by the band that reflects the
dependence on the Higgs (95 GeV $< M_H <$ 1000 GeV) and on the top
mass (169.2 GeV $< m_t <$ 179.4  GeV). We notice that the presence of
genuine electroweak correction is quite evident. 

\begin{figure}[ht]
\protect
\epsfxsize=10cm
\begin{center}
\mbox{\psfig{file=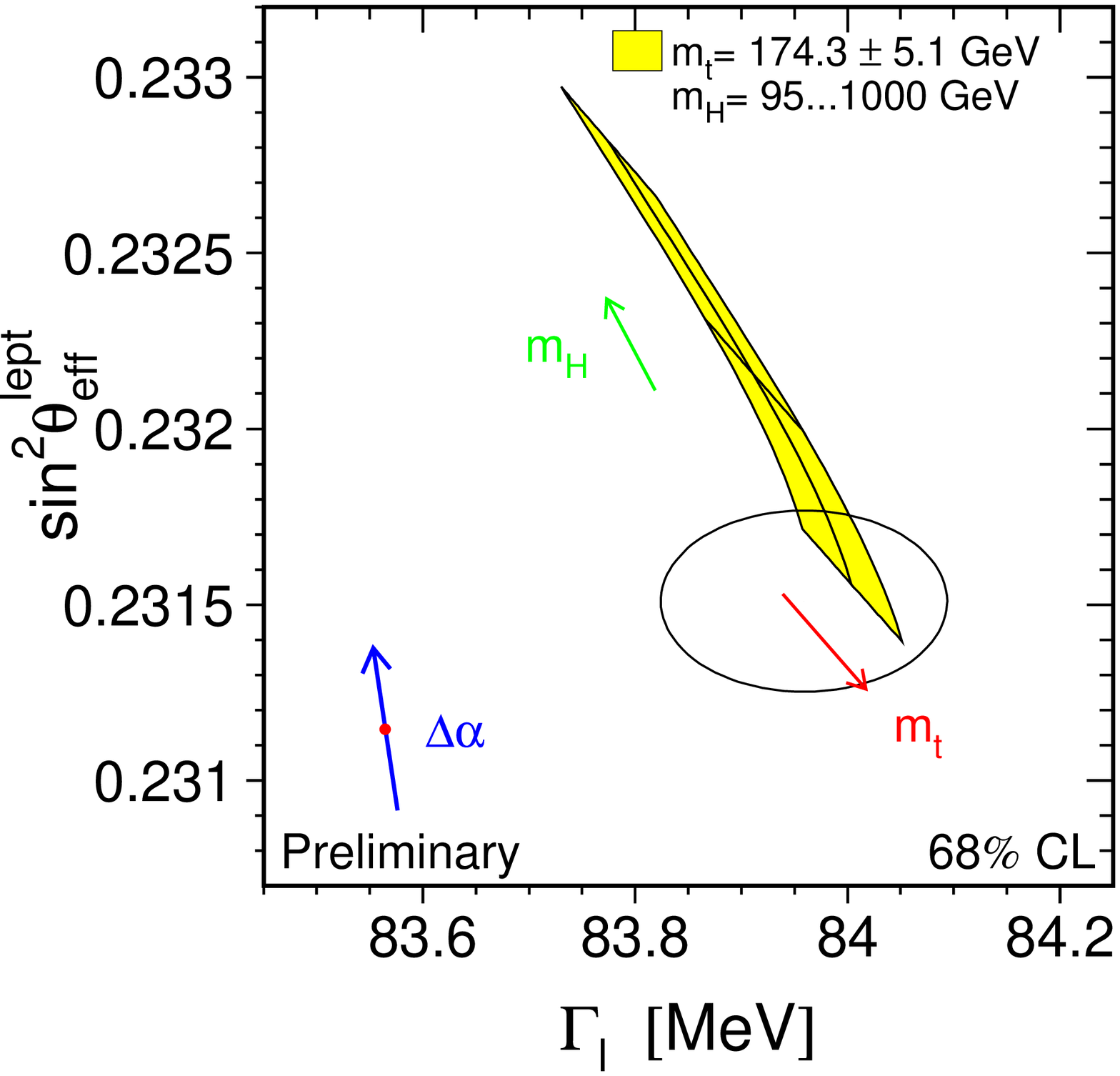,width=0.7\textwidth}}
\end{center}
\begin{center}
\begin{minipage}[h]{12cm}
\begin{center}
{\it Fig.\ 7: LEP + SLD measurements of $\sin^2
\theta_{\text{eff}}$ and $\Gamma_\ell$, compared to the Standard
Model prediction \cite{LepSite}.}
\end{center}
\end{minipage}
\end{center}
\end{figure}

Another important evidence for pure electroweak correction comes from
the radiative correction ${\Delta r_W}$ to the relation between $M_W$
and $G_F$,
\begin{equation}
\left(1 - \frac{M_W^2}{M_Z^2} \right)  \frac{M_W^2}{M_Z^2} =
\frac{\pi \alpha(M_Z^2)}{\sqrt{2} G_F M_Z^2 (1- {\Delta r_W})} \; ,
\label{mw:gf}
\end{equation}         
where $\alpha(M_Z^2)$ is given by (\ref{alp:mz}), and therefore, the
effect of the running of $\alpha$ was subtracted in the definition of
${\Delta r_W}$. Taking into account the value measured at LEP and
Tevatron, $M_W = 80.394 \pm 0.042$ GeV, we have  \cite{Sirlin:99}:
${(\Delta r)}_W^{\text{exp}} =  - 0.02507 \pm 0.00259$.  Thus, the
correction representing only the electroweak contribution, not
associated with the running of $\alpha$, is $\sim 10 \; \sigma$
different from zero.

\linha


\subsubsection{$g_V^\ell$, $g_A^\ell$, and the Lepton Universality}  \indent

The partial $Z$ width in the different lepton flavors is able to
provide a very important information on the universality of the
electroweak interactions. The values of $g_V^\ell$ and
$g_A^\ell$ can be plotted for $\ell = e$, $\mu$ and $\tau$.

\begin{figure}[ht]
\protect
\epsfxsize=10cm
\begin{center}
\mbox{\psfig{file=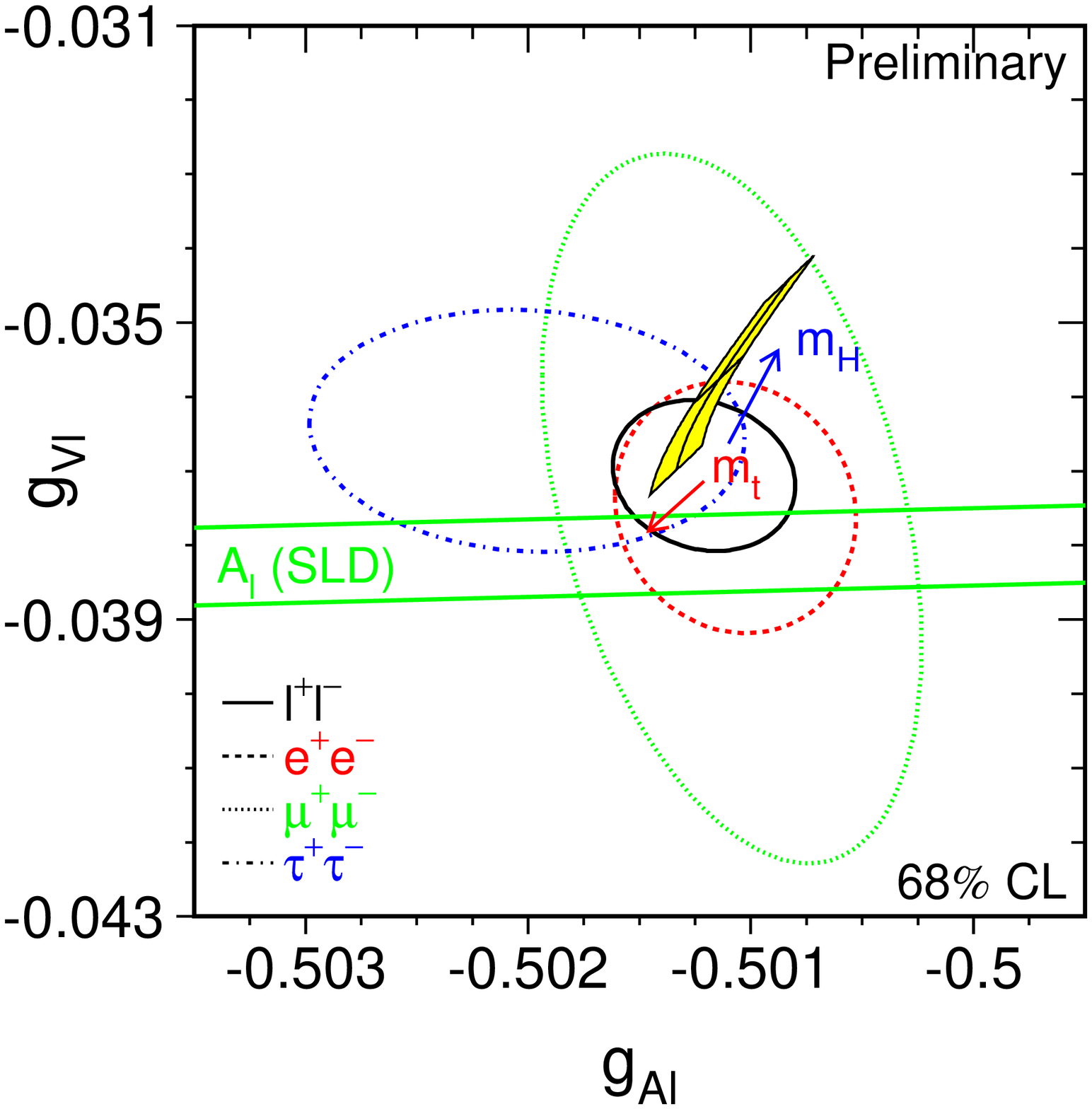,width=0.7\textwidth}}
\end{center}
\begin{center}
\begin{minipage}[ht]{12cm}
\begin{center}
{\it Fig.\ 8: 68\% C.L. contours in the $g_V^\ell \times g_A^\ell$
plane. The solid line is a fit assuming lepton universality. The band
corresponds to the SLD result from $A_{LR}$ (\ref{alr}) measurements 
\cite{LepSite}.}
\end{center}
\end{minipage}
\end{center}
\end{figure}

The result present in Fig.\ 8 shows that the measurements of
$g_V^\ell \times g_A^\ell$ are consistent with the hypothesis that the
electroweak interaction is universal and yields
\[
g_V^\ell = -0.03703 \pm 0.00068 \;\; , \;\;\;\;
g_A^\ell = -0.50105 \pm 0.00030 \; .
\]
Notice that the value of $g_A^\ell$ disagrees with the Born
prediction of $-0.5$ (\ref{ga}) by 3.5 $\sigma$. However they are  in
very good agreement with the Standard Model values \cite{pdg:98}: 
$(g_V^\ell)^{\text{SM}} = -0.0397 \pm 0.0003$ and
$(g_A^\ell)^{\text{SM}} = -0.5064 \pm 0.0001$.
This is another important evidence of the weak radiative corrections.


\subsubsection{Asymmetries} \indent

Since parity violation comes from the difference between the right
and left couplings of the $Z^0$ to fermions, it is convenient to
define the combination of the vector and axial  couplings of the
fermions as
\begin{equation}
A_f = \frac{2 g_V^f g_A^f}{(g_V^f)^2 + (g_A^f)^2} \; .
\label{af}
\end{equation}

The events $e^+ e^- \to f^+ f^-$ can be characterized by 
the momentum  direction of the emitted fermion. We say that the final
state fermion ($f^-$) travels forward ($F$) or backward ($B$) with
respect to the electron ($e^-$) beam. Therefore, we can define the
forward--backward  asymmetry by
\[
{A_{FB}} \equiv \frac{\sigma_F - \sigma_B}{\sigma_F + \sigma_B} \; ,
\]
and at the $Z$ pole, this asymmetry is given by
\begin{equation}
A_{FB}^{0\, , \, f} = \frac{3}{4} A_e A_f \; .
\label{afb}
\end{equation}

The measurement of $A_{FB}^{0\, , \, f}$ for charged leptons, and
$c$ and $b$ quarks give us information only on the product of
$A_e$ and $A_f$. On the other hand, the measurement of the $\tau$
lepton polarization is able to determine the values of $A_e$ and
$A_\tau$ separately. The longitudinal $\tau$ polarization is defined
as
\[
{\cal P}_\tau \equiv \frac{\sigma_R - \sigma_L}{\sigma_R + \sigma_L}
\; ,
\]
where $\sigma_{R(L)}$ is the cross section for tau--lepton pair
production of a right (left) handed $\tau^-$. At the $Z$ pole, ${\cal
P}_\tau$ can be written in terms of scattering $(e^-,\tau^-)$ angle
$\theta$ as,
\[
{\cal P}_\tau = - \frac{A_\tau (1 +\cos^2\theta) + 2 A_e \cos\theta}
               {1 + \cos^2\theta + 2 A_e A_\tau\cos\theta} \; .
\]
This yields \cite{Lep:99}
\[
A_e = 0.1479 \pm 0.0051 \;\; , \;\;\; A_\tau = 0.1431 \pm 0.0045 \; ,
\]
which are in agreement with the lepton universality ($A_\ell = 0.1469
\pm 0.0027$). They are also in agreement with the Standard Model
prediction: $A_\ell^{\text{SM}} = 0.1475 \pm 0.0013$. 

This result can be used to extract information on the heavy quark
couplings: $A_c = 0.646 \pm 0.043$ and $A_b = 0.899 \pm 0.025$, which
should be compared with the  Standard Model values of 
$A_c^{\text{SM}} = 0.6679 \pm 0.0006$ and $A_b^{\text{SM}} = 0.9348
\pm 0.0001$.

Another interesting asymmetry that can be measured by SLD is the
left--right cross section asymmetry,
\begin{equation}
A_{\text{LR}} = \frac{\sigma_L - \sigma_R}{\sigma_L + \sigma_R} \; ,
\label{alr}
\end{equation}
where $\sigma_{L(R)}$ is the cross section for (left--) right--handed
incident electron with the positron kept unpolarized.  Since, at the
$Z$ pole, $A_{\text{LR}} = A_e$, we can get the best measurement of the
electron couplings: $A_e = 0.1510 \pm 0.0025$ (see Fig.\ 8).  


\subsubsection{Higgs Mass Sensitivity} \indent

In order to give an idea of the sensitivity of the different
electroweak observables to the Higgs boson mass, we compare in Fig.\
9 the experimental values with the the Standard Model theoretical
predictions, as a function of $M_H$.

The vertical band represents the experimental measurement with the
respective error. The theoretical prediction includes the errors in 
$\alpha(M_Z^2)$, from $\Delta\alpha_{\text{had}}$ (\ref{del:alp}), 
$\alpha_s(M_Z^2)$, and $m_t$ (see Table I). 

\newpage
\null
\vspace{-2cm}
\includegraphics[width=13cm,height=9.8cm]{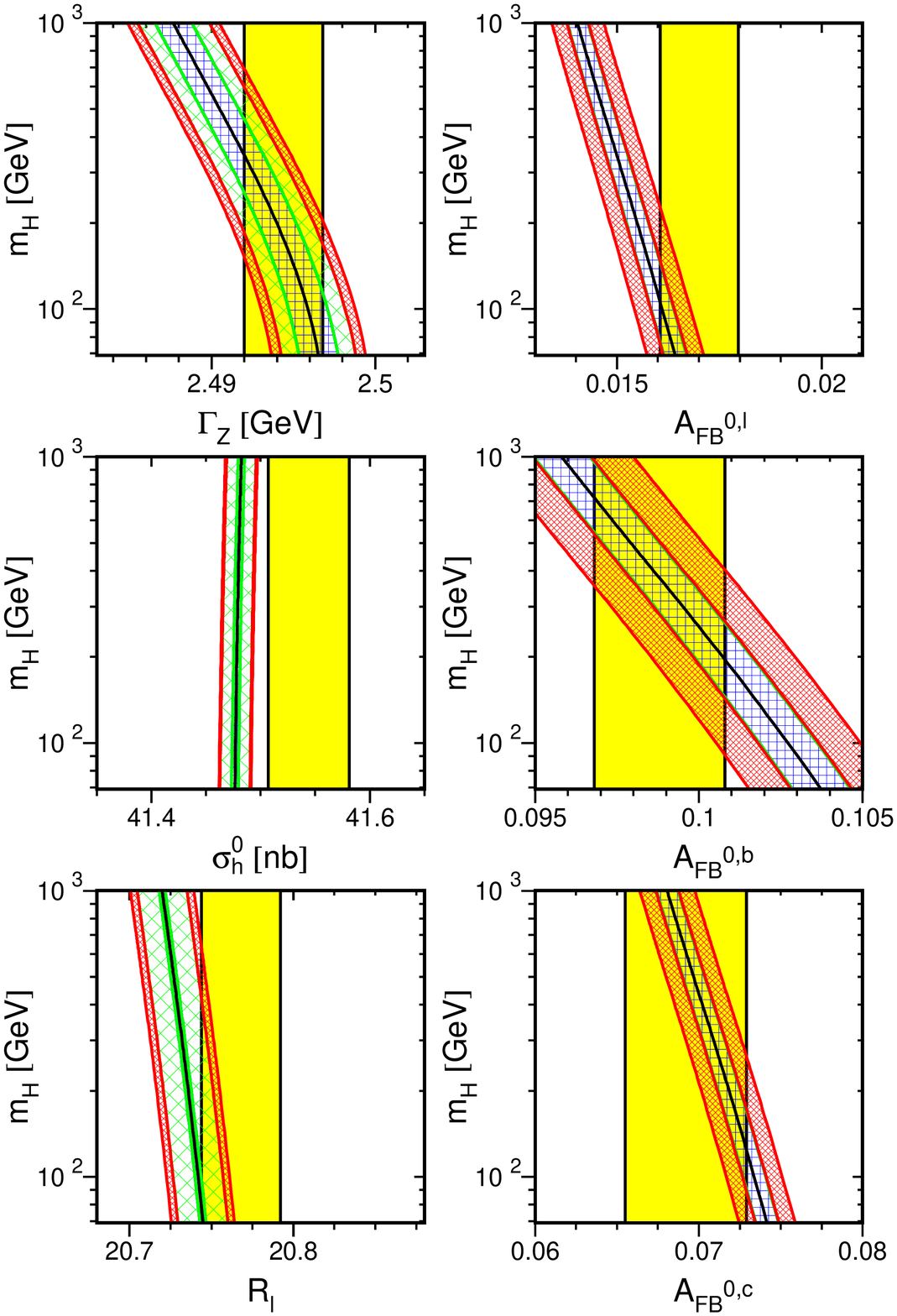}

\vspace{-0.5cm}
\hspace{1.8pt}
\includegraphics[width=13cm,height=9.8cm]{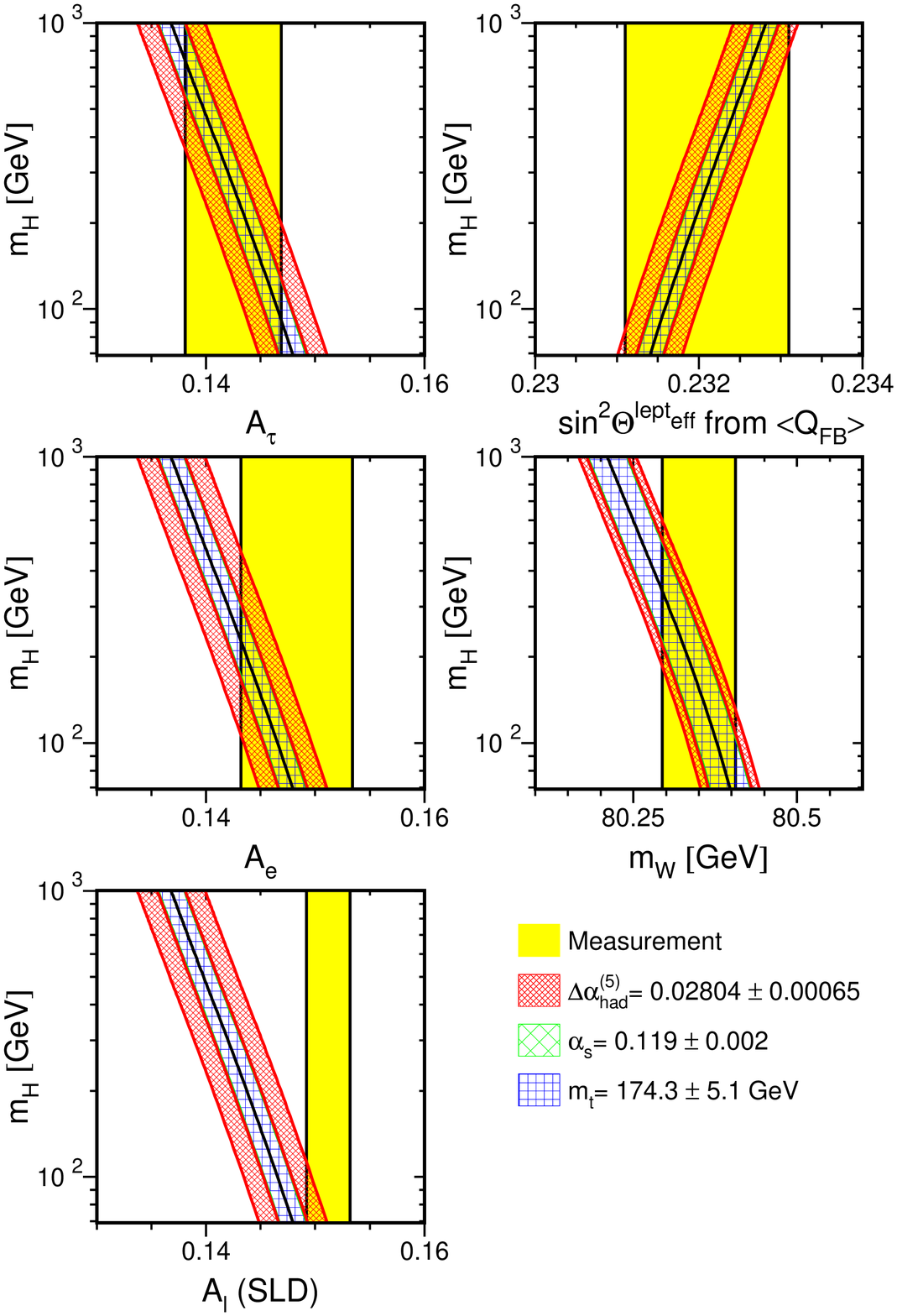}
\begin{center}
\begin{minipage}[h]{12cm}
\begin{center}
{\it Fig.\ 9: LEP measurements compared with the Standard Model
predictions, as a function of $M_H$ \cite{LepSite}.}
\end{center}
\end{minipage}
\end{center}

In this figure $\sigma_h^0$ is the hadronic cross section at the $Z$
pole, $R_\ell \equiv (\Gamma_{\text{had}}/\Gamma_\ell)$. $A_{FB}^{0\,
, \, f}$ is defined in (\ref{afb}) and $A_f$ in (\ref{af}).
$<Q_{FB}>$ is the average charge, which is related to the
forward--backward asymmetries by
\[
< Q_{FB} > = \sum_q \; \delta_q \; A_{FB}^q \;
\frac{ \Gamma_{q\bar{q}}}{\Gamma_{\text{hadr}}} \; ,
\]   
where $\delta_q$ is the average charge difference between the $q$ and
$\bar{q}$ hemispheres. For the sake of comparison $A_e$, extracted by
SLD from $A_{LR}$ (\ref{alr}), is also shown. We can see from Fig.\ 9
that dependence on the Higgs mass varying in the range 95 GeV $< M_H
<$ 1000 GeV is quite mild for all the observables, since the Higgs
effect enters only via $\log(M_H^2/M_Z^2)$ factors.

\begin{center}
\includegraphics[scale=0.45]{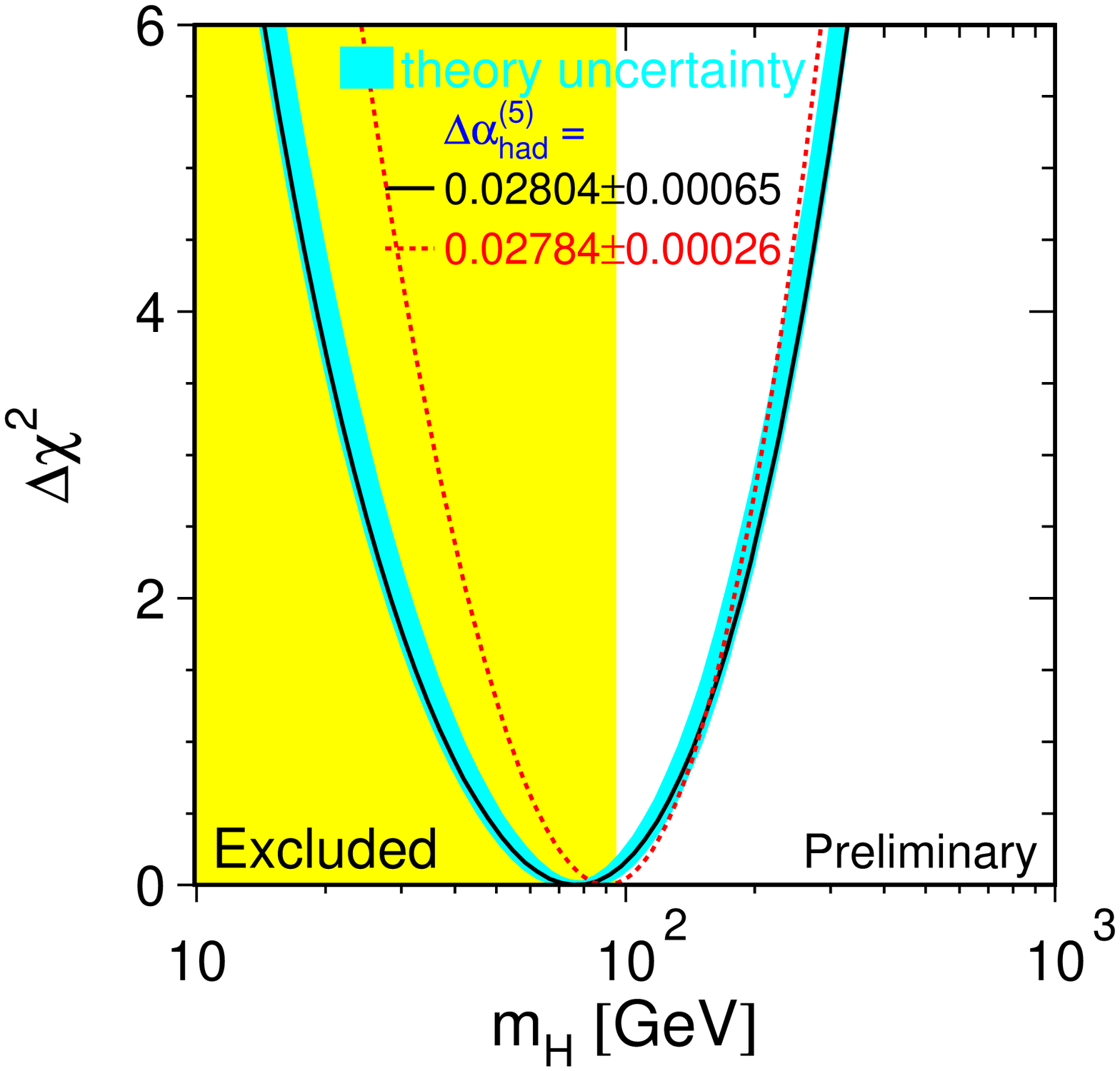}
\end{center}

\begin{center}
\begin{minipage}[h]{12cm}
\begin{center}
{\it Fig.\ 10: $\Delta \chi^2 \equiv \chi^2 - \chi^2_{\text{min}}$ as
a function of $M_H$ \cite{Lep:99,LepSite}.}
\end{center}
\end{minipage}
\end{center}

However we can extract information on $M_H$ from the global fit
including all data on the different observables. In Fig.\ 10 we show
the plot of $\Delta \chi^2 \equiv \chi^2 - \chi^2_{\text{min}}$
versus  $M_H$. The left vertical band represents the excluded region
due to the direct search for the Higgs ($M_H \gtrsim 95$ GeV). The
band represents an estimate of the theoretical error due to missing
higher order corrections. The global fit results in $M_H =
91^{+64}_{-41}$  GeV. 

\null
\vspace{-1cm}
\begin{center}
\includegraphics[scale=0.4]{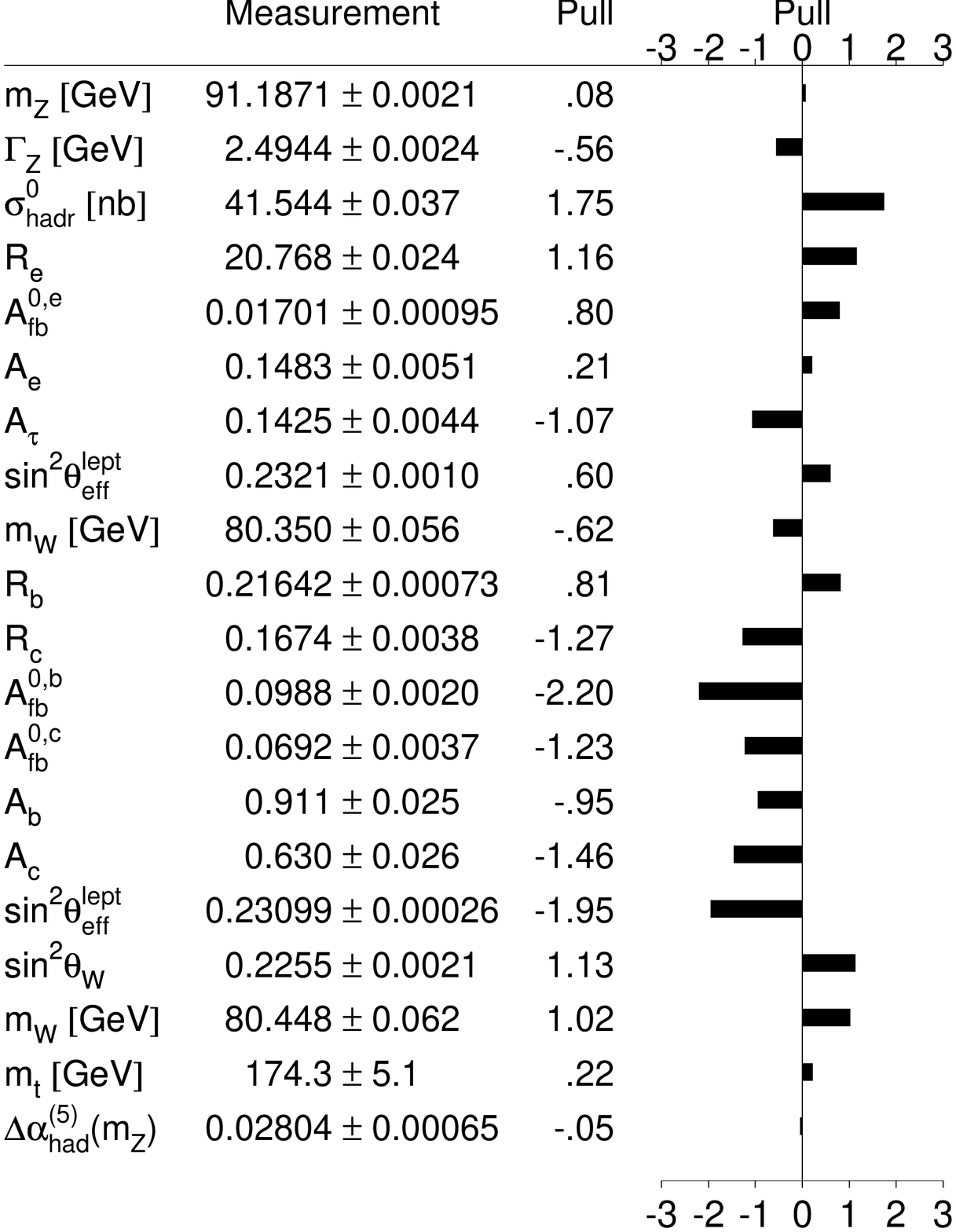}
\end{center}

\begin{minipage}[h]{12cm}
\begin{center}
{\it Fig.\ 11: Comparison of the precision electroweak measurements 
with the Standard Model predictions \cite{LepSite}.}
\end{center}
\end{minipage}

\vskip 0.5cm
As a final comparison, we present in Fig.\ 11 a list of several
electroweak observables. The experimental values are compared with
the Standard Model theoretical predictions. The Pull $\equiv
(O_{\text{meas}} - O_{\text{fit}})/{\sigma_{\text{meas}}}$, 
represents the number of standard deviations that separate the
central  values. This results show an impressive agreement with the
Standard Model expectations.


\chapter{~ The Higgs Boson Physics} \label{higgs:bos} \indent

\section{Introduction}\indent

The procedure of generating vector boson masses in a gauge invariant
way requires the introduction of a complex doublet of scalar fields
(\ref{phi}) which corresponds to four degrees of freedom. Three out
of these are ``eaten'' by the gauge bosons, $W^+$, $W^-$, and $Z^0$,
and become their longitudinal degree of freedom. Therefore, it
remains in the physical spectrum of the theory the combination
\[
\frac{(\phi^0 + \bar{\phi^0} )}{\sqrt{2}} = { H + v} \; ,
\]
where $v$ is given by (\ref{v}), and $H$ is the physical Higgs boson
field. 

The Higgs boson mass (\ref{mh}) can be written as
\begin{equation}
M_H = \sqrt{- 2 { \mu^2}} = \sqrt{2 { \lambda}} \; v =
\left(\frac{\sqrt{2}}{G_F}\right)^{1/2} \sqrt{{ \lambda}} \; .
\label{mass:higgs}
\end{equation}
Both Higgs potential parameters, $\mu^2$ and $\lambda$, are {\it a
priori} unknown --- just their ratio is fixed by the low energy data
[see (\ref{g:rel}) and (\ref{mw:mz})]. Therefore the Standard Model
does not provide any direct information on the Higgs boson mass.

The discovery of this particle is one of the challenges of the
high--energy colliders. This is the most important missing piece of the
Standard Model and its experimental verification could furnish very
important information on the spontaneous breaking of the electroweak
symmetry and on the mechanism for generating fermion masses. The
phenomenology of the Standard Model Higgs boson is covered in great
detail in reference \cite{Gunion:90}. Recent review articles include
Ref.\ \cite{Kniehl:94}, \cite{Djouadi:95}, \cite{Spira:98},
\cite{Quigg:99}. We intend to emphasize here the most relevant
properties of Higgs particle and make a brief summary of the prospects
for its search in the near future.  

\section{Higgs Boson Properties}

\subsubsection{The Higgs Couplings} \indent

The Higgs boson couples to all particles that get mass ($\propto v$)
through the spontaneous symmetry breaking of $SU(2)_L \otimes
U(1)_Y$. We collect in Table III the intensity of the coupling to
the different particles from (\ref{l:sca:1}),
(\ref{h:self}), (\ref{yuk:lep}), and (\ref{yuk:qua}), 
\begin{center}
\begin{tabular}{cc}
\hline\hline 
&\\[-0.2cm]
~~~ Coupling ~~~  & ~~~~~ Intensity ~~~ \\[0.2cm]
\hline 
&\\[-0.2cm]
$H f \bar{f}$  	&   $M_f/v$ 	\\
$H W^+ W^-$ 	&   $2 M_W^2/v$ \\
$H Z^0 Z^0$	&   $M_Z^2/v$ 	\\
$H H W^+ W^-$ 	&   $M_W^2/v^2$ \\
$H H  Z^0 Z^0$ 	&   $M_Z^2/2v^2$\\
$H H H$ 	&   $M_H^2/2v$ 	\\
$H H H H$  	&   $M_H^2/8v^2$\\[0.2cm]
\hline\hline 
\end{tabular}
\end{center}
\centerline{\it Table III: The Higgs coupling to different particles.}

\vskip 0.5cm
From the results of Table III it becomes evident that the Higgs
couples proportionally to the particle masses. Therefore we can
establish two general principles that should guide the search of the
Higgs boson: $(i)$ it will be produced in association with heavy
particles; $(ii)$ it will decay into the heaviest particles that
are accessible kinematically.

Besides the couplings presented in Table III, the Higgs can also
couple to $\gamma\gamma$ \cite{Ellis:76}, $Z\gamma$
\cite{Cahn:79,Barroso:86} and also to gluons
\cite{Georgi:78,Novaes:83}, at one loop level. The neutral and weak
interacting Higgs boson can interact with photons through loops of
charged particles that share the weak and electromagnetic
interactions: leptons, quarks and $W$ boson. In the same way the
Higgs couples indirectly with the gluons via loops of  (weak and
strong interacting) quarks. 

\subsubsection{Bounds on the Higgs Boson Mass} \indent

Since the Higgs boson mass is not predicted by the model we should
rely on some experimental and theoretical bounds to guide our future
searches. The most stringent lower bound was recently established by the LEP
Collaborations \cite{Higgs:99} and read
\[
M_H > 95.2 \; \mbox{GeV} \; . 
\]
at 95\% C.L..

It is also possible to obtain a theoretical lower bound on the Higgs
boson mass based on the stability of the Higgs potential when 
quantum corrections to the classical potential (\ref{hig:pot}) are
taken into account \cite{Sher:89}.  Requiring that the standard
electroweak minimum is stable ({\it i.e.} the vacuum is an absolute
minimum) up to the Planck scale, $\Lambda = 10^{19}$ GeV, the
following bound can be established \cite{Casas:95}:
\[
M_H \; \mbox {(in GeV)} > 133 + 1.92 (m_t - 175) - 
4.28 \left(\frac{\alpha_s - 0.12}{0.006} \right)\; .
\]
The behavior of $M_H$ as a function of the scale $\Lambda$ is given
in the lower curve of Fig.\ 13, for $m_t=175$ GeV and $\alpha_s =
0.118$. We see from this figure that, if a Higgs boson is discovered
with $M_H \simeq 100$ GeV, it would mean that the electroweak vacuum
is instable at $\Lambda \sim 10^5$ GeV \footnote{Reversing the
argument, since we live in a stable vacuum, this means that the
Standard Model must break down at this same scale.}.

There are also some theoretical upper bounds on the Higgs boson mass.
A bound can be obtained by requiring that unitarity is not violated in
the scattering of vector bosons \cite{Lee:77}. Let us take as an
example the $WW$ scattering represented in Fig.\ 12. 

\begin{figure}[ht]
\protect
\epsfxsize=10cm
\begin{center}
\mbox{\psfig{file=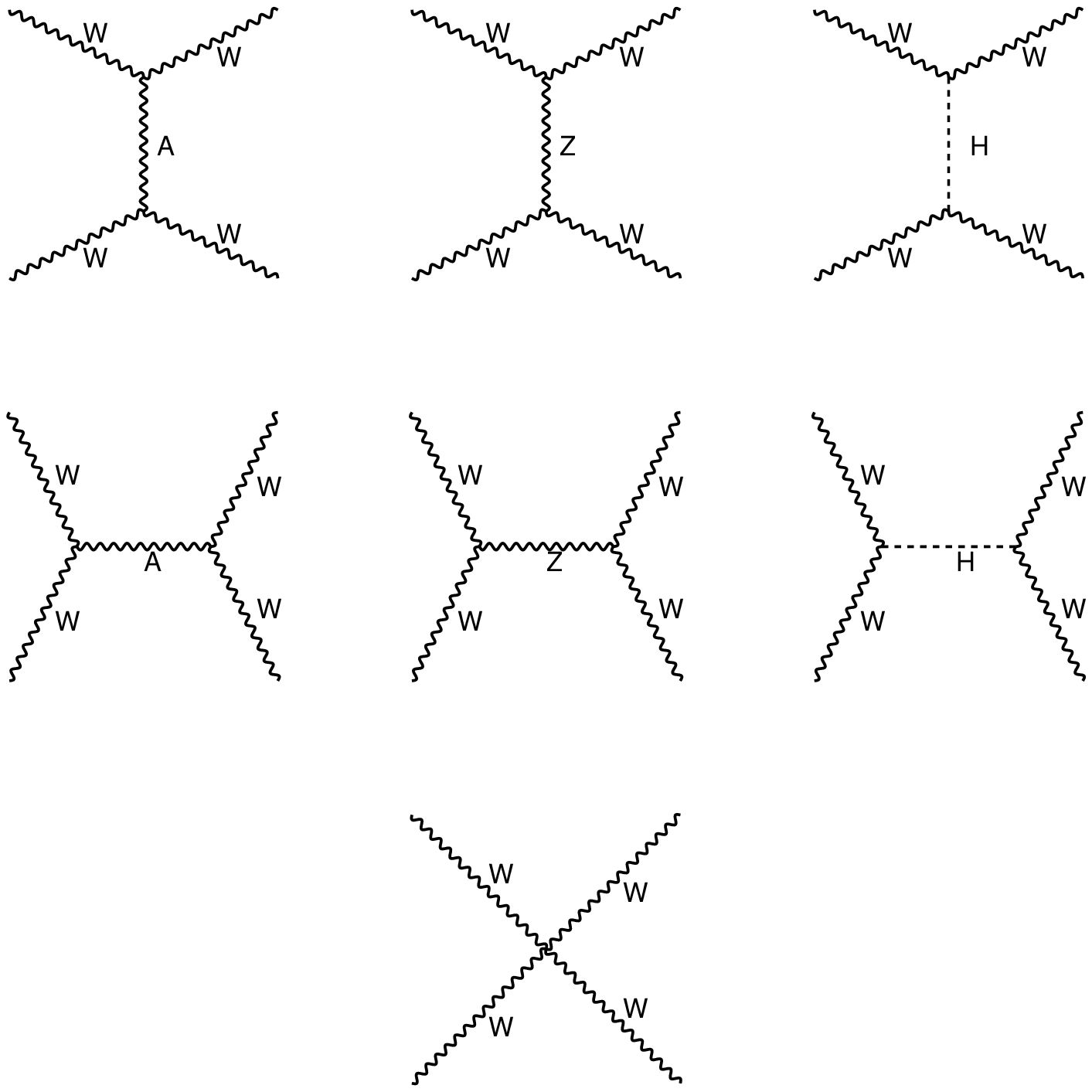,width=0.7\textwidth}}
\end{center}
\begin{center}
\begin{minipage}[h]{12cm}
\begin{center}
{\it Fig.\ 12: Feynman contributions to $W^+ W^- \to W^+ W^-$.}
\end{center}
\end{minipage}
\end{center}
\end{figure}

We should notice that if we exclude the Higgs boson contribution by
taking $M_H \to \infty$, we expect that the remaining amplitudes
would eventually violate unitarity, since the theory is not
renormalisable without the Higgs. Therefore, it is natural to expect
that the Higgs mass should play an important r\^ole in high energy
behaviour of the scattering amplitudes of longitudinally polarized
vector bosons. This is what happened for instance in the reaction
$e^+e^- \to W^+ W^-$ discussed in section \ref{remarks}.

A convenient way to estimate amplitudes involving longitudinal gau\-ge
bosons is through the use of the Goldstone Boson Equivalence
Theorem \cite{Lee:77,Chanowitz:85}.
This theorem states that at high energies, the amplitude ${\cal M}$
for emission or absorption of a longitudinally polarized gauge boson
is equal to the amplitude for emission or absorption of the
corresponding Goldstone boson, up to terms that fall like $1/E^2$,
{\it i.e.},
\begin{equation}
{\cal M}(W^\pm_L, Z^0_L) = {\cal M} (\omega^\pm, z^0) + 
{\cal O} \left( M_{W,Z}^2/E^2 \right) \; .
\label{equ:the}
\end{equation}

We can use an effective Lagrangian approach to describe the Goldstone
boson interactions. Starting from the Higgs doublet in terms of
$\omega^\pm$ and $z^0$,
\[
\Phi =
\frac{1}{\sqrt{2}}
                        \left( \ba{c}
                        i\sqrt{2} \omega^+\\
                        v + H - i z^0
                         \ea \right) \; ,
\]
we can write the Higgs potential as
\begin{eqnarray*} 
V(\Phi^\dagger\Phi) &=& \mu^2 \Phi^\dagger\Phi + \lambda (\Phi^\dagger\Phi)^2
\\
&=& \frac{1}{2} M_H^2 H^2 + \frac{g}{4} \frac{M_H^2}{M_W} H ( H^2
+ 2 \omega^+ \omega^- + z^{0^2}) \\
&& + \frac{g^2}{32} \frac{M_H^2}{M_W^2} ( H^2 + 2 \omega^+
\omega^- + z^{0^2})^2 \; .
\end{eqnarray*}

Therefore, with the aid of (\ref{equ:the}), the amplitude for $W^+_L
W^-_L  \to W^+_L  W^-_L$ is obtained as,
\begin{eqnarray*}
{\cal M} (W^+_L W^-_L  \to W^+_L  W^-_L) &\simeq&  
{\cal M} (\omega^+ \omega^- \to \omega^+ \omega^-) 
\\
&=& - i \; \frac{g^2}{4}
\frac{M_H^2}{M_W^2}  \left( 2 + \frac{M_H^2}{s - M_H^2} +
\frac{M_H^2}{t - M_H^2} \right) \; .
\end{eqnarray*}
and, at high energies, we have:
\[
{\cal M} (\omega^+ \omega^- \to \omega^+ \omega^-) \simeq 
- i \; \frac{g^2}{2} \frac{M_H^2}{M_W^2} = 
- i \; 2 \sqrt{2} G_F M_H^2 \; .
\]
Therefore, for $s$-wave, unitarity requires
\[
A_0 = \frac{1}{16\pi} |{\cal M}(\omega^+ \omega^- \to \omega^+
\omega^-)| = \frac{2 G_F}{8 \pi \sqrt{2}} M_H^2 < 1 \; .
\]

When this result is combined with the other possible channels ($z^0
z^0$, $z^0 h$, $hh$) it leads to the requirement that $\lambda \lesssim
8 \pi/3$ or, translated in terms of the Higgs mass,
\[
M_H \lesssim \left( \frac{8 \pi \sqrt{2}}{3 G_F} \right)^{1/2} \simeq
1 \, \mbox{TeV} \; .
\]

Another way of imposing a bound on the Higgs mass is provided by the
analysis of the triviality of the Higgs potential \cite{Sher:89}. The
renormalization group equation, at one loop, for the quartic coupling
$\lambda$ is
\[
\frac{d\lambda}{dt} = \frac{1}{16 \pi^2} \left(12 \lambda^2 \right) +
\; (\mbox{terms involving} \;  g, g^\prime, \mbox{Yukawa})  \; ,
\]
where $t=\log(Q^2/\mu^2)$. Therefore, for a pure $\phi^4$ potential,
{\it i.e.}, when the gauge and Yukawa couplings are neglected, we have
the solution
\[
\frac{1}{\lambda(\mu)} - \frac{1}{\lambda(Q)} = 
\frac{3}{4 \pi^2} \log\left(\frac{Q^2}{\mu^2}\right) \; .
\]

\begin{figure}[ht]
\protect
\epsfxsize=10cm
\begin{center}
\mbox{\psfig{file=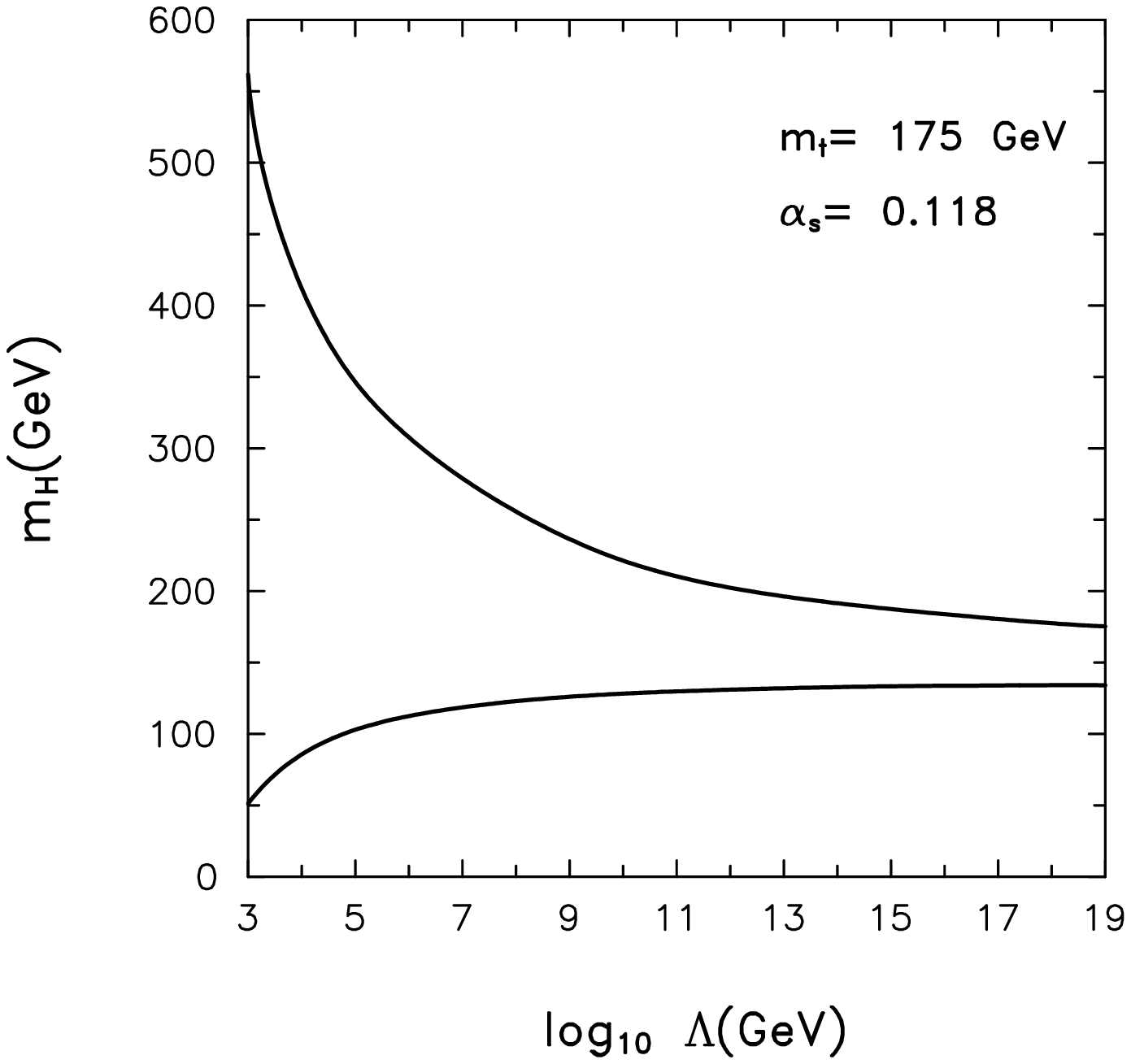,width=0.6\textwidth}}
\end{center}
\begin{center}
\begin{minipage}[h]{12cm}
\begin{center}
{\it Fig.\ 13: Perturbative and stability bound on $M_H$ as a
function of the scale $\Lambda$, from Ref.\ \cite{Casas:95}.}
\end{center}
\end{minipage}
\end{center}
\end{figure}

Since the stability of the Higgs potential requires that $\lambda(Q)
\geq  0$, we can write
\begin{equation}
\lambda(\mu) \leq \frac{4 \pi^2}{3\log(Q^2/\mu^2)} \; ,
\label{lambda:1}
\end{equation}
and, for large values of $Q^2$, we can see that $\lambda(\mu) \to 0$ and
the theory becomes trivial, that is, not interacting. The relation
(\ref{lambda:1}), can be written as
\[
Q^2 \leq \mu^2 \exp\left[\frac{4 \pi^2}{3 \lambda(\mu)} \right] \; .
\]

This result gives rise to a bound in the Higgs boson mass when we
consider the scale $\mu^2 = M_H^2$ and take into account 
(\ref{mass:higgs}),
\[
Q^2 \leq M_H^2 \exp\left[\frac{8 \pi^2 v^2}{3 M_H^2} \right]\; .
\]
Therefore, there is a maximum scale $Q^2 = \Lambda^2$, for a given
Higgs boson mass, up to where the Standard Model theory should be
valid.

In Fig.\ 13, we present the stability bound (lower curve) and the
triviality bound (upper curve) on the Higgs boson mass as a function
of the scale $\Lambda$. If we expect that the Standard Model is valid
up to a given scale --- let us say $\Lambda_{\text{GUT}} \sim
10^{16}$ GeV \cite{Ross:84} --- a bound on the Higgs mass should lie
between both curves, in this case $140$ GeV $\lesssim M_H \lesssim
180$ GeV.

\section{Production and Decay Modes}

\subsection{The Decay Modes of the Higgs Boson} \indent 

The possible decay modes of the Higgs boson are essentially determined
by the value of its mass. In Fig.\ 14 we present the Higgs branching
ratio for different $M_H$. 

\begin{figure}[ht]
\protect
\epsfxsize=10cm
\begin{center}
\mbox{\psfig{file=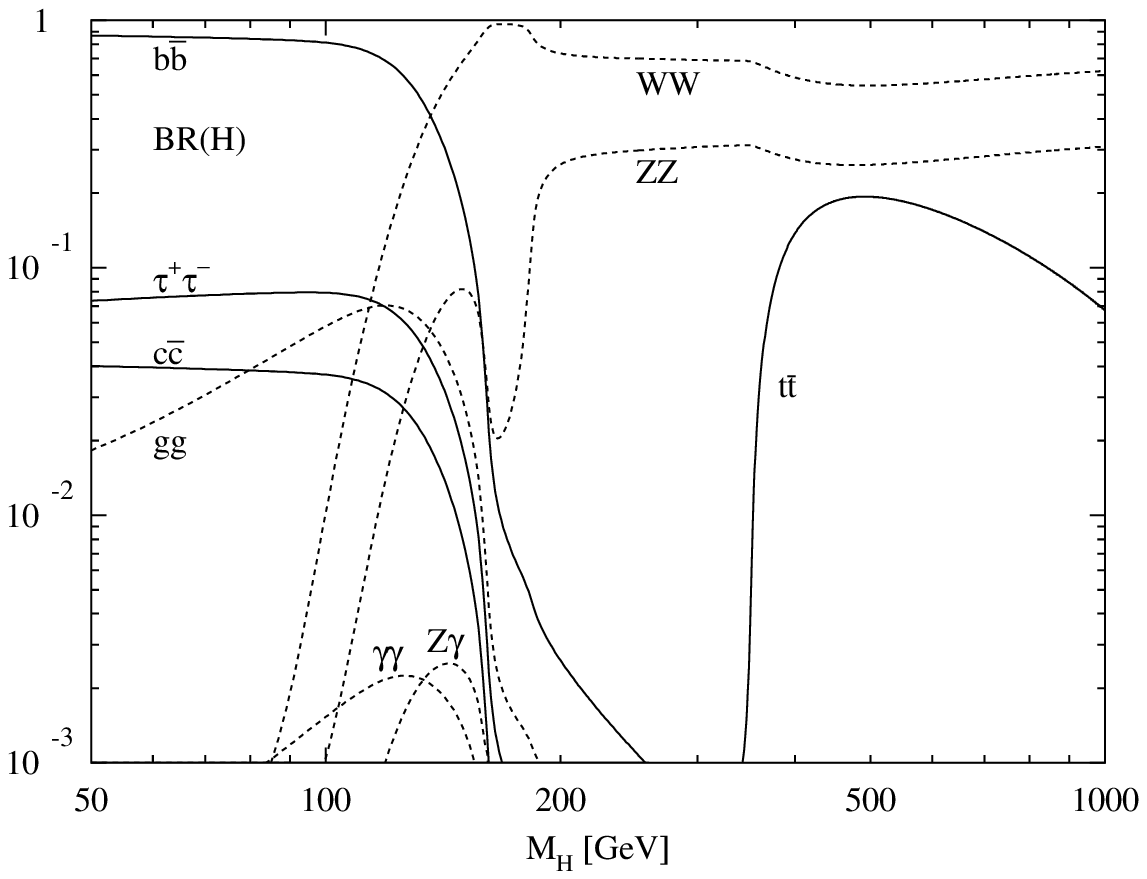,width=0.9\textwidth}}
\end{center}
\begin{center}
\begin{minipage}[h]{12cm}
\begin{center}
{\it Fig.\ 14: The branching ratios of the Higgs boson as a function of
its mass from Ref.\ \cite{Spira:98}.}
\end{center}
\end{minipage}
\end{center}
\end{figure}

When the Higgs boson mass lies in the range $95$ GeV $< M_H < 130$
GeV, the Higgs is quite narrow $\Gamma_H < 10$ MeV and the main
branching ratios come from the heaviest fermions that are accessible
kinematically:
\begin{eqnarray*}
BR(H \to b \bar{b}) &\sim &  90   \% \; , \\
BR(H \to c \bar{c}) &\simeq & BR(H \to \tau^+ \tau^-) \sim  5   \% \; .
\end{eqnarray*}
For $M_H \simeq 120$ GeV the gluon--gluon channel is important
giving  a contribution of $\sim 5 \%$ of the width. For a heavier
Higgs, {\it i.e.} $M_H > 130$ GeV, the vector boson channels $H \to V
V^\ast$, with $V = W$, $Z$, are dominant, 
\begin{eqnarray*}
BR(H \to W^+ W^-) & \sim & 65 \% \; ,\\
BR(H \to Z^0 Z^0) & \sim & 35 \% \; .
\end{eqnarray*}
For $M_H \simeq 500$ GeV the top quark pair production contributes
with $\sim \; 20 \%$ of the width. Note that the BR($H \to
\gamma\gamma$) is always small ${\cal O}(10^{-3})$. However, we can
think of some alternative models that give rise to larger
$H\gamma\gamma$ couplings (for a review see \cite{Concha:99} and
references therein).  For large values of $M_H$ the Higgs becomes a
very wide resonance: $\Gamma_H \sim [M_H \; \mbox{(in TeV)}]^3/2$.

\subsection{Production Mechanisms at Colliders}

\subsubsection{Electron--Positron Colliders}\indent

The Higgs boson can be produced in electron--positron collisions via
the Bjorken mechanism \cite{Bjorken:76} or vector boson fusion
\cite{Cahn:84},
\[
\ba{ll}
\mbox{($i$) Bjorken:} & ~~~ e^+ + e^- \to Z \to Z \; H \; , \\[0.2cm]
\mbox{($ii$) WW Fusion:}  & ~~~ e^+ + e^- \to \nu \bar{\nu}
(WW) \to \nu \bar{\nu} \;  H \; , \\[0.2cm]
\mbox{($iii$) ZZ Fusion:}		& ~~~ e^+ + e^- \to e^+ e^- (ZZ) \to 
e^+ e^- \;  H  \; .
\ea 
\]

At LEPI and II, where $\sqrt{s} \simeq M_Z$ or $2 \, M_W$ the Higgs
production is dominated by the Bjorken mechanism and they were able
to rule out from very small Higgs masses up to 95.2 GeV
\cite{Higgs:99}. Maybe, when the whole analysis is complete, they will
be able to rule out up to $M_H \sim 106$ GeV. 

At the future $e^+e^-$ accelerators,  like the Next Linear Collider
\cite{Nlc:Review}, where $\sqrt{s} = 500$ GeV, the production of a
Higgs with  $100 < M_H < 200$ GeV will be  dominated by the $WW$
fusion, since its cross section behaves like $\sigma \propto
\log(s/M_H^2)$ and therefore dominates at high energies. We can
expect around $2000$ events per year for an integrated luminosity of
${\cal L} \simeq 50$ fb$^{-1}$, and the Next Linear Collider should
be able to  explore up to $M_H \sim 350$ GeV.


\subsubsection{Hadron Colliders} \indent

At proton--(anti)proton collisions the Higgs boson can be produced via
the gluon fusion mechanism \cite{Georgi:78,Novaes:83}, through the vector
boson fusion and in association with a $W^\pm$ or a $Z^0$,

\[
\ba{ll}
\mbox{($i$) Gluon fusion:}		& 
~~~ p \bar{p} \to gg \to H \; , \\[0.2cm]
\mbox{($ii$) VV Fusion:}			& 
~~~ p \bar{p} \to VV \to H \; , \\[0.2cm]
\mbox{($iii$) Association with V:}		& 
~~~ p \bar{p} \to q\bar{q}^\prime \to V \; H \; .
\ea 
\]

At the Fermilab Tevatron \cite{Stange:94}, with $\sqrt{s} = 1.8$
($2$) TeV, the Higgs is better produced in association with vector
boson and they  look for the $V H (\to b \bar{b})$ signature. After the
improvement in the luminosity at TEV33 they will need ${\cal L} \sim 10$
fb$^{-1}$ to explore up to $M_H \sim 100$ GeV with $5 \, \sigma$.

At the CERN Large Hadron Collider \cite{Spira:95}, that will operate
with $\sqrt{s} = 14$ TeV, the dominant production mechanism is
through the gluon fusion  and the best signature will be $H \to ZZ
\to 4 \, \ell^\pm$ for $M_H > 130$ GeV. For  $M_H < 130$ GeV they
should rely on the small BR$(H \to \gamma\gamma) \sim 10^{-3}$. We
expect that the LHC can explore up to $M_H \sim 700$ GeV with an
integrated luminosity of ${\cal L} \sim 100$ fb$^{-1}$.

Once the Higgs boson is discovered it is important to establish with
precision several of its properties like mass, spin, parity and width.
The next step would be to search for processes involving multiple Higgs
production, like $VV \to HH$ or $gg \to HH$, which could give some
information on the Higgs self--coupling. 

\chapter{Closing Remarks} \indent

In the last 30 years, we have witnessed the striking success of a
gauge theory for the electroweak interactions. The Standard Model
made  some new and crucial predictions. The existence of a weak
neutral current and of intermediate vector bosons, with definite
relation between their masses, were confirmed by the experiments. 

Recently, a set of very precise tests were performed by Tevatron, LEP
and SLC colliders that were able to reach an accuracy of 0.1\% or
even better, in the measurement of the electroweak parameters. This
guarantees that even the quantum structure of the model was
successfully confronted with the experimental data. It was verified
that the $W$ and $Z$ couplings to leptons and quarks have exactly the
same values anticipated by the Standard Model. We already have some
strong hints that the triple--gauge--boson couplings respect the
structure prescribed by the $SU(2)_L \otimes U(1)_Y$ gauge symmetry.
The Higgs boson, remnant of the spontaneous symmetry breaking, has
not yet been discovered. However, important information, extracted
from the global fitting of data taking into account the loop effects
of the Higgs, assures that this particle is just around the corner.
The Higgs mass should be less than $\sim 260$ GeV at 95\% C.L., in
full agreement with the theoretical upper bounds for the Higgs mass.

These remarkable achievements let just a small room for the new
physics beyond the Standard Model. Nevertheless, we still have some
conceptual difficulties like the hierarchy problem, that may indicate
that the explanation provided by the Standard Model should not be
the end of the story. 

\linha

A series of alternative theories --- technicolour, grand unified
theory, supersymmetric extensions, superstrings, extra dimension
theories, etc --- have been proposed, but they all suffer from lack
of an experimental spark. Nevertheless, the physics beyond the
Standard Model is also beyond the scope of these lectures . . . 

\subsection*{Acknowledgments} \indent

We are grateful to M.\ C.\ Gonzalez--Garcia and T.\ L.\ Lungov for
the critical reading of the manuscript. This work was supported by
Conselho Nacional de Desenvolvimento Cien\-t\'{\i}\-fi\-co e
Tecnol\'ogico (CNPq), by Funda\c{c}\~ao de Amparo \`a Pesquisa do
Estado de S\~ao Paulo (FAPESP), by Programa de Apoio a N\'ucleos de
Excel\^encia (PRONEX), and by Funda\c{c}\~ao para o Desenvolvimento
da Universidade Estadual Paulista (FUNDUNESP).     



\end{document}